\documentclass[pra,aps,twocolumn,groupedaddress,nofootinbib,superscriptaddress]{revtex4-2}
\usepackage{graphicx}
\usepackage[nointegrals]{wasysym} %
\usepackage[export]{adjustbox}
\usepackage{amsmath,amsfonts,amssymb,latexsym}
\usepackage{hhline}
\usepackage{bm}
\usepackage{verbatim}
\usepackage{enumitem}
\usepackage{mathrsfs}
\usepackage{slashed}
\usepackage{empheq}
\usepackage{tabularx} 

\renewcommand{\Re}{{\rm \, Re\,}}
\renewcommand{\Im}{{\rm Im}}

\newcommand{\Tr}{\text{Tr}}

\newcommand{\wh}{\widehat}

\newcommand{\lan}{\langle}
\newcommand{\ran}{\rangle}

\newcommand{\unit}{\mathbf{1}}

\newcommand{\da}{{\dagger}}
\newcommand{\doa}{\downarrow}
\newcommand{\upa}{\uparrow}

\newcommand{\ra}{\rightarrow}

\newcommand{\wt}{\widetilde}

\newcommand{\uvx}{{\mathbf{\hat x}}}

\newcommand{\uvz}{{\mathbf{\hat z}}}
\newcommand{\uvth}{{\boldsymbol{\hat\theta}}}

\renewcommand{\(}{\left(}
\renewcommand{\)}{\right)}
\renewcommand{\[}{\left[}
\renewcommand{\]}{\right]}
\newcommand{\mt}{\mapsto}

\newcommand{\twp}{{2\pi}}
\newcommand{\fpi}{{4\pi}}

\newcommand{\D}{\nabla}

\newcommand\bpm            {\begin{pmatrix}}
	\newcommand\epm           {\end{pmatrix}}

\def\app#1#2{%
	\mathrel{%
		\setbox0=\hbox{$#1\sim$}%
		\setbox2=\hbox{%
			\rlap{\hbox{$#1\propto$}}%
			\lower1.1\ht0\box0%
		}%
		\raise0.25\ht2\box2%
	}%
}

\newcommand{\vp}{\varphi}
\newcommand{\vt}{\vartheta}
\renewcommand{\cp}{\Phi}
\newcommand{\ct}{\Theta}
\newcommand{\vs}{\varsigma}

\newcommand{\inv}{^{-1}}

\newcommand{\ope}\odot

\usepackage{manfnt}

\renewcommand{\prl}{{\, \parallel \,}}
\newcommand{\bi}{\begin{itemize}}
	\newcommand{\ei}{\end{itemize}}

\usepackage{amsthm}

\newtheorem{theorem}{Theorem}

\theoremstyle{definition}
\newtheorem{definition}{Definition}
\theoremstyle{definition}

\newcommand\bd            {\begin{definition}}
	\newcommand\ed            {\end{definition}}
\newcommand\bt            {\begin{theorem}}
	\newcommand\et            {\end{theorem}}
\newcommand\be            {\begin{equation}}
	\newcommand\ee            {\end{equation}}
\newcommand\ba            {\begin{aligned}}
	\newcommand\ea            {\end{aligned}}
\newcommand\bea{\begin{equation}\begin{aligned}}
		\newcommand\eea{\end{aligned}\end{equation}}

\usepackage{subcaption}

\usepackage{hyperref} 
\hypersetup{final}
\hypersetup{colorlinks, citecolor=red, linkcolor=red, urlcolor=red} 

\newcommand{\sss}{\subsubsection}
\renewcommand{\ss}{\subsection}

\renewcommand{\a}{\alpha}
\renewcommand{\b}{\beta}
\renewcommand{\d}{\delta}
\newcommand{\De}{\Delta}
\newcommand{\g}{\gamma}
\newcommand{\G}{\Gamma}
\newcommand{\s}{\sigma}
\renewcommand{\S}{\Sigma} 

\newcommand{\ep}{\varepsilon} %
\renewcommand{\l}{\lambda}
\renewcommand{\L}{\Lambda}
\renewcommand{\t}{\theta}

\renewcommand{\o}{\omega}

\renewcommand{\r}{\rho}

\newcommand{\U}{\Upsilon}

\newcommand{\z}{\zeta}

\newcommand{\bfth}{{\boldsymbol{\theta}}}

\newcommand{\bfeta}{{\boldsymbol{\eta}}}

\newcommand{\bfA}{\mathbf{A}}
\newcommand{\bfB}{\mathbf{B}}

\newcommand{\bfD}{\mathbf{D}}

\newcommand{\bfG}{\mathbf{G}}

\newcommand{\bfJ}{\mathbf{J}}
\newcommand{\bfK}{\mathbf{K}}

\newcommand{\bfd}{\mathbf{d}}

\newcommand{\bfk}{\mathbf{k}}

\newcommand{\bfn}{\mathbf{n}}

\newcommand{\bfq}{\mathbf{q}}

\newcommand{\bfs}{\mathbf{s}}

\newcommand{\bfv}{\mathbf{v}}

\newcommand{\rr}{\mathbb{R}}
\newcommand{\qq}{\qquad}

\newcommand{\zz}{\mathbb{Z}}

\newcommand{\mcf}{\mathcal{F}}

\newcommand{\mcd}{\mathcal{D}}

\newcommand{\mcg}{\mathcal{G}}

\newcommand{\mch}{\mathcal{H}}

\newcommand{\mcp}{\mathcal{P}}

\newcommand{\sfT}{\mathsf{T}}

\newcommand{\sft}{\mathsf{t}}

\usepackage[mathscr]{eucal} %

\newcommand{\scp}{\mathscr{P}}

\usepackage{braket}

\renewcommand{\k}{\ket}

\usepackage{dcolumn}
\usepackage{pifont}
\usepackage{ulem}

\newcommand{\pg}{D_6} %

\begin{document}
	
	\title{Pairing symmetry of twisted bilayer graphene: a phenomenological synthesis}
	
	\author{Ethan Lake, Adarsh S. Patri, and T. Senthil}
	\affiliation{Department of Physics, Massachusetts Institute of Technology, Cambridge, MA, 02139}
	
	\begin{abstract}
		
		One of the outstanding questions in the study of twisted bilayer graphene --- from both experimental and theoretical points of view --- is the nature of its superconducting phase. In this work we perform a comprehensive synthesis of existing experiments, and argue that experimental constraints are strong enough to allow the structure of the superconducting order parameter to be nearly uniquely determined. In particular, we argue that the order parameter is nodal, and is formed from an admixture of spin-singlet and spin-triplet Cooper pairs. This argument is made on phenomenological grounds, without committing to any particular microscopic model of the superconductor. Existing data is insufficient to determine the orbital parity of the order parameter, which could be either $p$-wave or $d$-wave. We propose a way in which the measurement of Andreev edge states can be used to distinguish between the two. 
		
	\end{abstract}
	
	\maketitle
	
	\section{Introduction }
	
	The discovery of strong correlation physics in magic-angle twisted bilayer graphene (TBG)  \cite{cao2018correlated,cao2018unconventional} and related systems \cite{yankowitz2019tuning,lu2019superconductors,chen2019evidence,sharpe2019emergent,chen2020tunable,serlin2020intrinsic,zondiner2020cascade,saito2020isospin,saito2021hofstadter,wu2021chern,chen2020electrically,cao2020nematicity,rozen2020entropic,zhou2021half,balents2020superconductivity,andrei2021marvels} has raised a number of fascinating questions. How should we think about the myriad correlated insulating states seen in these systems? What is the nature and mechanism of the superconductivity in TBG? What is the relationship, if any, between the superconductivity and the proximate correlated insulator? What is the physics of the strange metallic normal state observed in the vicinity of the correlated insulators in TBG? 

	In this paper we focus on the pairing symmetry of the prominent superconductivity seen in TBG at electron densities corresponding to a Moire superlattice filling $\nu$ between $-2$ and $-3$.  Why address the specific question of pairing symmetry? First, this question has a clear-cut answer unlike, say, the question of the mechanism of superconductivity, which may require a detailed understanding of the normal state and its instability. Second, as we argue below, as of fairly recently, the amount of accumulated experimental data  has become diverse and high-quality enough so as to allow the structure of the pairing in TBG to be very strongly constrained {(see table \ref{tab:expt_summary} for an overview)}. It is thus an opportune moment to discuss the pairing symmetry at a phenomenological level. Since the present state of microscopic theory is very much in flux --- with a theoretical proposal existing for essentially every imaginable pairing symmetry --- we view this experimentally-focused   approach, with some minimal theoretical input,  as a safer way to proceed. In this paper we will therefore  synthesize a variety of experimental observations in an attempt to pin down the pairing symmetry in TBG, without  committing to any detailed theory of the microscopic origin of the superconducting state.

	Our analysis leads us to a rather unusual pairing symmetry. 
	In particular, we will argue that the paired state features strong {\it spontaneously} generated spin-orbit coupling, where each of the graphene valleys contain only a {\it single} electron spin polarization, and with opposite valleys carrying opposite spins. Since in this scenario independent $SU(2)$ spin rotational symmetry is broken, the pairing --- which occurs between electrons in opposite valleys --- is neither spin singlet nor spin triplet, but an admixture of the two. As far as spin and valley are concerned, the pairing is thus similar to the `Ising superconductivity' which occurs in certain transition metal dichalcogenides (TMDs) \cite{lu2015evidence,xi2016ising,saito2016superconductivity}, in which the spin-valley locking is induced by strong spin-orbit coupling. One can thus say that TBG is in some sense a `spontaneous TMD'.  
	
	We will also argue that several distinct experimental factors point towards the order parameter being nodal.  The first indication comes from recent STM experiments \cite{oh2021evidence,kim2021spectroscopic}, which provide strong evidence of a nodal gap. We will corroborate this conclusion through a simple Ginzburg-Landau analysis showing that the independent observation of nematicity in the superconducting state \cite{cao2020nematicity,park2021magic} also implies a nodal gap. The orbital component of the order parameter can have either odd orbital parity ($p$-wave) or even orbital parity ($d$-wave). While existing experimental data does not seem to be sufficient to unambiguously distinguish between the two, we propose a concrete way of resolving this issue in future experiments by looking for Andreev bound states at normal-superconductor interfaces, or by performing a $c$-axis Josephson experiment between two rotated copies of TBG (see section \ref{sec:future}). We also point out that forming a $c$-axis Josephson junction between TBG and a {\it conventional} $s$-wave SC can provide an easy consistency check on our proposal: if TBG is nodal as suggested, the Josephson current in such a junction should be heavily suppressed (and will receive contributions only from $4e$ tunneling events).
	
	We note that superconductivity in TBG occurs at a variety of filling ranges punctuated by correlated insulators at integer $\nu$. In addition, evidence for superconductivity has been reported in some situations in which the correlated insulator is absent, either due to the twist angle being significantly less than the magic angle, or due to the effects of a proximate screening gate \cite{stepanov2020untying,saito2020independent,siriviboon2021abundance}. Our analysis will have little to say about these other situations which are not as well studied experimentally (although some further comments are provided in sec. \ref{sec:tuning_screening}). Thus we focus exclusively on the commonly observed prominent superconductor at fillings $-3< \nu < -2$ in devices where a well-developed  correlated insulator is present at $\nu = -2$. 
	In such devices the measured carrier density (as extracted through either Hall effect or Shubnikov-deHaas experiments) is determined by the deviation of $\nu$ from $-2$. The superconductivity then descends from a normal state with this small carrier density.  %
	All indications \cite{kim2021spectroscopic,park2021tunable,hao2021electric} are that this SC is fairly strongly coupled near $\nu=-2$, and becomes relatively weakly coupled when hole-doped to near $\nu=-3$ (both due to coherence-length measurements and due to the well-defined quantum oscillations that occur near $\nu=-3$). 

	We will freely draw on experiments on both TBG and magic-angle twisted trilayer graphene (TTG) (as well as four- and higher-layer devices \cite{park2021magic,zhang2021ascendance}). These systems are very closely related \cite{khalaf2019magic} and almost certainly possess the same pairing symmetry, meaning that (for the most part) we will not separate them in our analysis. 
	
	The remainder of this paper is structured as follows. In section \ref{sec:int} we discuss experimental constraints on the internal (spin and valley) structure of the pairing; it is here that we argue for the aforementioned spin-valley locking. In section \ref{sec:orb} we discuss the orbital structure of the pairing, and give several arguments for the existence of a nodal gap. Section \ref{sec:future} is devoted to a discussion of future experiments that could bolster or refute our proposal for the pairing, and we conclude in \ref{sec:disc}.

	\begin{table}
		\def\arraystretch{1.25}
		\begin{tabularx}{.5\textwidth}{l|X}
			\quad Pairing Constraint \quad &\quad Experimental Evidence\quad  \\
			\hline 
			Spins locked to valleys & Landau fan degeneracy \cite{cao2018correlated,yankowitz2019tuning}, Dirac revivals  \cite{zondiner2020cascade,wong2020cascade}, isospin ordering quenched by $B_{\parallel}$ \cite{rozen2020entropic,saito2020isospin}\\ 
			Not pure spin singlet & Anisotropic $B_{c,\parallel}$ \cite{cao2020nematicity}, strong Pauli violation in TTG \cite{inplane_fields,kim2021spectroscopic,liu2022isospin} \\ 
			Not pure $\k{\upa\upa}$ / $\k{\doa\doa}$ triplet & $T_c,I_c$ decrease monotonically with $B$ \cite{cao2018unconventional}, large $I_c$ \cite{hao2021electric} \\ 
			Not $\k{\upa\doa} + \k{\doa\upa}$ triplet & Incompatible with spin-valley locking, SC enhanced by SOC \cite{arora2020superconductivity}, positive intervalley Hunds coupling $J_H >0$ \cite{morissette2022electron}  \\ 
			Nodal gap & STM studies \cite{oh2021evidence,kim2021spectroscopic}, anisotropic $B_{c,\parallel}$ \cite{cao2020nematicity}			
		\end{tabularx}
		\caption{A summary of select experimental observations and their role in constraining the pairing symmetry for the SC near $\nu=-2$.}
		\label{tab:expt_summary}
	\end{table}

	\section{Internal structure: spin and valley}\label{sec:int} 

	We will write the order parameter matrix of the SC as $[\wh \De_\bfk]_{\a\b}$,
	where the indices $\a,\b$ run over spin and valley, and where $\wh\De^T_{-\bfk} = - \wh\De_\bfk$ by Fermi statistics. We will argue that the combination of several key observations from experiment allow the matrix structure of $\wh\De_\bfk$ to be uniquely determined. 
	
	Early observations of  %
	a critical in-plane field $B_{c,\prl}$ in TBG  approximately equal to its Pauli-limiting value \cite{cao2018unconventional} naturally suggested spin singlet pairing.  However, strong evidence against this assumption has subsequently emerged. $B_{c,\prl}$ is actually strongly angle-dependent \cite{cao2020nematicity}. Such an anisotropic critical field cannot be produced by Zeeman effects of the in-plane field alone, as spin-orbit coupling is negligible in TBG. In fact, an in-plane field also has a significant orbital coupling in TBG owing to the large moire lattice constant (so that the magnetic flux through an inter-layer  unit cell  is not small). Indeed a simple estimate \cite{cao2020nematicity} shows that the strength of this orbital coupling is comparable to the Zeeman coupling. Furthermore, this orbital effect is pair-breaking, as it leads to a difference in dispersion between single particle states related by time reversal.  Thus the value of $B_{c,\prl}$ in TBG does not offer a clear-cut probe of the spin structure of the pairing. 
	
	Crucial input comes from comparing with a different system, namely alternately twisted trilayer graphene (TTG) near its magic angle, which has also been shown to have robust superconductivity for $-3 < \nu < -2$ \cite{hao2021electric,park2021tunable}. The essential physics of the flat bands of TBG and TTG can reasonably be expected to be similar, however, TTG has a mirror symmetry which simplifies some aspects of its physics \cite{khalaf2019magic}. The band structure of TTG consists of a mirror-even flat band sector which is essentially the same as in TBG, and a mirror-odd sector with a dispersing Dirac cone (for each valley) that is essentially the same as that in familiar monolayer graphene.  Thus we may say that TTG = TBG + MLG at the band structure level. These two sectors will be coupled together by the interactions, and by perturbations that break the mirror symmetry. Nevertheless we may hope to obtain important clues into the physics of TBG by studying TTG. In contrast to TBG, in TTG, the mirror symmetry ensures that in the presence of the in-plane field, the flat band dispersions of time reversal related single particle states are degenerate. Thus the orbital depairing effects of an in-plane field are expected to be strongly suppressed, so that the main effects of such a field occur through Zeeman coupling \cite{lake2021reentrant,qin2021plane}. Remarkably, the SC in TTG was shown \cite{inplane_fields} to strongly violate the Pauli limit in the entire range of doping between $\nu = -2$ and $\nu = -3$. Crucially, this violation was seen even near $\nu=-3$, where the SC is weakly coupled \cite{inplane_fields,hao2021electric,park2021tunable} (since the Pauli-limit violation is calculated assuming a BCS relation between $T_c$ and the gap, this violation is significant only in the weak coupling regime). This observation has been confirmed in subsequent studies on TTG \cite{kim2021spectroscopic,liu2022isospin}, and strong Pauli-limit violation has also been demonstrated in magic angle quadruple- and quintuple-layer devices, for which the story is similar to that of TTG \cite{park2021magic,ledwith2021tb}. On the face of the above evidence, we are thus lead to the conclusion that the superconductivity in TBG --- despite first appearances --- is actually not spin singlet.

	Given this, the natural next step is to examine order parameters with spin triplet pairing. First consider $S^z\neq 0$ triplet pairing $(\k{\upa\upa}$ or $\k{\doa\doa})$, in which the Cooper pairs have nonzero magnetization. We claim that this type of pairing is in conflict with several different experimental observations. First, the finite magnetization of the Cooper pairs would mean that $T_c$ and the critical current $I_c$ would {\it increase} in a small applied magnetic field (either in-plane or out of plane). Indeed, since the Cooper pair magnetization couples to the field via a term proportional $\bfB \cdot \Tr[\wh\De^\da \bfs \wh\De]$, the gain in $T_c$ and $I_c$ from this coupling is linear in $B$, which at small $B$ will always win out over the suppression of $T_c,I_c$ due to orbital effects (which enter at order $B^2$).  This is in direct tension with the fact that both $T_c$ and $I_c$ have been observed to decrease monotonically with small applied fields in every existing experiment.  
	
	A second factor supporting this claim comes from Ref. \cite{cornfeld2020spin}, where it was pointed out that in zero field, the critical current density $J_c$ of a $S^z\neq0$ triplet superconductor is bounded from above by the current density induced by a $4\pi$ phase winding across the sample: $J_c \leq 8\pi e \r_s / \hbar L_x$, where $\r_s$ is the superconducting phase stiffness and $L_x$ is the linear size of the sample in the direction along which the current is measured (this fact is due to the topology of the order parameter manifold being so as to not admit any well-defined vortices).
	This gives a critical current bounded from above by 
	\be I_c \leq I_\fpi = 8\pi e \r_s \a / \hbar \approx 530 \times  \a \frac{\r_s}{ 1\, {\rm K}}\,\,  {\rm nA},\ee 
	where $\a=L_y/L_x$ is the aspect ratio. As in \cite{cornfeld2020spin}, let us estimate $\r_s$ by way of $\r_s \approx 2T_{BKT}/\pi$, so that $I_\fpi \approx 330 \times T_{BKT}/(1\, {\rm K})$. If we take $T_{BKT} = 2$K and $\a\approx1/ 6$ to reflect the measurement in \cite{hao2021electric}, this gives a critical current of $I_\fpi \approx100\, {\rm nA}$, which is about a factor of 7 {\it smaller} than the current observed in \cite{hao2021electric}.\footnote{Further tests can be performed by investigating the dependence of $I_c$ on system size. If the critical current is set by $I_\fpi$, which depends only on the aspect ratio $\a$, $I_c$ should not scale with system size. In a more conventional scenario where the critical current density is size-independent, $I_c\propto L$.} 

	For these reasons, we will regard $S^z\neq 0$ triplet pairing as being ruled out by experiment. At the very least, any theories proposing such a pairing channel \cite{qin2021plane} will need to explain how to resolve the existing tension with the measurements of $T_c$ and $I_c$.

	Further input into the nature of the superconductor is provided by considering the normal state from which it is born. As already mentioned, at filling $-3<\nu<-2$, this normal state has a carrier density $\nu + 2$ per unit cell, {\it i.e}, only the excess doped holes of the correlated insulator state are mobile \cite{cao2018correlated}.  Crucially it has long been observed that the Landau fan that emanates from $\nu = -2$ in the hole doped side has a 2-fold degeneracy \cite{cao2018correlated,yankowitz2019tuning}, in contrast to the natural 4-fold degeneracy expected due to the presence of 4 flavor degrees of freedom (2 spin and 2 valley). This observation is naturally explained if there is  flavor polarization already in the normal state such that the number of available electron flavors is reduced compared to charge neutrality by a factor of 2. This observation is also suggested by the series of `resets' in the chemical potential seen at integer fillings \cite{zondiner2020cascade,wong2020cascade}, indicating flavor polarization which survives up to large temperatures $T_{pol} \approx 25 K \gg T_c$. Theoretically zero-momentum flavor ordering (flavor ferromagnetism) occurs naturally in strong coupling treatments of TBG \cite{bultinck2020ground,kang2019strong,lian2021twisted} and related flat band systems \cite{zhang2019nearly,bultinck2020mechanism,zhang2019twisted,repellin2020ferromagnetism,liu2021theories}, and is encouraged by the band topology present in many of these systems. Such zero momentum ordering may also characterize the correlated insulator states of TBG, in contrast with the antiferromagnetism typical of Mott-Hubbard models of correlated insulators in systems with trivial band topology. 
	
	Thus we will assume that the normal metallic state from which the superconductivity descends has zero momentum flavor ordering that is responsible for the reduction of  the flavor degeneracy to 2. The precise direction of flavor ordering may be sensitive to details of different systems. Indeed we will not need to assume that the direction of any flavor ordering in the correlated insulator is necessarily the same as in the doped metallic state that obtains for $-3 < \nu < -2$. We note also that there is direct experimental support for flavor ferromagnetism in TBG aligned with a hexagonal Boron Nitride (hBN) substrate at $\nu = 3$ \cite{sharpe2019emergent,serlin2020intrinsic}.\footnote{Flavor polarization --- in the specific form of valley polarization --- is also seen in ABC trilayer graphene aligned with a hBN substrate \cite{chen2020tunable}, and leads to time reversal breaking ferromagnetic order.} In that system the experiments support valley polarization (so that one valley is occupied preferentially relative to other), leading to spontaneous breaking of time reversal symmetry. For TBG unaligned with hBN, which is the situation we are concerned with in this paper, we will discard the possibility of this kind of flavor ordering: time reversal breaking would have lead to hysteresis at a non-zero temperature, which is not seen (furthermore, such valley polarization is not easily compatible with superconductivity, which generally involves pairing of time-reversal related single particle states).

	Thus we are lead to consider pairing that takes place in a flavor-polarized state, in which the number of available electrons is reduced by half from the 4 degenerate flavors (2 spin, 2 valley) present near $\nu=0$. The SC cannot involve any additional flavor polarization on top of this, as additional polarization is ruled out by the STM measurements of \cite{oh2021evidence}, which find a zero-bias conductance $dI/dV(0)$ less than 50\% of the normal-state value at weak tunneling strengths (where $dI/dV$ probes the SC DOS), and a $dI/dV(0)$ of more than 150\% of the normal-state value at strong tunneling strenghts (where $dI/dV$ is dominated by Andreev reflection processes).\footnote{ This observation alone is not quite enough to rule out $S^z\neq0$ triplet pairing; see Sec. \ref{sec:stm}.} 
	Letting $\mcp$ denote the projector onto the polarized subspace, the pairing function must then satisfy 
	\be \label{deltaproj} \mcp \wh \De_\bfk =  \wh \De_\bfk \mcp^T = \wh\De_\bfk,\ee
	with ${\rm rank\, }[\mcp] = 2$. In the following we will write $\mcp$ in terms of valley-space Pauli matrices $\tau^\mu$ and spin-space Pauli matrices $s^\mu$. 
	
	One constraint which we view as being rather safe (both theoretically and experimentally) is that the pairing occurs between electronic states related by time reversal: thus we assume that the pairing is intervalley, and occurs at zero center-of-mass momentum. As already mentioned, this assumption rules out valley polarization (i.e. $\mcp = (\unit \pm \tau^z)/2$). This leaves us with the options of either polarizing spin, or else polarizing a linear combination of $K$ and $K'$ valleys (by taking $\mcp = |v\ran\lan v|$, with $|v\ran$ the $+1$ eigenvector of the matrix $\cos(\t)\tau^x + \sin(\t)\tau^y$). Distinguishing between these possibilities requires a few more pieces of experimental information. 
	
	Given that the pairing cannot be spin singlet, suppose first that the polarized degree of freedom is a linear combination of valleys, but that spin remains unpolarized. {In this case, the matrix structure of $\wh \De$ can be written as 
		\be \wh\De \propto |v^*\ran\lan v| \bfd \cdot \bfs \,is^y\ee  
		for some vector $\k v$ in valley space and some complex vector $\bfd$. Since $SU(2)$ spin symmetry is unbroken in this scenario, we can choose $\bfd = \uvz$ without loss of generality, giving $\wh\De \propto |v^*\ran\lan v| s^x$, so that the spins pair in the $S^z=0$ component of the triplet.  We claim however that such an order parameter is rather unlikely, for a variety of reasons. }
	
	One reason comes from the experiment of \cite{arora2020superconductivity}, which studied superconductivity in TBG placed on top of a layer of WSe$_2$. The presence of WSe$_2$ serves to induce SOC in TBG, both of Ising ($\l_I \tau^z s^z$) and Rashba ($\l_R (\tau^z \s^x s^y -\s^y s^x$)) type. While the strengths of $\l_I,\l_R$ are unknown, best estimates place both parameters at the level of a few mev and positive \cite{arora2020superconductivity}, which should be large enough to have an appreciable effect on the pairing. In particular, the Ising term acts to favor anti-parallel spin-valley locking, and would suppress the $\wh \De \propto |v^*\ran\lan v| s^x$ state currently under discussion as then $[\mcp,\tau^zs^z]\neq0$. However, \cite{arora2020superconductivity} found that the SC at $\nu=-2-\d$ was not suppressed in the presence of SOC, and indeed was even made more robust, surviving down to lower twist angles than in devices without the WSe$_2$ layer.\footnote{This includes devices with twist angles low enough such that the correlated insulator at $\nu=-2$ was absent. 
	} Another piece of evidence comes from \cite{rozen2020entropic,saito2020isospin}, which observed a large entropy present at most fillings away from charge neutrality, which was attributed to (soft) fluctuations in isospin order. This entropy was seen to be fairly strongly quenched by the application of an {\it in-plane} magnetic field, pointing to the isospin ordering as occurring in spin (rather than valley) space. Finally, while our aim in this work is to draw solely on existing experimental data, there is also a theoretical reason for disfavoring $S^z=0$ triplet pairing. This comes from the strong coupling analysis of TBG \cite{bultinck2020ground}, where it can be shown that Coulomb interactions favor that the flavor polarized at the highest energy scales be spin (either as a simple spin ferromagnet or anti-aligned in opposite valleys). 
	
	We are then led to consider polarizations involving spin. In this scenario, the low-energy electrons in each valley possess only a single spin flavor, and we may write the projector $\mcp$ as 
	\be \mcp  = (1+\tau^z)|\eta\ran\lan \eta| + (1-\tau^z) |\eta'\ran \lan \eta'|,\ee 
	where $\k\eta$ ($\k{\eta'}$) is the direction in which the spin is polarized in the $K$ ($K'$) valley. Since we are in a situation with spin-valley locking (SVL), the SC forms in an environment in which $SU(2)$ spin symmetry has been broken, and therefore it not need be either triplet or singlet. To make this explicit, we may write \eqref{deltaproj} as 	
	\bea \label{delta_d}\wh \De_\bfk & = 
	\frac{\De_\bfk}{\sqrt2} (i (\tau^+ -p\tau^-) \bfd \cdot \bfs + (\tau^++p\tau^-) d_0 ) is^y,\eea 
	where $p$ denotes the orbital parity (defined by $\De_\bfk = p \De_{-\bfk}$), and the 4-vector $d^\mu = (d^0,\bfd)$ is defined by $d^\mu \equiv \lan \eta'^*| is^y s^\mu | \eta\ran/\sqrt2$. In the following we will take $\k\eta, \k{\eta'}$ to be real, as their phases can simply be absorbed into that of $\De_\bfk$. 
	
	The magnitudes $|\bfd|^2$ and $|d^0|^2$ respectively determine the amount of triplet and singlet pairing present in $\wh \De_\bfk$, and are given in terms of $\k\eta, \k{\eta'}$ as 
	\be |\bfd|^2 = \frac14(3+\bfeta\cdot\bfeta'),\qq |d^0|^2 = \frac14(1-\bfeta \cdot \bfeta'),\ee 
	where $\bfeta = \lan \eta | \bfs|\eta\ran$ (and likewise for $\bfeta'$). Unless the spins in the two valleys are ferromagnetically aligned (in which case the pairing is pure triplet), both singlet and triplet components are present. 
	
	{Microscopically, the relative orientation of the spins in the two valleys is determined by a term $J_H \bfeta \cdot \bfeta'$, where $J_H$ is a parameter known as the intervalley Hunds coupling. Determining the sign of $J_H$ from first principles is difficult, as effects from phonons and Coulomb interactions push $J_H$ in opposite directions. We note however a recent electron spin resonance experiment \cite{morissette2022electron} which observed an {\it anti}-ferromagnetic $J_H > 0$. This favors anti-parallel alignment, and is consistent with the above arguments based on the phenomenology of the SC \cite{lake2021reentrant,khalaf2020symmetry}.}
	
	With parallel alignment ruled out, we conclude that the spins must be aligned in {\it anti-parallel} directions in the two valleys, as advocated for in \cite{lake2021reentrant} on the basis of experiments in TTG (and suggested as a possibility in \cite{christos2021correlated} and examined earlier in \cite{scheurer2020pairing}). 
	In an in-plane magnetic field, $\k\eta,\k{\eta'}$ slowly cant to point along the field direction, which they do without inducing any pair-breaking effects (for details see e.g. \cite{lake2021reentrant,scheurer2020pairing}). This allows the SC to survive in in-plane fields well in excess of the Pauli limit, explaining the observed Pauli limit violation  \cite{inplane_fields,liu2022isospin} in a manner similar to the (stronger) violation seen in monolayer TMD superconductors \cite{lu2015evidence,xi2016ising,saito2016superconductivity}. Anti-parallel alignment is also suggested by the aforementioned experiment \cite{arora2020superconductivity} showing that the SC was made {\it more} robust in the presence of a sizable Ising SOC term $\l_I s^z \tau^z$, which favors anti-parallel SVL (by time reversal symmetry, Rashba SOC also favors anti-parallel alignment).

	\section{Orbital component and nodal superconductivity} \label{sec:orb}  

	Having settled the matrix structure of $\wh \De_\bfk$, we now turn to constraining its $\bfk$ dependence. 
	
	One set of experiments which have immediate relevance for the form of $\De_\bfk$ are the STM studies performed on TBG \cite{oh2021evidence} and TTG \cite{kim2021spectroscopic}. These studies have shown strong spectroscopic evidence of a nodal (`V'-shaped) gap over most of the doping range in which the SC occurs, although there is evidence of a flatter `U'-shaped  gap near $\nu=-2$ in TTG (where the SC is most strongly coupled). Tunneling spectra have also been obtained using gate-defined tunneling junctions in a single TBG device \cite{de2021gate,rodan2021highly}, but these experiments were unable to unambiguously distinguish between a V-shaped gap and a U-shaped gap that had been `smeared' by finite $T$ and nonzero quasiparticle broadening effects.\footnote{A devil's advocate might try to explain the V-shaped gap in STM in the same way, although doing so requires unphysically large quasiparticle broadening $\G$ \cite{oh2021evidence}. Additionally, the fact that the spectra in TTG appear to be {\it more} V-shaped at {\it weaker} coupling \cite{kim2021spectroscopic} suggests that an $s$-wave gap with large $\G$ is rather unlikely.}

	Before discussing the nodal character of $\De_\bfk$ further, we should note that \cite{oh2021evidence,kim2021spectroscopic} found evidence for a very large optimal-doping single-particle gap of order $\De_{sp} \sim 1$ meV in TBG and $1.6$ meV in TTG  --- which persists well into the normal state --- and \cite{oh2021evidence} a comparatively smaller SC gap (as measured by Andreev reflection) of $\De_{sc} \sim 0.3$ meV. 
	There are various ways to interpret the large separation between $\De_{sp}$ and $\De_{sc}$. The point of view we will adopt in most of this paper (although it is not essential in much of our analysis) is that $\De_{sp}$ is due to the flavor polarization which we have argued must exist in the normal state. Indeed, in the vicinity of $\nu=-2$, and in the context of a simple Hartree-Fock treatment, flavor polarization can fix the chemical potential to be near the charge neutrality point of the Dirac cones present in the flavor-polarized subspace, therefore enabling a Dirac-like density of states to persist up to the scale of the flavor ordering, which we consequently take to be $\De_{sp}$. 

	Returning to the orbital character of the pairing, a further experiment which we will argue constrains $\De_\bfk$ is that of \cite{cao2020nematicity}, in which it was found that the SC phase is robustly nematic, with the in-plane critical field $\bfB_{\prl,c} = B_{\prl,c}(\cos\t_B,\sin\t_B)$ having a strong 2-fold anisotropy as a function of $\t_B$ (with this finding appearing again in Ref. \cite{park2021magic}).
	The results of \cite{cao2020nematicity} strongly suggest that the nematic director is weakly pinned by strain, and that in the absence of strain the system has an appreciable nematic susceptibility. Indeed, the nematic director varies {\it continuously} as a function of doping. This is expected if the nematic director is pinned by a combination of strain and a large nematic susceptibility, but is hard to understand in the absence of strain (as then there would only be 3 inequivalent choices for the director). Furthermore, the direction of nematicity in \cite{cao2020nematicity} is also changed upon heating and re-cooling the sample, indicating that the 
	strain plays the role of selecting out (i.e. pinning) a domain of the order parameter, rather than being the sole driving force behind the anisotropy.
	In the following we will argue that these observations are only consistent with a nodal order parameter (and {\it not} simply a $|\De_\bfk|$ which is an anisotropic, but everywhere nonzero, function of $\wh\bfk$). This lets us argue for the presence of nodes independently of STM, bolstering the case for their existence.

	To explore the consequences of nematicity in detail, we will perform a Ginzburg-Landau analysis of the SC in the presence of applied strain and in-plane magnetic fields (see the supplementary of \cite{cao2020nematicity} for a similar treatment). %

	We first remind the reader of the spatial symmetries present in the low-energy theory of TBG (see e.g. \cite{po2018origin}). The point group we will focus our attention on is $D_6$. \footnote{$D_6$ is isomorphic to $C_{6v}$ indicated in \cite{po2018origin}.} This group is generated by a sixfold rotation $C_3C_2$ and a reflection $C_{2y}$. These generators are conventionally taken to act on the band annihilation operators $c_\bfk$ as $C_3:c_\bfk\mt c_{R_{\twp/3}\bfk}$, $C_2: c_\bfk \mt \tau^x c_{-\bfk}$, and $C_{2y} : c_\bfk \mt c_{(k_x,-k_y)}$.
	However, since we are considering superconductors that form in an environment with SVL, the above forms of $C_2,C_{2y}$ are not symmetries in general.\footnote{Unless $\k\eta \propto \k{\eta'}$ so that the pairing is an $S^z\neq0$ triplet --- but we have already suggested that such a situation is unlikely.} Instead, the actions of $C_2,C_{2y}$ must be combined with spin rotations (due to the aforementioned spontaneously generated spin-orbit coupling) . For example, if we let $\k\eta = \k\upa, \k{\eta'} = \k\doa$, then $C_2, C_{2y}$ are modified by an addition action of $s^x$.
	
	We will assume that the normal state of the system is {\it not} nematic, and preserves the full $\pg$ rotational symmetry of the BM model. This is motivated in part by the fact that across the range of twist angles where SC is observed, only the SC (and not the normal state) was found to be {\it robustly} nematic \cite{cao2020nematicity}, suggesting that the analysis of the superconductor can be safely carried out in the setting of a $\pg$-symmetric normal state. Further comments to this effect will be given at the end of this subsection. 

	Assuming then that the normal state preserves $\pg$, the orbital part of the order parameter $\De_\bfk$ can be characterized in terms of the irreps of $\pg$. Each of these irreps in principle contain order parameters transforming as linear combinations of an infinite number of angular momentum channels $l$. All told there are six irreps: the one-dimensional $A_1,A_2$ ($l = 0 \mod  6$) and $B_1,B_2$ ($l=3 \mod 6$) irreps, and the two-dimensional $E_1$ ($l = 1,5 \mod 6$) and $E_2$ ($l = 2,4\mod 6$) irreps. Here $A_2$ ($B_1$) differs from $A_1$ ($B_2$) by the action of $C_{2y}$ differing by a factor of $-1$, while $C_{2y}$ is represented as a nontrivial matrix for $E_1$ and $E_2$).

	Since only one superconducting transition is observed, we will only consider pairing involving a single irrep. We will also only consider order parameters that contain only the single lowest angular momentum $l$ harmonic in their respective irrep (e.g. only $l=1$ for the $E_1$ representation). We will thus refer to the the $A_{1/2},B_{1/2},E_1,E_2$ irreps as $s_\pm,f_\pm,p,d$ pairing channels respectively, where the $\pm$ subscripts on $s,f$ refer to $C_{2y}$ eigenvalues. Keeping only the lowest harmonic simplifies things to some extent, and is also experimentally motivated: order parameters that combine more than one value of $l$ generically have a density of states which is non-analytic at multiple different nonzero frequencies, in contrast to the relatively distinct V-shaped gaps seen experimentally. 
	
	For pairing in the $p$ and $d$ wave irreps, we can decompose the function $\De_\bfk$ into chiral components as 
	\be \De_\bfk = \De^+_\bfk + \De^-_\bfk,\ee 
	where $\De^\pm_\bfk$ are parametrized in terms of a complex number $\De$ and two angles $\psi,\vp$:
	\be \De^+_\bfk = \cos(\psi) \De  e^{il(\t_\bfk +\vp)} \qq \De^-_\bfk = \sin(\psi) \De e^{-il(\t_\bfk + \vp)} \ee 
	Note that for these irreps, $\De_\bfk$ has nodes iff $\cos(\psi) = \pm \sin(\psi)$; for all other choices of $\psi$, $\De_\bfk$ is fully gapped (the magnitude of the gap is uniform in momentum space iff $\psi\in \frac\pi2 \zz$). 
	Under the action of the generating elements of $\pg$, rotation by an angle $\d$ about the $\hat{z}$-axis and the $C_{2y}$ reflection act respectively as 
	\bea  & R_\d \,:\, \vp \mt \vp + \d  \\ 
	&  C_{2y}\,:\, \psi\mt\psi+\pi/2, \, \, \vp \mt -\vp + \pi/2l,\,\, \De \mt i \De.\eea
	From a symmetry perspective, it is straightforward to enumerate the ways in which heterostrain and the applied in-plane field enter into the Landau-Ginzburg free energy at leading order. 
	For the $s_\pm$ and $f_\pm$ irreps, one only has the trivial couplings $ | \Delta | ^2 B^2$ and $| \Delta | ^2 (\ep_{xx} + \ep_{yy})$, where $\ep_{xx} + \ep_{yy}$ is the bulk-area expansion of the two-dimensional crystal. This follows simply from the fact that these irreps are one-dimensional. The fact that the $s_\pm, f_\pm$ irreps have no nematic couplings to the field and strain at leading order means that they are likely not compatible with the experimentally observed nematicity; hence we will focus on the $p$ and $d$ wave irreps in what follows. 

	For the $p$ and $d$ irreps, the form of the coupling follows from the tensor product $E_{1,2} \otimes E_{1,2} = A_1 \oplus E_2$.
	The produced product of $A_1$ is the isotropic coupling to the magnetic field and the bulk-area expansion strain tensor, just as with the $s_\pm,f_\pm$ irreps. The produced product of $E_2$ permits a coupling 
	to the magnetic field and strain. Defining the vectors
	\bea \bfD_B & \equiv (B_x^2 -B_y^2,2 B_xB_y) \equiv B^2(\cos(2\t_B),\sin(2\t_B)) \\ 
	\bfD_\ep &\equiv (\ep_{xx} - \ep_{yy},2\ep_{xy}) \equiv \ep (\cos(2\t_\ep),\sin(2\t_\ep)),\eea 
	the gauge-invariant couplings between the external fields and the order parameter are, to quadratic order in $\De$,  
	\bea 
	\frac12|\De|^2 \sin(2 \psi) \bfD_{B/\ep} \cdot (\cos(2l\varphi),-p \sin(2l\varphi)),\eea 
	where again $p = (-1)^l$ is the orbital parity. These terms are invariant under all $SO(2)$ rotations in the $p$-wave case, but only under $C_6$ rotations in the $d$-wave case. Thus in the $d$-wave case, any nematicity in the in-plane critical field owes its existence entirely to the fact that the rotational symmetry group of the SC is discrete, instead of continuous. 

	Equipped with this symmetry knowledge, to sixth order in $\De$ and lowest order in the strain and magnetic fields, the Landau-Ginzburg free energy $F$ for the gap function may be written as (see also \cite{cao2020nematicity})
	\bea \label{lgdelta} \, F & = \frac K2 |\De|^2 (\D \phi)^2+ r|\De|^2  \\
	& + \frac12 |\De|^2 \sin(2\psi)\( \ep \cos(2\t_\ep+2lp\vp) + B^2 \cos(2\t_B + 2lp\vp)\) \\ & +\frac g4 |\De|^4 \sin^2(2\psi) + \frac{u}2|\De|^4 \\ & + \frac w8 \sin^3(2\psi) |\De|^6 \cos(6l\vp)  + v(\cos^6(\psi) + \sin^6(\psi)) |\De|^6,   \eea 
	where $\De = |\De|e^{i\phi}$, and $r$ is a function of $B^2$ and $\ep_{xx} + \ep_{yy}$ which is negative at $B=0$. Note that here we have kept only gradients in $\phi$, since $|\De|,\psi,\vp$ will all be made massive by the quadratic and quartic couplings. 
	The term proportional to $g$ favors either a nodal order parameter ($| \sin(2\psi) | = 1$) if $g<0$, or a chiral uniformly gapped order parameter ($\sin(2\psi)=0$) if $g>0$. 
	Minimizing over $\psi$ at small $|\De|$, where the sextic terms can be neglected, gives 
	\be \sin(2\psi) = \begin{dcases} \frac{|\wt r|}{g|\De|^2} \quad & {\rm if} \, \, 0 < \frac{|\wt r|}{g|\De|^2} < 1 \\ 
		\pm 1 \quad & {\rm else} \end{dcases} \ee 
	where we have defined the field contribution to $r$ as 
	\be \wt r \equiv \ep \cos(2\t_\ep + 2lp\vp) + B^2 \cos(2\t_B +2lp\vp),\ee
	with $\vp$ always able to be chosen such that $\wt r <0$. 
	
	Note that a nodal order parameter ($|\sin(2\psi)| =1$) is always preferred if $g<0$, while it is preferred even if $g>0$, provided $|\wt r|/(2g|\De|^2)>1$ (which will always be satisfied close enough to $T_c$). On the other hand, if the order parameter is fully gapped ($|\sin(2\psi)|\neq1$), we must have $\sin(2\psi) = |\wt r|/g|\De|^2$. Upon inserting this into $F$, we find that in this case, the dependence of $F$ on $\wt r$, which contains the dependence of $F$ on $\t_\ep,\t_B$, is entirely contained in the term $- |\wt r|^2/2g$,
	which is in fact {\it independent} of $\De$. This means that provided the order parameter is fully gapped, the critical current and in-plane critical fields computed within Landau-Ginzburg will not be sensitive to $\t_\ep,\t_B$, since the value of $\De$ which minimizes $F$ will be independent of these quantities. Since the critical current and field {\it do} experimentally show dependence on at least $\t_B$, we conclude that the order parameter is very likely to be nodal, independently of information from STM studies. 
	
	This still leaves open the question of whether the pairing is $p$-wave or $d$-wave, which is more subtle to address and cannot be determined from the existence of nematicity alone within the above framework. We will have more to say about distinguishing $p$ and $d$ in section \ref{sec:future}. 
	
	We now briefly come back to our assumption that the order parameter transforms in an irrep of $\pg$. In the experiments of \cite{cao2020nematicity}, nematicity in the normal state was found only for devices very close to the magic angle, while nematicity in the SC was found over a wider range of twist angles. Furthermore, even in the devices with a nematic normal state, the nematicity was observed to be strongest only over a narrow doping range near $\nu=-2$. As in \cite{cao2020nematicity}, we interpret this as indicating that the normal state breaks $C_2$ in the absence of strain only for a certain narrow range of twist angles and doping, that strain merely weakly selects out a nematic direction and does not enter as a significant $C_2$-breaking field, and that the normal-state nematic order parameter is not directly related to the $\sin(2\psi)(\cos(2l\vp),-p\sin(2l\vp))$ nematic order parameter of the SC. 
	
	Finally, we should mention that several STM studies have observed nematicity in the normal state \cite{choi2019electronic,jiang2019charge,kerelsky2019maximized}. These works found evidence of a large electronic nematicity susceptibility near $\nu=0$, near $\nu=\pm1$ in Ref. \cite{jiang2019charge}, and near the correlated insulators in Ref.  \cite{kerelsky2019maximized}. This is consistent with the transport measurements of \cite{cao2020nematicity}, where nematicity in the $\nu<0$ normal state --- when it exists --- is strongest near $\nu=-2$, and weak / absent near $\nu=-3$. Since by all accounts there is only one superconducting phase for $-3<\nu<-2$, we view a treatment which uses unbroken $\pg$ symmetry to classify the order parameter as being legitimate, with nematicity in the normal state not playing an {\it essential} role in determining the pairing symmetry. 

	\section{Future experiments} \label{sec:future} 

	In this section we will discuss ways in which our proposal for the pairing symmetry can be investigated in future experiments, possible ways to distinguish $p$-wave from $d$-wave, and some more general features of the superconducting phase diagram.  Most of this discussion will be couched in the language of BCS theory, and as such is perhaps not a priori directly applicable for fillings close to $\nu=-2$, where the superconductor is strongly coupled. In appendix \ref{sec:strong} we will discuss an alternate approach which does not rely on BCS theory and which is more well-suited for the strong-coupling regime. In any case, assuming that the structure of the pairing (which is what we are interested in) is unchanged between $\nu=-2$ and $\nu=-3$, a weak-coupling analysis valid only near $\nu=-3$ is sufficient. Given the discussion of the previous section, we will assume throughout that $\De_\bfk$ as a function of angle $\t$ on the Fermi surface takes the form 
	\be \De_\t = \De e^{i\phi} \cos(l[\t + \g]),\ee 
	with $l=1,2$ ($p$-wave or $d$-wave), $\De$ real, and $\g$ some angular offset picked out by strain. 
	
	\begin{figure}
		\includegraphics[scale=1.2]{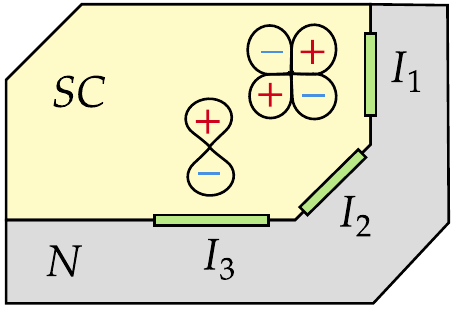}
		\caption{\label{fig:andreev} A schematic of the Andreev bound state measurement needed to distinguish between $p$ and $d$-wave pairing. The grey region marked $N$ is a normal metal, while the green bars $I_{i}$ ($i=1,2,3$) denote different interfaces at which the zero-bias tunneling conductance $G_{0,i}$ is measured. A comparison of the values of $G_{0,i}$ can distinguish between $p$-wave pairing (where all of the $G_{0,i}$ are distinct) and $d$-wave pairing (where $G_{0,1} = G_{0,3}$). For the orientation of the nodes as shown, the $d$-wave order parameter would exhibit large zero bias peaks at $I_{1}$ and $I_{3}$ and no peak at $I_2$, while the $p$-wave order parameter would exhibit no peak at $I_1$, a medium peak at $I_2$, and a large peak at $I_3$.  
		} 
	\end{figure} 
	
	\ss{Andreev bound states} 
	
	Andreev bound states, which occur at the edges of a superconductor, are a manifestation of the quantum interference effects of electron and hole quasiparticles scattering off the superconducting order parameter.
	These bound states manifest themselves through the appearance of a sharp peak in the zero-bias conductance $G_0$, as measured e.g. by the tunneling current between the superconductor and an adjacent normal metal. Crucially, for nodal gaps, the size of the peak in $G_0$ depends sensitively on the relative orientation between the nodes and the edge unit normal. This consequently allows a measurement of $G_0$ to distinguish between different orbital pairing symmetries \cite{hu1994midgap}, a phenomenon which has been studied very extensively in the context of $d$-wave pairing in the cuprates \cite{lofwander2001andreev}. In the present setting, a careful study of $G_0$ for various different edge geometries is able to distinguish between $p$ and $d$-wave pairing. 
	
	To understand this, consider a sample of superconducting TBG which is adjacent to a region of normal metal, as depicted in Fig. \ref{fig:andreev}. In the illustration of Fig. \ref{fig:andreev}, the superconducting and normal regions are drawn as being part of the same TBG device, with the boundary between the two being defined using electrostatic gates (as in the experiments of Refs.  \cite{de2021gate,rodan2021highly}). Note that in order to clearly resolve the features of the zero-bias conductance, it may be desirable to separate the superconducting and normal regions by thin insulating barriers, so that the current at the interface is in the tunneling regime.  Another possibility is to simply terminate the superconducting region of the sample with vacuum, and to measure $G_0$ at the interface using STM. 
	
	We will assume that the boundary between the SC and normal metal is broken into multiple well-defined interfaces $I_i$ $(I_{1,2,3}$ in Fig. \ref{fig:andreev}), with distinct unit normal vectors $\bfn_i$ at each interface. We require that each interface be relatively smooth at the Moire scale, so that to a good approximation, electron tunneling at $I_i$ conserves momentum in the direction orthogonal to $\bfn_i$. 
	The tunneling current at a given interface $I_i$ can be analyzed by an application of an extended version of the Blonder–Tinkham–Klapwijk tunneling theory \cite{stm_slab}. Adapting the approach proposed in Ref. \cite{Kashiwaya_spin_triplet_tunneling} and choosing coordinates such that $\bfn_i = \uvx$, electrons with momentum ($k_x, k_y$) are either: (i) normal reflected as electrons with momentum ($-k_x, k_y$), (ii) Andreev reflected as a hole with momentum ($k_x, k_y$), (iii) transmitted into the superconductor as an electron-quasiparticle with momentum ($k_x, k_y$), or (iv) transmitted into the superconductor as a hole-quasiparticle with momentum ($-k_x, k_y$).
	Since the electron and hole quasiparticles in the SC live at different momenta, they experience different pairing order parameters $\Delta_+ = \Delta(k_x, k_y)$ and $\Delta_- = \Delta(-k_x, k_y)$, respectively.
	For an $s$-wave order parameter, these potentials are identical. For a nodal order parameter however, $\De_+$ need not equal $\De_-$, and depending on the scattering geometry the two potentials may even have opposite signs. If indeed ${\rm sgn}(\De_+) = - {\rm sgn}(\De_-)$, quasiparticles scattering at the interface experience a $\pi$ phase shift, which results in the formation of edge-localized bound states; this in turn produces a peak in the zero-bias conductance $G_0$ \cite{stm_slab}. 
	More details are provided in appendix \ref{app_andreev_bound_states}.
	
	Let the angle between the first node and the interface normal $\bfn_i$ be denoted as $\vt_i$. Both $p$ and $d$-wave order parameters can produce peaks in $G_0$ for certain ranges of $\vt_i$. The key to distinguishing the two cases comes from the fact that $G_0$ is $\pi/2$-periodic in $\vt_i$ in the $d$-wave case, while it is only $\pi$-periodic in the $p$-wave case (as can be seen by finding the angles $\vt_i$ for which ${\rm sgn}(\De_+) = - {\rm sgn}(\De_-)$). This can be summarized by saying that as a function of $\vt_i$, we have 
	\be \label{g0_period} G_0(\vt_i) \propto \cos(2l \vt_i).\ee 
	
	Since the alignment of the nodes in TBG is presumably controlled by non-universal aspects like strain, there is no way to know a priori the value of $\vt_i$ (this is in contrast to e.g. the cuprates, where the orientation of the nodes is determined by the lattice). However, by measuring $G_0$ at multiple interfaces possessing different orientations $\bfn_i$, it is possible to sample a range of $\vt_i$ values, and to thereby distinguish between $p$ and $d$-wave pairing by examining the dependence of $G_0$ on interface angle. 
	As an example, consider the setup in Fig. \ref{fig:andreev}, where three different interfaces are formed at values of $\vt_i$ differing by $\pi/2$. For the interface $I_1$ and the orientations of the nodes as shown in the figure, the $d$-wave gap will display a strong peak in $G_0$, while the $p$-wave gap will display no peak. For interface $I_2$ the $d$-wave gap will display no peak, while the $p$-wave gap will display a small peak. Finally, for interface $I_3$, both the $p$ and $d$-wave gaps will display strong peaks. While these statements were made referencing the particular orientation of nodes drawn in Fig. \ref{fig:andreev}, the exact orientation is unimportant: the important thing is only to sample a number of interface orientations $\bfn_i$ which is large enough to allow one to measure the periodicity of $G_0$ with respect to the interface angle. Wtih this information, the angular momentum of the order parameter can then be read off from \eqref{g0_period}.

	\ss{Josephson experiments}
	
	Josephson experiments are a standard way of revealing phase-sensitive information about the pairing symmetry. 
	It turns out however that SVL and the extremely tiny size of the Moire BZ lead to complications that frustrate many attempts to use such experiments to distinguish between $p$ and $d$-wave gaps. 
	
	First consider a single Josephson junction formed by placing TBG and a conventional $s$-wave SC side-by-side. In other contexts, the properties of such Josephson junctions have been investigated in exhaustive detail in the literature \cite{rev_mod_phys_pairing_sym}, but in the present setting the presence of valley degrees of freedom and SVL provides a few novelties. 
	
	Let $\phi^S$ and $\phi^L$ denote the phases of the TBG SC and $s$-wave lead, respectively. For a junction with no appreciable intrinsic SOC and whose interface has unit normal an angle $\t_n$ from the $\uvx$ axis, we show in appendix \ref{app:josephson} that the free energy associated with $2e$ tunneling events at the junction takes the form 
	\be F_J \propto  \cos(\phi^S-\phi^L) \cos(2\t_{n}-(-1)^ll\g).\ee
	Note that the Josepshon current $I_{2e}$ is nonzero for both $p$ and $d$-wave pairing; this is made possible by the fact that the pairing is an admixture of both spin-singlet and spin-triplet. Furthermore, the dependence of $I_{2e}$ on $\t_n$ is the {\it same} for both pairing symmetries (although in the $p$-wave case, $I_{2e}$ vanishes in the limit of a circular Fermi surface, where the dispersions in the $K$ and $K'$ valleys are identical).  
	
	\begin{figure}
		\includegraphics[scale=1.2]{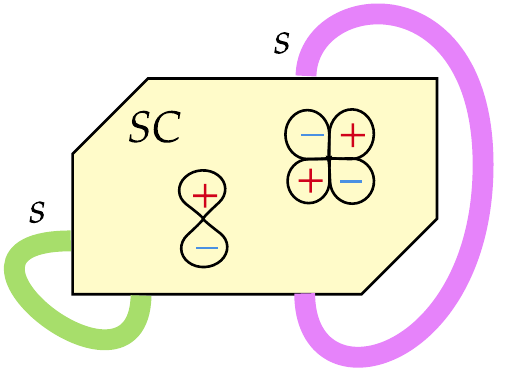}
		\caption{\label{fig:squid} A schematic of a squid experiment one might imagine using to distinguish different orbital pairing channels.	The green and purple lines represent conventional $s$-wave superconducting leads. A SQUID experiment using the green loop will be able to distinguish $s$ from $d$ or $p$ (but cannot differentiate $p$ from $d$), while one along the purple loop will generically fail to distinguish between any of $s,p,$ and $d$.} 
	\end{figure} 
	
	{The dependence of $F_J$ on $\t_n$ allows a SQUID-type experiment to provide further evidence against $s$-wave pairing. In such an experiment, one connects loops of conventional $s$-wave lead to the TBG device, forming two Josephson junctions with unit normals $\t_{n,1},\t_{n,2}$ (see the purple and green curves in Fig. \ref{fig:squid}). For $\t_{n,1}-\t_{n,2} = \pi/2$, the critical current as a function of a magnetic flux $\cp$ threaded through the junction will exhibit a maximum at $\cp=0$ if TBG has $s$-wave pairing, and a maximum at $\cp = \pm \pi$ if the pairing is either $p$ or $d$-wave.\footnote{ Of course given the 2d nature of the problem, it will not be possible to restrict the threaded flux to the junction region. This is not an issue however, as for the present purposes it is enough simply to determine whether or not the current is maximal at $\cp = 0$.}  since the $\t_n$ dependence of $F_J^l$ is the same for both $p$ and $d$-wave, such an experiment cannot generically be used to distinguish the two.
		
		Finally, we point out that such Josephson experiments have the ability to provide further evidence against pure triplet pairing. Indeed, an observed nonzero $2e$ Josephson current into an $s$-wave lead would rule out pure triplet pairing, with the Josephson current in that case being proportional to $\bfd \cdot \Tr[\bfs] = 0$.

		A different (and more difficult) phase-sensitive experiment for probing the orbital nature of the pairing is Josephson scanning tunneling microscopy \cite{kimura2009josephson,hamidian2016detection}. In such an experiment one measures the Josephson effect in a superconducting STM tip brought close to the superconducting sample. Since the Josephson current is sensitive to the angular momenta of both the tip and sample order parameters, performing this experiment with both $s$-wave and $d$-wave tips (using e.g. BSCCO for the latter \cite{hamidian2016detection}) offers the potential to distinguish between pairing channels in different angular momenta. 
		
		Unfortunately this technique is unlikely to be able to distinguish $p$ and $d$-wave pairing in TBG. To see this, note that after averaging over the tip and sample Fermi surfaces, a nonzero tunneling current $I_{2e}$ is only possible if the angular momenta $l_{tip}, l_{sample}$ of the tip and sample SCs are equal. As realistically we will have either $l_{tip} = 0$ or $l_{tip}= 2$, we will definitely have $I_{2e}=0$ if TBG has $p$-wave pairing. However, 
		we claim that $I_{2e}$ will vanish even in the case where TBG has $d$-wave pairing, and a $d$-wave STM tip is used. Indeed, due to the huge difference in the sizes of the tip and TBG Brillouin zones, TBG electrons tunneling into the tip will only tunnel into small regions near the projections of the monolayer $K,K'$ points. This means the tunneling current is in fact not sensitive to the full $d$-wave nature of the tip, with the tip effectively behaving as an $s$-wave gap. Therefore in all of the scenarios we have considered, $I_{2e}=0$. An observation of a sizable $I_{2e}\neq0$ in such a tunneling experiment would then point to an $s$-wave order parameter, and force us to re-examine our priors about the gap being nodal. 
		
		Lastly, one (rather ambitious) Josephson experiment which {\it could} unambiguously distinguish between $p$ and $d$-wave pairing would be to assemble a heterostructure consisting of {\it two} vertically-stacked TBG superconductors, with the top SC being able to be rotated relative to the bottom SC by an arbitrary angle $\vp$. In this case, the Josephson contribution to the free energy is 
		\be F_J^l(\vp) \propto  \De^t\De^b \cos(\phi^t-\phi^b) \cos(l[\g^t - \g^b +\vp])\ee
		where the order parameters on the top / bottom TBG layers are $\De^{t/b}_\t = \De^{t/b} e^{i\phi^{t/b}} \cos(l[\t + \g^{t/b} \pm \vp/2])$. 
		For $d$-wave pairing, $I_{2e}(\vp)$ would therefore vanish at four locations as $\vp$ is varied from $0$ to $\twp$, while for $p$-wave pairing $I_{2e}(\vp)$ would only vanish twice. Note that the ability to continuously rotate the relative angle between the two SCs is necessary, as $\g^t,\g^b$ are fixed by non-universal details, and as such the relative orientation between the nodes on the two layers is not known a priori.

		\ss{STM experiments} \label{sec:stm} %

		One possible extension of the existing STM experiments \cite{kim2021spectroscopic,oh2021evidence} --- which one might imagine would be capable of distinguishing $p$ and $d$-wave pairing --- is to perform STM in the presence of an in-plane magnetic field $\bfB_\prl$, and to examine the dependence of the tunneling signal on the orientation of $\bfB_\prl$. Interestingly however, in appendix \ref{app:stm}
		we calculate the tunneling conductance $dI/dV$ using Keldysh techniques and show that such information actually does {\it not} generically suffice to distinguish $p$ and $d$-wave pairing, both in the weak and strong (Andreev) tunneling regimes. 
		
		{Our calculation of $dI/dV$ is done both in the limit where the tunneling conserves momentum in the TBG Brillouin zone, and in the limit where the tunneling matrix element is completely independent of momentum. In the latter limit---which models the STM tip as a quantum dot---we find that the zero-bias conductance $dI/dV(0)$ in the strong-tunneling limit (appropriate for analyzing point-contact spectroscopy experiments) goes as 
			\be \frac{dI}{dV}(0) \propto | \int \frac{d\t}\twp m_\t\, {\rm sgn}(\De_\t)|^2,\ee
			where we have taken $\De_\t\in \rr$ without loss of generality, and where $m_\t$ is the effective mass in the $K$ valley, with $C_3$ symmetry imposing $m_\t = m_{\t+\twp/3}$. For nodal gaps where $\De_\t$ changes sign around the Fermi surface, the angular integral thus leads to a significant suppression of $dI/dV(0)$. Such a suppression is not present in the momentum-conserving limit, which yields results similar to those obtained within the BTK analysis of Ref. \cite{oh2021evidence}.  
			
			This suppression could very well be an explanation for the dips in $dI/dV(0)$ seen in some of the devices studied in Ref. \cite{oh2021evidence}. However, Ref. \cite{oh2021evidence} observes at least one device that exhibits a strong peak in $dI/dV(0)$ in the strong-tunneling limit. Whether this says something about the nature of the gap or is simply due to the tunneling being approximately momentum-conserving in that device is unclear at present.} 
		
		While we are unable to distinguish $p$ from $d$ using STM, STM experiments can still provide further insight into the internal (spin and valley) structure of the order parameter. 
		For example, a further experimental test to rule out $S^z \neq 0$ triplet pairing would be to perform spin-polarized STM: in such an experiment $S^z\neq0$ pairing would produce a nonzero Andreev conductance, while our proposed anti-parallel spin-valley locked state would not. 
		Note that Ref. \cite{oh2021evidence} argued that the observation of a sample with $G_0/G_N \approx 1.6$ (with $G_N$ the normal state conductance) already rules out $S^z\neq 0$ triplet pairing, since this ratio is impossible to achieve in a setting where the SC has a larger degree of flavor polarization than the normal state. 
		While the latter statement is correct, it cannot in general be used to argue against $S^z\neq0$ pairing, since it is possible for the normal state to itself be spin polarized.

		\ss{Tuning correlations and the evolution from a small to large Fermi surface} \label{sec:tuning_screening}

		So far we have mostly been focused on the physics of isolated TBG close to the magic angle. In this setting the Coulomb interactions between electrons are very strong, and much of the physics is controlled by the competition between these interactions and electron kinetic energy. It is then interesting to ask what would happen if one were able to gradually reduce the strength of interactions. In particular, it is natural to wonder about what happens to the SC as this occurs, given that the SC is formed out of an interaction-driven flavor-polarized state.
		Does the SVL nature of the pairing change as the interactions are weakened? Could there a phase transition as a function of interaction strength, where the orbital character of the pairing changes? What is the dependence of $T_c$ on the interaction strength? Understanding the answers to these kinds of questions could help us understand at a more fundamental level {\it why} the pairing symmetry in TBG is what it is, in a way which goes beyond the phenomenological analysis we have focused on in this paper.
		
		There have already been attempts to partially address these questions experimentally, where the effective Coulomb interaction in TBG is reduced by way of proximitizing TBG with various metallic states \cite{stepanov2020untying,saito2020independent,liu2021tuning}. Consider first the weak-screening limit, where the normal state is still flavor-polarized. In this limit the SC still forms out of a small flavor-polarized Fermi surface, whose area $A$ is determined by hole doping away from $-2$: at $\nu = -2 - \d$,  $A_{weak}\propto \d$. The normal state in this limit furthermore exhibits $T$-linear resistivity \cite{cao2020strange,jaoui2021quantum} and hints of a Hall angle varying as $\cot\t_H \sim T^2$ \cite{lyu2021strange}, just as in the strange metal phase of the cuprates. In \cite{liu2021tuning}, a small amount of screening was added, which had the effect of (very) slightly boosting $T_c$. While this measurement indicates that the properties of the SC carry a nontrivial dependence on $\ep$, the limitation on the strength of screening means that only a very small range of $\ep$ was able to be explored. 
		
		On the other hand, in \cite{stepanov2020untying,saito2020independent} it was  suggested that a `large' amount of screening was added (large $\ep$); in this case the SC was found to survive, while the normal state was argued to have been transformed into a conventional flavor-unpolarized Fermi liquid. It is still unclear to what extent these experiments provide a completely accurate picture of the strongly screened limit.
		However, it is nevertheless in principle possible to consider tuning to a regime in which the SC forms out of a `large' Fermi surface,  whose area $A_{large} \propto 2+\d$ is set by the doping away from charge neutrality.
		
		\begin{figure}
			\includegraphics{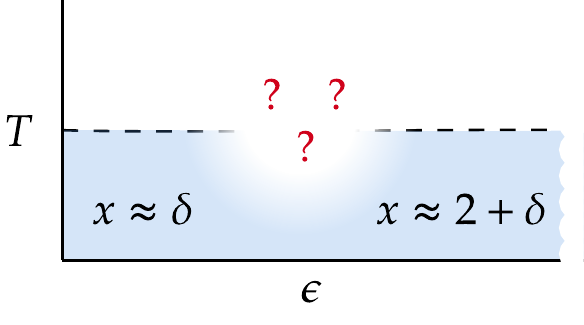}
			\caption{\label{fig:screening_fig} A schematic depiction of the evolution of the SC at filling $\nu = -2 - \d$ as a function of the dielectric constant $\ep$. In the unscreened pure system the SC is formed out of a density of doped holes $x \approx \d$, while for strong screening the SC it is $x\approx 2+\d$. How this evolution takes place at intermediate $\ep$, and whether or not $T_c$ remains nonzero at intermediate $\ep$ (the regions marked with ?s), is an interesting question we leave to the future.} 
		\end{figure}
		
		In this picture then, the system therefore evolves with increasing $\ep$ from a `small Fermi surface' SC formed out of a strange metal, to a `large Fermi surface' SC formed out of a Fermi liquid -- see figure \ref{fig:screening_fig} for an illustration. 
		This again immediately brings to mind comparisons with (certain classes of) cuprates, where as a function of doping $p$ a small Fermi surface SC (with area $p$) evolves into a large Fermi surface SC (with area $1+p$).
		We find it fascinating that certain cuprate compounds and TBG seem to be so qualitatively similar in this sense, given that microscopically the two materials could not be more different. 
		
		While in the cuprates the SC happens to evolve smoothly with $p$, in TBG there is no a priori reason why the superconductors in the two limits should have the same pairing symmetry. Indeed, if superconductivity in the strong-screening limit is driven by a conventional phonon-based mechanism, it would seem rather unlikely for the strongly-screened SC to possess a nodal gap, as we have argued is likely the case for the weakly-screened SC. We therefore believe that it would be interesting for future experiments to study the evolution of the SC with screening in more detail, and in particular to examine what happens in the intermediate-screening regime. 
		
		Finally, note that this experiment does not necessarily have to be done by adding a screening metallic layer; it should also be possible to use the twist angle $\t$ as a proxy for interaction strength, with the evolution between the strongly- and weakly-correlated SCs happening as $\t$ is reduced from the magic angle. Indeed, there are already preliminary hints \cite{siriviboon2021abundance} of an interesting evolution of $T_c$ with decreasing $\t$.  
		
		\section{Discussion } \label{sec:disc} 
		
		In this paper we have taken a phenomenological approach to understanding the pairing symmetry in twisted bilayer graphene. We have shown that by drawing on various pieces of evidence in the existing experimental literature --- and taking minimal input from theory --- we are led inexorably to a rather exotic scenario, wherein spontaneous spin-orbit coupling leads to an order parameter consisting of an admixture of spin singlet and spin triplet. While we argue the pairing is likely to be nodal, current experiments are unable to unambiguously distinguish between even ($d$-wave) and odd ($p$-wave) orbital parity. We have suggested studying interfacial Andreev bound states as a way of resolving this remaining issue. We have also argued that TBG will {\it not} display an appreciable $c$-axis Josephson current when proximitized with a conventional $s$-wave superconductor, a claim which can readily be tested in future experiments. 
		
		We should point out that our discussion has focused on the properties of the SC at $\nu = -2-\d$, and we have deliberately avoided directing much attention to the properties of the correlated insulator at $\nu=-2$. Assuming that the flavor polarization leading to anti-parallel SVL is present in the insulator does not uniquely fix the nature of the insulating gap, and there are various candidate orders whose competition is decided by comparatively delicate effects \cite{bultinck2020ground,khalaf2020charged} (it is even possible in some samples to obtain an insulating state with a $\pm 2 e^2/h$ quantized anomalous Hall effect at $\nu=-2$ \cite{tseng2022anomalous}). While it is natural to assume that the SC is obtained by doping holes into the correlated insulator (with this scenario apparently being born out in the STM study of \cite{kim2021spectroscopic}), this does not necessarily always need to be the case, and it is possible that different types of correlated insulators (likely all with the same type of flavor polarization) are present in different samples that superconduct at $\nu=-2-\d$. This possibility is suggested by the fact that in \cite{oh2021evidence} the gap was seen to possibly close between the SC and insulator, and by the Josephson study of \cite{diez2021magnetic}, which found indications of a correlated insulator that strongly breaks time reversal being present in a device which also superconducts.  All of this discussion goes to show that flavor polarization --- rather than the existence of a correlated insulator --- is the more fundamental phenomenon to take into account when trying to address the nature of the SC, and indeed only flavor polarization has played a role in shaping some aspects of our phenomenological analysis. This means that our conclusions about the order parameter should apply equally well in samples that exhibit superconductivity and flavor polarization, but possess no correlated insulator (of course, whether or not it is possible to create such a sample is a separate question).

		\section*{Acknowledgments} 
		
		We are grateful to Erez Berg, Yuan Cao, Pablo Jarillo-Herrero, and Jane Park for discussions. 	EL is supported by a Hertz Fellowship. TS was supported by US Department of Energy grant DE- SC0008739, and partially through a Simons Investigator Award from the Simons Foundation. This work was also partly supported by the Simons Collaboration on Ultra-Quantum Matter, which is a grant from the Simons Foundation (651446, TS).
		
		\bibliography{pairing_symmetry_bib}

\begin{thebibliography}{78}%
\makeatletter
\providecommand \@ifxundefined [1]{%
 \@ifx{#1\undefined}
}%
\providecommand \@ifnum [1]{%
 \ifnum #1\expandafter \@firstoftwo
 \else \expandafter \@secondoftwo
 \fi
}%
\providecommand \@ifx [1]{%
 \ifx #1\expandafter \@firstoftwo
 \else \expandafter \@secondoftwo
 \fi
}%
\providecommand \natexlab [1]{#1}%
\providecommand \enquote  [1]{``#1''}%
\providecommand \bibnamefont  [1]{#1}%
\providecommand \bibfnamefont [1]{#1}%
\providecommand \citenamefont [1]{#1}%
\providecommand \href@noop [0]{\@secondoftwo}%
\providecommand \href [0]{\begingroup \@sanitize@url \@href}%
\providecommand \@href[1]{\@@startlink{#1}\@@href}%
\providecommand \@@href[1]{\endgroup#1\@@endlink}%
\providecommand \@sanitize@url [0]{\catcode `\\12\catcode `\$12\catcode
  `\&12\catcode `\#12\catcode `\^12\catcode `\_12\catcode `\%12\relax}%
\providecommand \@@startlink[1]{}%
\providecommand \@@endlink[0]{}%
\providecommand \url  [0]{\begingroup\@sanitize@url \@url }%
\providecommand \@url [1]{\endgroup\@href {#1}{\urlprefix }}%
\providecommand \urlprefix  [0]{URL }%
\providecommand \Eprint [0]{\href }%
\providecommand \doibase [0]{https://doi.org/}%
\providecommand \selectlanguage [0]{\@gobble}%
\providecommand \bibinfo  [0]{\@secondoftwo}%
\providecommand \bibfield  [0]{\@secondoftwo}%
\providecommand \translation [1]{[#1]}%
\providecommand \BibitemOpen [0]{}%
\providecommand \bibitemStop [0]{}%
\providecommand \bibitemNoStop [0]{.\EOS\space}%
\providecommand \EOS [0]{\spacefactor3000\relax}%
\providecommand \BibitemShut  [1]{\csname bibitem#1\endcsname}%
\let\auto@bib@innerbib\@empty
\bibitem [{\citenamefont {Cao}\ \emph {et~al.}(2018{\natexlab{a}})\citenamefont
  {Cao}, \citenamefont {Fatemi}, \citenamefont {Demir}, \citenamefont {Fang},
  \citenamefont {Tomarken}, \citenamefont {Luo}, \citenamefont
  {Sanchez-Yamagishi}, \citenamefont {Watanabe}, \citenamefont {Taniguchi},
  \citenamefont {Kaxiras} \emph {et~al.}}]{cao2018correlated}%
  \BibitemOpen
  \bibfield  {author} {\bibinfo {author} {\bibfnamefont {Y.}~\bibnamefont
  {Cao}}, \bibinfo {author} {\bibfnamefont {V.}~\bibnamefont {Fatemi}},
  \bibinfo {author} {\bibfnamefont {A.}~\bibnamefont {Demir}}, \bibinfo
  {author} {\bibfnamefont {S.}~\bibnamefont {Fang}}, \bibinfo {author}
  {\bibfnamefont {S.~L.}\ \bibnamefont {Tomarken}}, \bibinfo {author}
  {\bibfnamefont {J.~Y.}\ \bibnamefont {Luo}}, \bibinfo {author} {\bibfnamefont
  {J.~D.}\ \bibnamefont {Sanchez-Yamagishi}}, \bibinfo {author} {\bibfnamefont
  {K.}~\bibnamefont {Watanabe}}, \bibinfo {author} {\bibfnamefont
  {T.}~\bibnamefont {Taniguchi}}, \bibinfo {author} {\bibfnamefont
  {E.}~\bibnamefont {Kaxiras}}, \emph {et~al.},\ }\bibfield  {title} {\bibinfo
  {title} {Correlated insulator behaviour at half-filling in magic-angle
  graphene superlattices},\ }\href@noop {} {\bibfield  {journal} {\bibinfo
  {journal} {Nature}\ }\textbf {\bibinfo {volume} {556}},\ \bibinfo {pages}
  {80} (\bibinfo {year} {2018}{\natexlab{a}})}\BibitemShut {NoStop}%
\bibitem [{\citenamefont {Cao}\ \emph {et~al.}(2018{\natexlab{b}})\citenamefont
  {Cao}, \citenamefont {Fatemi}, \citenamefont {Fang}, \citenamefont
  {Watanabe}, \citenamefont {Taniguchi}, \citenamefont {Kaxiras},\ and\
  \citenamefont {Jarillo-Herrero}}]{cao2018unconventional}%
  \BibitemOpen
  \bibfield  {author} {\bibinfo {author} {\bibfnamefont {Y.}~\bibnamefont
  {Cao}}, \bibinfo {author} {\bibfnamefont {V.}~\bibnamefont {Fatemi}},
  \bibinfo {author} {\bibfnamefont {S.}~\bibnamefont {Fang}}, \bibinfo {author}
  {\bibfnamefont {K.}~\bibnamefont {Watanabe}}, \bibinfo {author}
  {\bibfnamefont {T.}~\bibnamefont {Taniguchi}}, \bibinfo {author}
  {\bibfnamefont {E.}~\bibnamefont {Kaxiras}},\ and\ \bibinfo {author}
  {\bibfnamefont {P.}~\bibnamefont {Jarillo-Herrero}},\ }\bibfield  {title}
  {\bibinfo {title} {Unconventional superconductivity in magic-angle graphene
  superlattices},\ }\href@noop {} {\bibfield  {journal} {\bibinfo  {journal}
  {Nature}\ }\textbf {\bibinfo {volume} {556}},\ \bibinfo {pages} {43}
  (\bibinfo {year} {2018}{\natexlab{b}})}\BibitemShut {NoStop}%
\bibitem [{\citenamefont {Yankowitz}\ \emph {et~al.}(2019)\citenamefont
  {Yankowitz}, \citenamefont {Chen}, \citenamefont {Polshyn}, \citenamefont
  {Zhang}, \citenamefont {Watanabe}, \citenamefont {Taniguchi}, \citenamefont
  {Graf}, \citenamefont {Young},\ and\ \citenamefont
  {Dean}}]{yankowitz2019tuning}%
  \BibitemOpen
  \bibfield  {author} {\bibinfo {author} {\bibfnamefont {M.}~\bibnamefont
  {Yankowitz}}, \bibinfo {author} {\bibfnamefont {S.}~\bibnamefont {Chen}},
  \bibinfo {author} {\bibfnamefont {H.}~\bibnamefont {Polshyn}}, \bibinfo
  {author} {\bibfnamefont {Y.}~\bibnamefont {Zhang}}, \bibinfo {author}
  {\bibfnamefont {K.}~\bibnamefont {Watanabe}}, \bibinfo {author}
  {\bibfnamefont {T.}~\bibnamefont {Taniguchi}}, \bibinfo {author}
  {\bibfnamefont {D.}~\bibnamefont {Graf}}, \bibinfo {author} {\bibfnamefont
  {A.~F.}\ \bibnamefont {Young}},\ and\ \bibinfo {author} {\bibfnamefont
  {C.~R.}\ \bibnamefont {Dean}},\ }\bibfield  {title} {\bibinfo {title} {Tuning
  superconductivity in twisted bilayer graphene},\ }\href@noop {} {\bibfield
  {journal} {\bibinfo  {journal} {Science}\ }\textbf {\bibinfo {volume}
  {363}},\ \bibinfo {pages} {1059} (\bibinfo {year} {2019})}\BibitemShut
  {NoStop}%
\bibitem [{\citenamefont {Lu}\ \emph {et~al.}(2019)\citenamefont {Lu},
  \citenamefont {Stepanov}, \citenamefont {Yang}, \citenamefont {Xie},
  \citenamefont {Aamir}, \citenamefont {Das}, \citenamefont {Urgell},
  \citenamefont {Watanabe}, \citenamefont {Taniguchi}, \citenamefont {Zhang}
  \emph {et~al.}}]{lu2019superconductors}%
  \BibitemOpen
  \bibfield  {author} {\bibinfo {author} {\bibfnamefont {X.}~\bibnamefont
  {Lu}}, \bibinfo {author} {\bibfnamefont {P.}~\bibnamefont {Stepanov}},
  \bibinfo {author} {\bibfnamefont {W.}~\bibnamefont {Yang}}, \bibinfo {author}
  {\bibfnamefont {M.}~\bibnamefont {Xie}}, \bibinfo {author} {\bibfnamefont
  {M.~A.}\ \bibnamefont {Aamir}}, \bibinfo {author} {\bibfnamefont
  {I.}~\bibnamefont {Das}}, \bibinfo {author} {\bibfnamefont {C.}~\bibnamefont
  {Urgell}}, \bibinfo {author} {\bibfnamefont {K.}~\bibnamefont {Watanabe}},
  \bibinfo {author} {\bibfnamefont {T.}~\bibnamefont {Taniguchi}}, \bibinfo
  {author} {\bibfnamefont {G.}~\bibnamefont {Zhang}}, \emph {et~al.},\
  }\bibfield  {title} {\bibinfo {title} {Superconductors, orbital magnets and
  correlated states in magic-angle bilayer graphene},\ }\href@noop {}
  {\bibfield  {journal} {\bibinfo  {journal} {Nature}\ }\textbf {\bibinfo
  {volume} {574}},\ \bibinfo {pages} {653} (\bibinfo {year}
  {2019})}\BibitemShut {NoStop}%
\bibitem [{\citenamefont {Chen}\ \emph {et~al.}(2019)\citenamefont {Chen},
  \citenamefont {Jiang}, \citenamefont {Wu}, \citenamefont {Lyu}, \citenamefont
  {Li}, \citenamefont {Chittari}, \citenamefont {Watanabe}, \citenamefont
  {Taniguchi}, \citenamefont {Shi}, \citenamefont {Jung} \emph
  {et~al.}}]{chen2019evidence}%
  \BibitemOpen
  \bibfield  {author} {\bibinfo {author} {\bibfnamefont {G.}~\bibnamefont
  {Chen}}, \bibinfo {author} {\bibfnamefont {L.}~\bibnamefont {Jiang}},
  \bibinfo {author} {\bibfnamefont {S.}~\bibnamefont {Wu}}, \bibinfo {author}
  {\bibfnamefont {B.}~\bibnamefont {Lyu}}, \bibinfo {author} {\bibfnamefont
  {H.}~\bibnamefont {Li}}, \bibinfo {author} {\bibfnamefont {B.~L.}\
  \bibnamefont {Chittari}}, \bibinfo {author} {\bibfnamefont {K.}~\bibnamefont
  {Watanabe}}, \bibinfo {author} {\bibfnamefont {T.}~\bibnamefont {Taniguchi}},
  \bibinfo {author} {\bibfnamefont {Z.}~\bibnamefont {Shi}}, \bibinfo {author}
  {\bibfnamefont {J.}~\bibnamefont {Jung}}, \emph {et~al.},\ }\bibfield
  {title} {\bibinfo {title} {Evidence of a gate-tunable mott insulator in a
  trilayer graphene moir{\'e} superlattice},\ }\href@noop {} {\bibfield
  {journal} {\bibinfo  {journal} {Nature Physics}\ }\textbf {\bibinfo {volume}
  {15}},\ \bibinfo {pages} {237} (\bibinfo {year} {2019})}\BibitemShut
  {NoStop}%
\bibitem [{\citenamefont {Sharpe}\ \emph {et~al.}(2019)\citenamefont {Sharpe},
  \citenamefont {Fox}, \citenamefont {Barnard}, \citenamefont {Finney},
  \citenamefont {Watanabe}, \citenamefont {Taniguchi}, \citenamefont
  {Kastner},\ and\ \citenamefont {Goldhaber-Gordon}}]{sharpe2019emergent}%
  \BibitemOpen
  \bibfield  {author} {\bibinfo {author} {\bibfnamefont {A.~L.}\ \bibnamefont
  {Sharpe}}, \bibinfo {author} {\bibfnamefont {E.~J.}\ \bibnamefont {Fox}},
  \bibinfo {author} {\bibfnamefont {A.~W.}\ \bibnamefont {Barnard}}, \bibinfo
  {author} {\bibfnamefont {J.}~\bibnamefont {Finney}}, \bibinfo {author}
  {\bibfnamefont {K.}~\bibnamefont {Watanabe}}, \bibinfo {author}
  {\bibfnamefont {T.}~\bibnamefont {Taniguchi}}, \bibinfo {author}
  {\bibfnamefont {M.}~\bibnamefont {Kastner}},\ and\ \bibinfo {author}
  {\bibfnamefont {D.}~\bibnamefont {Goldhaber-Gordon}},\ }\bibfield  {title}
  {\bibinfo {title} {Emergent ferromagnetism near three-quarters filling in
  twisted bilayer graphene},\ }\href@noop {} {\bibfield  {journal} {\bibinfo
  {journal} {Science}\ }\textbf {\bibinfo {volume} {365}},\ \bibinfo {pages}
  {605} (\bibinfo {year} {2019})}\BibitemShut {NoStop}%
\bibitem [{\citenamefont {Chen}\ \emph
  {et~al.}(2020{\natexlab{a}})\citenamefont {Chen}, \citenamefont {Sharpe},
  \citenamefont {Fox}, \citenamefont {Zhang}, \citenamefont {Wang},
  \citenamefont {Jiang}, \citenamefont {Lyu}, \citenamefont {Li}, \citenamefont
  {Watanabe}, \citenamefont {Taniguchi} \emph {et~al.}}]{chen2020tunable}%
  \BibitemOpen
  \bibfield  {author} {\bibinfo {author} {\bibfnamefont {G.}~\bibnamefont
  {Chen}}, \bibinfo {author} {\bibfnamefont {A.~L.}\ \bibnamefont {Sharpe}},
  \bibinfo {author} {\bibfnamefont {E.~J.}\ \bibnamefont {Fox}}, \bibinfo
  {author} {\bibfnamefont {Y.-H.}\ \bibnamefont {Zhang}}, \bibinfo {author}
  {\bibfnamefont {S.}~\bibnamefont {Wang}}, \bibinfo {author} {\bibfnamefont
  {L.}~\bibnamefont {Jiang}}, \bibinfo {author} {\bibfnamefont
  {B.}~\bibnamefont {Lyu}}, \bibinfo {author} {\bibfnamefont {H.}~\bibnamefont
  {Li}}, \bibinfo {author} {\bibfnamefont {K.}~\bibnamefont {Watanabe}},
  \bibinfo {author} {\bibfnamefont {T.}~\bibnamefont {Taniguchi}}, \emph
  {et~al.},\ }\bibfield  {title} {\bibinfo {title} {Tunable correlated chern
  insulator and ferromagnetism in a moir{\'e} superlattice},\ }\href@noop {}
  {\bibfield  {journal} {\bibinfo  {journal} {Nature}\ }\textbf {\bibinfo
  {volume} {579}},\ \bibinfo {pages} {56} (\bibinfo {year}
  {2020}{\natexlab{a}})}\BibitemShut {NoStop}%
\bibitem [{\citenamefont {Serlin}\ \emph {et~al.}(2020)\citenamefont {Serlin},
  \citenamefont {Tschirhart}, \citenamefont {Polshyn}, \citenamefont {Zhang},
  \citenamefont {Zhu}, \citenamefont {Watanabe}, \citenamefont {Taniguchi},
  \citenamefont {Balents},\ and\ \citenamefont {Young}}]{serlin2020intrinsic}%
  \BibitemOpen
  \bibfield  {author} {\bibinfo {author} {\bibfnamefont {M.}~\bibnamefont
  {Serlin}}, \bibinfo {author} {\bibfnamefont {C.}~\bibnamefont {Tschirhart}},
  \bibinfo {author} {\bibfnamefont {H.}~\bibnamefont {Polshyn}}, \bibinfo
  {author} {\bibfnamefont {Y.}~\bibnamefont {Zhang}}, \bibinfo {author}
  {\bibfnamefont {J.}~\bibnamefont {Zhu}}, \bibinfo {author} {\bibfnamefont
  {K.}~\bibnamefont {Watanabe}}, \bibinfo {author} {\bibfnamefont
  {T.}~\bibnamefont {Taniguchi}}, \bibinfo {author} {\bibfnamefont
  {L.}~\bibnamefont {Balents}},\ and\ \bibinfo {author} {\bibfnamefont
  {A.}~\bibnamefont {Young}},\ }\bibfield  {title} {\bibinfo {title} {Intrinsic
  quantized anomalous hall effect in a moir{\'e} heterostructure},\ }\href@noop
  {} {\bibfield  {journal} {\bibinfo  {journal} {Science}\ }\textbf {\bibinfo
  {volume} {367}},\ \bibinfo {pages} {900} (\bibinfo {year}
  {2020})}\BibitemShut {NoStop}%
\bibitem [{\citenamefont {Zondiner}\ \emph {et~al.}(2020)\citenamefont
  {Zondiner}, \citenamefont {Rozen}, \citenamefont {Rodan-Legrain},
  \citenamefont {Cao}, \citenamefont {Queiroz}, \citenamefont {Taniguchi},
  \citenamefont {Watanabe}, \citenamefont {Oreg}, \citenamefont {von Oppen},
  \citenamefont {Stern} \emph {et~al.}}]{zondiner2020cascade}%
  \BibitemOpen
  \bibfield  {author} {\bibinfo {author} {\bibfnamefont {U.}~\bibnamefont
  {Zondiner}}, \bibinfo {author} {\bibfnamefont {A.}~\bibnamefont {Rozen}},
  \bibinfo {author} {\bibfnamefont {D.}~\bibnamefont {Rodan-Legrain}}, \bibinfo
  {author} {\bibfnamefont {Y.}~\bibnamefont {Cao}}, \bibinfo {author}
  {\bibfnamefont {R.}~\bibnamefont {Queiroz}}, \bibinfo {author} {\bibfnamefont
  {T.}~\bibnamefont {Taniguchi}}, \bibinfo {author} {\bibfnamefont
  {K.}~\bibnamefont {Watanabe}}, \bibinfo {author} {\bibfnamefont
  {Y.}~\bibnamefont {Oreg}}, \bibinfo {author} {\bibfnamefont {F.}~\bibnamefont
  {von Oppen}}, \bibinfo {author} {\bibfnamefont {A.}~\bibnamefont {Stern}},
  \emph {et~al.},\ }\bibfield  {title} {\bibinfo {title} {Cascade of phase
  transitions and dirac revivals in magic-angle graphene},\ }\href@noop {}
  {\bibfield  {journal} {\bibinfo  {journal} {Nature}\ }\textbf {\bibinfo
  {volume} {582}},\ \bibinfo {pages} {203} (\bibinfo {year}
  {2020})}\BibitemShut {NoStop}%
\bibitem [{\citenamefont {Saito}\ \emph
  {et~al.}(2020{\natexlab{a}})\citenamefont {Saito}, \citenamefont {Ge},
  \citenamefont {Watanabe}, \citenamefont {Taniguchi}, \citenamefont {Berg},\
  and\ \citenamefont {Young}}]{saito2020isospin}%
  \BibitemOpen
  \bibfield  {author} {\bibinfo {author} {\bibfnamefont {Y.}~\bibnamefont
  {Saito}}, \bibinfo {author} {\bibfnamefont {J.}~\bibnamefont {Ge}}, \bibinfo
  {author} {\bibfnamefont {K.}~\bibnamefont {Watanabe}}, \bibinfo {author}
  {\bibfnamefont {T.}~\bibnamefont {Taniguchi}}, \bibinfo {author}
  {\bibfnamefont {E.}~\bibnamefont {Berg}},\ and\ \bibinfo {author}
  {\bibfnamefont {A.~F.}\ \bibnamefont {Young}},\ }\bibfield  {title} {\bibinfo
  {title} {Isospin pomeranchuk effect and the entropy of collective excitations
  in twisted bilayer graphene},\ }\href@noop {} {\bibfield  {journal} {\bibinfo
   {journal} {arXiv preprint arXiv:2008.10830}\ } (\bibinfo {year}
  {2020}{\natexlab{a}})}\BibitemShut {NoStop}%
\bibitem [{\citenamefont {Saito}\ \emph {et~al.}(2021)\citenamefont {Saito},
  \citenamefont {Ge}, \citenamefont {Rademaker}, \citenamefont {Watanabe},
  \citenamefont {Taniguchi}, \citenamefont {Abanin},\ and\ \citenamefont
  {Young}}]{saito2021hofstadter}%
  \BibitemOpen
  \bibfield  {author} {\bibinfo {author} {\bibfnamefont {Y.}~\bibnamefont
  {Saito}}, \bibinfo {author} {\bibfnamefont {J.}~\bibnamefont {Ge}}, \bibinfo
  {author} {\bibfnamefont {L.}~\bibnamefont {Rademaker}}, \bibinfo {author}
  {\bibfnamefont {K.}~\bibnamefont {Watanabe}}, \bibinfo {author}
  {\bibfnamefont {T.}~\bibnamefont {Taniguchi}}, \bibinfo {author}
  {\bibfnamefont {D.~A.}\ \bibnamefont {Abanin}},\ and\ \bibinfo {author}
  {\bibfnamefont {A.~F.}\ \bibnamefont {Young}},\ }\bibfield  {title} {\bibinfo
  {title} {Hofstadter subband ferromagnetism and symmetry-broken chern
  insulators in twisted bilayer graphene},\ }\href@noop {} {\bibfield
  {journal} {\bibinfo  {journal} {Nature Physics}\ ,\ \bibinfo {pages} {1}}
  (\bibinfo {year} {2021})}\BibitemShut {NoStop}%
\bibitem [{\citenamefont {Wu}\ \emph {et~al.}(2021)\citenamefont {Wu},
  \citenamefont {Zhang}, \citenamefont {Watanabe}, \citenamefont {Taniguchi},\
  and\ \citenamefont {Andrei}}]{wu2021chern}%
  \BibitemOpen
  \bibfield  {author} {\bibinfo {author} {\bibfnamefont {S.}~\bibnamefont
  {Wu}}, \bibinfo {author} {\bibfnamefont {Z.}~\bibnamefont {Zhang}}, \bibinfo
  {author} {\bibfnamefont {K.}~\bibnamefont {Watanabe}}, \bibinfo {author}
  {\bibfnamefont {T.}~\bibnamefont {Taniguchi}},\ and\ \bibinfo {author}
  {\bibfnamefont {E.~Y.}\ \bibnamefont {Andrei}},\ }\bibfield  {title}
  {\bibinfo {title} {Chern insulators, van hove singularities and topological
  flat bands in magic-angle twisted bilayer graphene},\ }\href@noop {}
  {\bibfield  {journal} {\bibinfo  {journal} {Nature Materials}\ ,\ \bibinfo
  {pages} {1}} (\bibinfo {year} {2021})}\BibitemShut {NoStop}%
\bibitem [{\citenamefont {Chen}\ \emph
  {et~al.}(2020{\natexlab{b}})\citenamefont {Chen}, \citenamefont {He},
  \citenamefont {Zhang}, \citenamefont {Hsieh}, \citenamefont {Fei},
  \citenamefont {Watanabe}, \citenamefont {Taniguchi}, \citenamefont {Cobden},
  \citenamefont {Xu}, \citenamefont {Dean} \emph
  {et~al.}}]{chen2020electrically}%
  \BibitemOpen
  \bibfield  {author} {\bibinfo {author} {\bibfnamefont {S.}~\bibnamefont
  {Chen}}, \bibinfo {author} {\bibfnamefont {M.}~\bibnamefont {He}}, \bibinfo
  {author} {\bibfnamefont {Y.-H.}\ \bibnamefont {Zhang}}, \bibinfo {author}
  {\bibfnamefont {V.}~\bibnamefont {Hsieh}}, \bibinfo {author} {\bibfnamefont
  {Z.}~\bibnamefont {Fei}}, \bibinfo {author} {\bibfnamefont {K.}~\bibnamefont
  {Watanabe}}, \bibinfo {author} {\bibfnamefont {T.}~\bibnamefont {Taniguchi}},
  \bibinfo {author} {\bibfnamefont {D.~H.}\ \bibnamefont {Cobden}}, \bibinfo
  {author} {\bibfnamefont {X.}~\bibnamefont {Xu}}, \bibinfo {author}
  {\bibfnamefont {C.~R.}\ \bibnamefont {Dean}}, \emph {et~al.},\ }\bibfield
  {title} {\bibinfo {title} {Electrically tunable correlated and topological
  states in twisted monolayer--bilayer graphene},\ }\href@noop {} {\bibfield
  {journal} {\bibinfo  {journal} {Nature Physics}\ ,\ \bibinfo {pages} {1}}
  (\bibinfo {year} {2020}{\natexlab{b}})}\BibitemShut {NoStop}%
\bibitem [{\citenamefont {Cao}\ \emph {et~al.}(2020{\natexlab{a}})\citenamefont
  {Cao}, \citenamefont {Rodan-Legrain}, \citenamefont {Park}, \citenamefont
  {Yuan}, \citenamefont {Watanabe}, \citenamefont {Taniguchi}, \citenamefont
  {Fernandes}, \citenamefont {Fu},\ and\ \citenamefont
  {Jarillo-Herrero}}]{cao2020nematicity}%
  \BibitemOpen
  \bibfield  {author} {\bibinfo {author} {\bibfnamefont {Y.}~\bibnamefont
  {Cao}}, \bibinfo {author} {\bibfnamefont {D.}~\bibnamefont {Rodan-Legrain}},
  \bibinfo {author} {\bibfnamefont {J.~M.}\ \bibnamefont {Park}}, \bibinfo
  {author} {\bibfnamefont {F.~N.}\ \bibnamefont {Yuan}}, \bibinfo {author}
  {\bibfnamefont {K.}~\bibnamefont {Watanabe}}, \bibinfo {author}
  {\bibfnamefont {T.}~\bibnamefont {Taniguchi}}, \bibinfo {author}
  {\bibfnamefont {R.~M.}\ \bibnamefont {Fernandes}}, \bibinfo {author}
  {\bibfnamefont {L.}~\bibnamefont {Fu}},\ and\ \bibinfo {author}
  {\bibfnamefont {P.}~\bibnamefont {Jarillo-Herrero}},\ }\bibfield  {title}
  {\bibinfo {title} {Nematicity and competing orders in superconducting
  magic-angle graphene},\ }\href@noop {} {\bibfield  {journal} {\bibinfo
  {journal} {arXiv preprint arXiv:2004.04148}\ } (\bibinfo {year}
  {2020}{\natexlab{a}})}\BibitemShut {NoStop}%
\bibitem [{\citenamefont {Rozen}\ \emph {et~al.}(2020)\citenamefont {Rozen},
  \citenamefont {Park}, \citenamefont {Zondiner}, \citenamefont {Cao},
  \citenamefont {Rodan-Legrain}, \citenamefont {Taniguchi}, \citenamefont
  {Watanabe}, \citenamefont {Oreg}, \citenamefont {Stern}, \citenamefont {Berg}
  \emph {et~al.}}]{rozen2020entropic}%
  \BibitemOpen
  \bibfield  {author} {\bibinfo {author} {\bibfnamefont {A.}~\bibnamefont
  {Rozen}}, \bibinfo {author} {\bibfnamefont {J.~M.}\ \bibnamefont {Park}},
  \bibinfo {author} {\bibfnamefont {U.}~\bibnamefont {Zondiner}}, \bibinfo
  {author} {\bibfnamefont {Y.}~\bibnamefont {Cao}}, \bibinfo {author}
  {\bibfnamefont {D.}~\bibnamefont {Rodan-Legrain}}, \bibinfo {author}
  {\bibfnamefont {T.}~\bibnamefont {Taniguchi}}, \bibinfo {author}
  {\bibfnamefont {K.}~\bibnamefont {Watanabe}}, \bibinfo {author}
  {\bibfnamefont {Y.}~\bibnamefont {Oreg}}, \bibinfo {author} {\bibfnamefont
  {A.}~\bibnamefont {Stern}}, \bibinfo {author} {\bibfnamefont
  {E.}~\bibnamefont {Berg}}, \emph {et~al.},\ }\bibfield  {title} {\bibinfo
  {title} {Entropic evidence for a pomeranchuk effect in magic angle
  graphene},\ }\href@noop {} {\bibfield  {journal} {\bibinfo  {journal} {arXiv
  preprint arXiv:2009.01836}\ } (\bibinfo {year} {2020})}\BibitemShut {NoStop}%
\bibitem [{\citenamefont {Zhou}\ \emph {et~al.}(2021)\citenamefont {Zhou},
  \citenamefont {Xie}, \citenamefont {Ghazaryan}, \citenamefont {Holder},
  \citenamefont {Ehrets}, \citenamefont {Spanton}, \citenamefont {Taniguchi},
  \citenamefont {Watanabe}, \citenamefont {Berg}, \citenamefont {Serbyn} \emph
  {et~al.}}]{zhou2021half}%
  \BibitemOpen
  \bibfield  {author} {\bibinfo {author} {\bibfnamefont {H.}~\bibnamefont
  {Zhou}}, \bibinfo {author} {\bibfnamefont {T.}~\bibnamefont {Xie}}, \bibinfo
  {author} {\bibfnamefont {A.}~\bibnamefont {Ghazaryan}}, \bibinfo {author}
  {\bibfnamefont {T.}~\bibnamefont {Holder}}, \bibinfo {author} {\bibfnamefont
  {J.}~\bibnamefont {Ehrets}}, \bibinfo {author} {\bibfnamefont {E.~M.}\
  \bibnamefont {Spanton}}, \bibinfo {author} {\bibfnamefont {T.}~\bibnamefont
  {Taniguchi}}, \bibinfo {author} {\bibfnamefont {K.}~\bibnamefont {Watanabe}},
  \bibinfo {author} {\bibfnamefont {E.}~\bibnamefont {Berg}}, \bibinfo {author}
  {\bibfnamefont {M.}~\bibnamefont {Serbyn}}, \emph {et~al.},\ }\bibfield
  {title} {\bibinfo {title} {Half and quarter metals in rhombohedral trilayer
  graphene},\ }\href@noop {} {\bibfield  {journal} {\bibinfo  {journal} {arXiv
  preprint arXiv:2104.00653}\ } (\bibinfo {year} {2021})}\BibitemShut {NoStop}%
\bibitem [{\citenamefont {Balents}\ \emph {et~al.}(2020)\citenamefont
  {Balents}, \citenamefont {Dean}, \citenamefont {Efetov},\ and\ \citenamefont
  {Young}}]{balents2020superconductivity}%
  \BibitemOpen
  \bibfield  {author} {\bibinfo {author} {\bibfnamefont {L.}~\bibnamefont
  {Balents}}, \bibinfo {author} {\bibfnamefont {C.~R.}\ \bibnamefont {Dean}},
  \bibinfo {author} {\bibfnamefont {D.~K.}\ \bibnamefont {Efetov}},\ and\
  \bibinfo {author} {\bibfnamefont {A.~F.}\ \bibnamefont {Young}},\ }\bibfield
  {title} {\bibinfo {title} {Superconductivity and strong correlations in
  moir{\'e} flat bands},\ }\href@noop {} {\bibfield  {journal} {\bibinfo
  {journal} {Nature Physics}\ }\textbf {\bibinfo {volume} {16}},\ \bibinfo
  {pages} {725} (\bibinfo {year} {2020})}\BibitemShut {NoStop}%
\bibitem [{\citenamefont {Andrei}\ \emph {et~al.}(2021)\citenamefont {Andrei},
  \citenamefont {Efetov}, \citenamefont {Jarillo-Herrero}, \citenamefont
  {MacDonald}, \citenamefont {Mak}, \citenamefont {Senthil}, \citenamefont
  {Tutuc}, \citenamefont {Yazdani},\ and\ \citenamefont
  {Young}}]{andrei2021marvels}%
  \BibitemOpen
  \bibfield  {author} {\bibinfo {author} {\bibfnamefont {E.~Y.}\ \bibnamefont
  {Andrei}}, \bibinfo {author} {\bibfnamefont {D.~K.}\ \bibnamefont {Efetov}},
  \bibinfo {author} {\bibfnamefont {P.}~\bibnamefont {Jarillo-Herrero}},
  \bibinfo {author} {\bibfnamefont {A.~H.}\ \bibnamefont {MacDonald}}, \bibinfo
  {author} {\bibfnamefont {K.~F.}\ \bibnamefont {Mak}}, \bibinfo {author}
  {\bibfnamefont {T.}~\bibnamefont {Senthil}}, \bibinfo {author} {\bibfnamefont
  {E.}~\bibnamefont {Tutuc}}, \bibinfo {author} {\bibfnamefont
  {A.}~\bibnamefont {Yazdani}},\ and\ \bibinfo {author} {\bibfnamefont {A.~F.}\
  \bibnamefont {Young}},\ }\bibfield  {title} {\bibinfo {title} {The marvels of
  moir{\'e} materials},\ }\href@noop {} {\bibfield  {journal} {\bibinfo
  {journal} {Nature Reviews Materials}\ ,\ \bibinfo {pages} {1}} (\bibinfo
  {year} {2021})}\BibitemShut {NoStop}%
\bibitem [{\citenamefont {Lu}\ \emph {et~al.}(2015)\citenamefont {Lu},
  \citenamefont {Zheliuk}, \citenamefont {Leermakers}, \citenamefont {Yuan},
  \citenamefont {Zeitler}, \citenamefont {Law},\ and\ \citenamefont
  {Ye}}]{lu2015evidence}%
  \BibitemOpen
  \bibfield  {author} {\bibinfo {author} {\bibfnamefont {J.}~\bibnamefont
  {Lu}}, \bibinfo {author} {\bibfnamefont {O.}~\bibnamefont {Zheliuk}},
  \bibinfo {author} {\bibfnamefont {I.}~\bibnamefont {Leermakers}}, \bibinfo
  {author} {\bibfnamefont {N.~F.}\ \bibnamefont {Yuan}}, \bibinfo {author}
  {\bibfnamefont {U.}~\bibnamefont {Zeitler}}, \bibinfo {author} {\bibfnamefont
  {K.~T.}\ \bibnamefont {Law}},\ and\ \bibinfo {author} {\bibfnamefont
  {J.}~\bibnamefont {Ye}},\ }\bibfield  {title} {\bibinfo {title} {Evidence for
  two-dimensional ising superconductivity in gated mos2},\ }\href@noop {}
  {\bibfield  {journal} {\bibinfo  {journal} {Science}\ }\textbf {\bibinfo
  {volume} {350}},\ \bibinfo {pages} {1353} (\bibinfo {year}
  {2015})}\BibitemShut {NoStop}%
\bibitem [{\citenamefont {Xi}\ \emph {et~al.}(2016)\citenamefont {Xi},
  \citenamefont {Wang}, \citenamefont {Zhao}, \citenamefont {Park},
  \citenamefont {Law}, \citenamefont {Berger}, \citenamefont {Forr{\'o}},
  \citenamefont {Shan},\ and\ \citenamefont {Mak}}]{xi2016ising}%
  \BibitemOpen
  \bibfield  {author} {\bibinfo {author} {\bibfnamefont {X.}~\bibnamefont
  {Xi}}, \bibinfo {author} {\bibfnamefont {Z.}~\bibnamefont {Wang}}, \bibinfo
  {author} {\bibfnamefont {W.}~\bibnamefont {Zhao}}, \bibinfo {author}
  {\bibfnamefont {J.-H.}\ \bibnamefont {Park}}, \bibinfo {author}
  {\bibfnamefont {K.~T.}\ \bibnamefont {Law}}, \bibinfo {author} {\bibfnamefont
  {H.}~\bibnamefont {Berger}}, \bibinfo {author} {\bibfnamefont
  {L.}~\bibnamefont {Forr{\'o}}}, \bibinfo {author} {\bibfnamefont
  {J.}~\bibnamefont {Shan}},\ and\ \bibinfo {author} {\bibfnamefont {K.~F.}\
  \bibnamefont {Mak}},\ }\bibfield  {title} {\bibinfo {title} {Ising pairing in
  superconducting nbse2 atomic layers},\ }\href@noop {} {\bibfield  {journal}
  {\bibinfo  {journal} {Nature Physics}\ }\textbf {\bibinfo {volume} {12}},\
  \bibinfo {pages} {139} (\bibinfo {year} {2016})}\BibitemShut {NoStop}%
\bibitem [{\citenamefont {Saito}\ \emph {et~al.}(2016)\citenamefont {Saito},
  \citenamefont {Nakamura}, \citenamefont {Bahramy}, \citenamefont {Kohama},
  \citenamefont {Ye}, \citenamefont {Kasahara}, \citenamefont {Nakagawa},
  \citenamefont {Onga}, \citenamefont {Tokunaga}, \citenamefont {Nojima} \emph
  {et~al.}}]{saito2016superconductivity}%
  \BibitemOpen
  \bibfield  {author} {\bibinfo {author} {\bibfnamefont {Y.}~\bibnamefont
  {Saito}}, \bibinfo {author} {\bibfnamefont {Y.}~\bibnamefont {Nakamura}},
  \bibinfo {author} {\bibfnamefont {M.~S.}\ \bibnamefont {Bahramy}}, \bibinfo
  {author} {\bibfnamefont {Y.}~\bibnamefont {Kohama}}, \bibinfo {author}
  {\bibfnamefont {J.}~\bibnamefont {Ye}}, \bibinfo {author} {\bibfnamefont
  {Y.}~\bibnamefont {Kasahara}}, \bibinfo {author} {\bibfnamefont
  {Y.}~\bibnamefont {Nakagawa}}, \bibinfo {author} {\bibfnamefont
  {M.}~\bibnamefont {Onga}}, \bibinfo {author} {\bibfnamefont {M.}~\bibnamefont
  {Tokunaga}}, \bibinfo {author} {\bibfnamefont {T.}~\bibnamefont {Nojima}},
  \emph {et~al.},\ }\bibfield  {title} {\bibinfo {title} {Superconductivity
  protected by spin--valley locking in ion-gated mos2},\ }\href@noop {}
  {\bibfield  {journal} {\bibinfo  {journal} {Nature Physics}\ }\textbf
  {\bibinfo {volume} {12}},\ \bibinfo {pages} {144} (\bibinfo {year}
  {2016})}\BibitemShut {NoStop}%
\bibitem [{\citenamefont {Oh}\ \emph {et~al.}(2021)\citenamefont {Oh},
  \citenamefont {Nuckolls}, \citenamefont {Wong}, \citenamefont {Lee},
  \citenamefont {Liu}, \citenamefont {Watanabe}, \citenamefont {Taniguchi},\
  and\ \citenamefont {Yazdani}}]{oh2021evidence}%
  \BibitemOpen
  \bibfield  {author} {\bibinfo {author} {\bibfnamefont {M.}~\bibnamefont
  {Oh}}, \bibinfo {author} {\bibfnamefont {K.~P.}\ \bibnamefont {Nuckolls}},
  \bibinfo {author} {\bibfnamefont {D.}~\bibnamefont {Wong}}, \bibinfo {author}
  {\bibfnamefont {R.~L.}\ \bibnamefont {Lee}}, \bibinfo {author} {\bibfnamefont
  {X.}~\bibnamefont {Liu}}, \bibinfo {author} {\bibfnamefont {K.}~\bibnamefont
  {Watanabe}}, \bibinfo {author} {\bibfnamefont {T.}~\bibnamefont
  {Taniguchi}},\ and\ \bibinfo {author} {\bibfnamefont {A.}~\bibnamefont
  {Yazdani}},\ }\bibfield  {title} {\bibinfo {title} {Evidence for
  unconventional superconductivity in twisted bilayer graphene},\ }\href@noop
  {} {\bibfield  {journal} {\bibinfo  {journal} {Nature}\ }\textbf {\bibinfo
  {volume} {600}},\ \bibinfo {pages} {240} (\bibinfo {year}
  {2021})}\BibitemShut {NoStop}%
\bibitem [{\citenamefont {Kim}\ \emph {et~al.}(2021)\citenamefont {Kim},
  \citenamefont {Choi}, \citenamefont {Lewandowski}, \citenamefont {Thomson},
  \citenamefont {Zhang}, \citenamefont {Polski}, \citenamefont {Watanabe},
  \citenamefont {Taniguchi}, \citenamefont {Alicea},\ and\ \citenamefont
  {Nadj-Perge}}]{kim2021spectroscopic}%
  \BibitemOpen
  \bibfield  {author} {\bibinfo {author} {\bibfnamefont {H.}~\bibnamefont
  {Kim}}, \bibinfo {author} {\bibfnamefont {Y.}~\bibnamefont {Choi}}, \bibinfo
  {author} {\bibfnamefont {C.}~\bibnamefont {Lewandowski}}, \bibinfo {author}
  {\bibfnamefont {A.}~\bibnamefont {Thomson}}, \bibinfo {author} {\bibfnamefont
  {Y.}~\bibnamefont {Zhang}}, \bibinfo {author} {\bibfnamefont
  {R.}~\bibnamefont {Polski}}, \bibinfo {author} {\bibfnamefont
  {K.}~\bibnamefont {Watanabe}}, \bibinfo {author} {\bibfnamefont
  {T.}~\bibnamefont {Taniguchi}}, \bibinfo {author} {\bibfnamefont
  {J.}~\bibnamefont {Alicea}},\ and\ \bibinfo {author} {\bibfnamefont
  {S.}~\bibnamefont {Nadj-Perge}},\ }\bibfield  {title} {\bibinfo {title}
  {Spectroscopic signatures of strong correlations and unconventional
  superconductivity in twisted trilayer graphene},\ }\href@noop {} {\bibfield
  {journal} {\bibinfo  {journal} {arXiv preprint arXiv:2109.12127}\ } (\bibinfo
  {year} {2021})}\BibitemShut {NoStop}%
\bibitem [{\citenamefont {Park}\ \emph
  {et~al.}(2021{\natexlab{a}})\citenamefont {Park}, \citenamefont {Cao},
  \citenamefont {Xia}, \citenamefont {Sun}, \citenamefont {Watanabe},
  \citenamefont {Taniguchi},\ and\ \citenamefont
  {Jarillo-Herrero}}]{park2021magic}%
  \BibitemOpen
  \bibfield  {author} {\bibinfo {author} {\bibfnamefont {J.~M.}\ \bibnamefont
  {Park}}, \bibinfo {author} {\bibfnamefont {Y.}~\bibnamefont {Cao}}, \bibinfo
  {author} {\bibfnamefont {L.}~\bibnamefont {Xia}}, \bibinfo {author}
  {\bibfnamefont {S.}~\bibnamefont {Sun}}, \bibinfo {author} {\bibfnamefont
  {K.}~\bibnamefont {Watanabe}}, \bibinfo {author} {\bibfnamefont
  {T.}~\bibnamefont {Taniguchi}},\ and\ \bibinfo {author} {\bibfnamefont
  {P.}~\bibnamefont {Jarillo-Herrero}},\ }\bibfield  {title} {\bibinfo {title}
  {Magic-angle multilayer graphene: A robust family of moir$\backslash$'e
  superconductors},\ }\href@noop {} {\bibfield  {journal} {\bibinfo  {journal}
  {arXiv preprint arXiv:2112.10760}\ } (\bibinfo {year}
  {2021}{\natexlab{a}})}\BibitemShut {NoStop}%
\bibitem [{\citenamefont {Stepanov}\ \emph {et~al.}(2020)\citenamefont
  {Stepanov}, \citenamefont {Das}, \citenamefont {Lu}, \citenamefont
  {Fahimniya}, \citenamefont {Watanabe}, \citenamefont {Taniguchi},
  \citenamefont {Koppens}, \citenamefont {Lischner}, \citenamefont {Levitov},\
  and\ \citenamefont {Efetov}}]{stepanov2020untying}%
  \BibitemOpen
  \bibfield  {author} {\bibinfo {author} {\bibfnamefont {P.}~\bibnamefont
  {Stepanov}}, \bibinfo {author} {\bibfnamefont {I.}~\bibnamefont {Das}},
  \bibinfo {author} {\bibfnamefont {X.}~\bibnamefont {Lu}}, \bibinfo {author}
  {\bibfnamefont {A.}~\bibnamefont {Fahimniya}}, \bibinfo {author}
  {\bibfnamefont {K.}~\bibnamefont {Watanabe}}, \bibinfo {author}
  {\bibfnamefont {T.}~\bibnamefont {Taniguchi}}, \bibinfo {author}
  {\bibfnamefont {F.~H.}\ \bibnamefont {Koppens}}, \bibinfo {author}
  {\bibfnamefont {J.}~\bibnamefont {Lischner}}, \bibinfo {author}
  {\bibfnamefont {L.}~\bibnamefont {Levitov}},\ and\ \bibinfo {author}
  {\bibfnamefont {D.~K.}\ \bibnamefont {Efetov}},\ }\bibfield  {title}
  {\bibinfo {title} {Untying the insulating and superconducting orders in
  magic-angle graphene},\ }\href@noop {} {\bibfield  {journal} {\bibinfo
  {journal} {Nature}\ }\textbf {\bibinfo {volume} {583}},\ \bibinfo {pages}
  {375} (\bibinfo {year} {2020})}\BibitemShut {NoStop}%
\bibitem [{\citenamefont {Saito}\ \emph
  {et~al.}(2020{\natexlab{b}})\citenamefont {Saito}, \citenamefont {Ge},
  \citenamefont {Watanabe}, \citenamefont {Taniguchi},\ and\ \citenamefont
  {Young}}]{saito2020independent}%
  \BibitemOpen
  \bibfield  {author} {\bibinfo {author} {\bibfnamefont {Y.}~\bibnamefont
  {Saito}}, \bibinfo {author} {\bibfnamefont {J.}~\bibnamefont {Ge}}, \bibinfo
  {author} {\bibfnamefont {K.}~\bibnamefont {Watanabe}}, \bibinfo {author}
  {\bibfnamefont {T.}~\bibnamefont {Taniguchi}},\ and\ \bibinfo {author}
  {\bibfnamefont {A.~F.}\ \bibnamefont {Young}},\ }\bibfield  {title} {\bibinfo
  {title} {Independent superconductors and correlated insulators in twisted
  bilayer graphene},\ }\href@noop {} {\bibfield  {journal} {\bibinfo  {journal}
  {Nature Physics}\ }\textbf {\bibinfo {volume} {16}},\ \bibinfo {pages} {926}
  (\bibinfo {year} {2020}{\natexlab{b}})}\BibitemShut {NoStop}%
\bibitem [{\citenamefont {Siriviboon}\ \emph {et~al.}(2021)\citenamefont
  {Siriviboon}, \citenamefont {Lin}, \citenamefont {Scammell}, \citenamefont
  {Liu}, \citenamefont {Rhodes}, \citenamefont {Watanabe}, \citenamefont
  {Taniguchi}, \citenamefont {Hone}, \citenamefont {Scheurer},\ and\
  \citenamefont {Li}}]{siriviboon2021abundance}%
  \BibitemOpen
  \bibfield  {author} {\bibinfo {author} {\bibfnamefont {P.}~\bibnamefont
  {Siriviboon}}, \bibinfo {author} {\bibfnamefont {J.-X.}\ \bibnamefont {Lin}},
  \bibinfo {author} {\bibfnamefont {H.~D.}\ \bibnamefont {Scammell}}, \bibinfo
  {author} {\bibfnamefont {S.}~\bibnamefont {Liu}}, \bibinfo {author}
  {\bibfnamefont {D.}~\bibnamefont {Rhodes}}, \bibinfo {author} {\bibfnamefont
  {K.}~\bibnamefont {Watanabe}}, \bibinfo {author} {\bibfnamefont
  {T.}~\bibnamefont {Taniguchi}}, \bibinfo {author} {\bibfnamefont
  {J.}~\bibnamefont {Hone}}, \bibinfo {author} {\bibfnamefont {M.~S.}\
  \bibnamefont {Scheurer}},\ and\ \bibinfo {author} {\bibfnamefont
  {J.}~\bibnamefont {Li}},\ }\bibfield  {title} {\bibinfo {title} {Abundance of
  density wave phases in twisted trilayer graphene on wse $ \_2$},\ }\href@noop
  {} {\bibfield  {journal} {\bibinfo  {journal} {arXiv preprint
  arXiv:2112.07127}\ } (\bibinfo {year} {2021})}\BibitemShut {NoStop}%
\bibitem [{\citenamefont {Park}\ \emph
  {et~al.}(2021{\natexlab{b}})\citenamefont {Park}, \citenamefont {Cao},
  \citenamefont {Watanabe}, \citenamefont {Taniguchi},\ and\ \citenamefont
  {Jarillo-Herrero}}]{park2021tunable}%
  \BibitemOpen
  \bibfield  {author} {\bibinfo {author} {\bibfnamefont {J.~M.}\ \bibnamefont
  {Park}}, \bibinfo {author} {\bibfnamefont {Y.}~\bibnamefont {Cao}}, \bibinfo
  {author} {\bibfnamefont {K.}~\bibnamefont {Watanabe}}, \bibinfo {author}
  {\bibfnamefont {T.}~\bibnamefont {Taniguchi}},\ and\ \bibinfo {author}
  {\bibfnamefont {P.}~\bibnamefont {Jarillo-Herrero}},\ }\bibfield  {title}
  {\bibinfo {title} {Tunable strongly coupled superconductivity in magic-angle
  twisted trilayer graphene},\ }\href@noop {} {\bibfield  {journal} {\bibinfo
  {journal} {Nature}\ ,\ \bibinfo {pages} {1}} (\bibinfo {year}
  {2021}{\natexlab{b}})}\BibitemShut {NoStop}%
\bibitem [{\citenamefont {Hao}\ \emph {et~al.}(2021)\citenamefont {Hao},
  \citenamefont {Zimmerman}, \citenamefont {Ledwith}, \citenamefont {Khalaf},
  \citenamefont {Najafabadi}, \citenamefont {Watanabe}, \citenamefont
  {Taniguchi}, \citenamefont {Vishwanath},\ and\ \citenamefont
  {Kim}}]{hao2021electric}%
  \BibitemOpen
  \bibfield  {author} {\bibinfo {author} {\bibfnamefont {Z.}~\bibnamefont
  {Hao}}, \bibinfo {author} {\bibfnamefont {A.}~\bibnamefont {Zimmerman}},
  \bibinfo {author} {\bibfnamefont {P.}~\bibnamefont {Ledwith}}, \bibinfo
  {author} {\bibfnamefont {E.}~\bibnamefont {Khalaf}}, \bibinfo {author}
  {\bibfnamefont {D.~H.}\ \bibnamefont {Najafabadi}}, \bibinfo {author}
  {\bibfnamefont {K.}~\bibnamefont {Watanabe}}, \bibinfo {author}
  {\bibfnamefont {T.}~\bibnamefont {Taniguchi}}, \bibinfo {author}
  {\bibfnamefont {A.}~\bibnamefont {Vishwanath}},\ and\ \bibinfo {author}
  {\bibfnamefont {P.}~\bibnamefont {Kim}},\ }\bibfield  {title} {\bibinfo
  {title} {Electric field--tunable superconductivity in alternating-twist
  magic-angle trilayer graphene},\ }\href@noop {} {\bibfield  {journal}
  {\bibinfo  {journal} {Science}\ }\textbf {\bibinfo {volume} {371}},\ \bibinfo
  {pages} {1133} (\bibinfo {year} {2021})}\BibitemShut {NoStop}%
\bibitem [{\citenamefont {Zhang}\ \emph {et~al.}(2021)\citenamefont {Zhang},
  \citenamefont {Polski}, \citenamefont {Lewandowski}, \citenamefont {Thomson},
  \citenamefont {Peng}, \citenamefont {Choi}, \citenamefont {Kim},
  \citenamefont {Watanabe}, \citenamefont {Taniguchi}, \citenamefont {Alicea}
  \emph {et~al.}}]{zhang2021ascendance}%
  \BibitemOpen
  \bibfield  {author} {\bibinfo {author} {\bibfnamefont {Y.}~\bibnamefont
  {Zhang}}, \bibinfo {author} {\bibfnamefont {R.}~\bibnamefont {Polski}},
  \bibinfo {author} {\bibfnamefont {C.}~\bibnamefont {Lewandowski}}, \bibinfo
  {author} {\bibfnamefont {A.}~\bibnamefont {Thomson}}, \bibinfo {author}
  {\bibfnamefont {Y.}~\bibnamefont {Peng}}, \bibinfo {author} {\bibfnamefont
  {Y.}~\bibnamefont {Choi}}, \bibinfo {author} {\bibfnamefont {H.}~\bibnamefont
  {Kim}}, \bibinfo {author} {\bibfnamefont {K.}~\bibnamefont {Watanabe}},
  \bibinfo {author} {\bibfnamefont {T.}~\bibnamefont {Taniguchi}}, \bibinfo
  {author} {\bibfnamefont {J.}~\bibnamefont {Alicea}}, \emph {et~al.},\
  }\bibfield  {title} {\bibinfo {title} {Ascendance of superconductivity in
  magic-angle graphene multilayers},\ }\href@noop {} {\bibfield  {journal}
  {\bibinfo  {journal} {arXiv preprint arXiv:2112.09270}\ } (\bibinfo {year}
  {2021})}\BibitemShut {NoStop}%
\bibitem [{\citenamefont {Khalaf}\ \emph {et~al.}(2019)\citenamefont {Khalaf},
  \citenamefont {Kruchkov}, \citenamefont {Tarnopolsky},\ and\ \citenamefont
  {Vishwanath}}]{khalaf2019magic}%
  \BibitemOpen
  \bibfield  {author} {\bibinfo {author} {\bibfnamefont {E.}~\bibnamefont
  {Khalaf}}, \bibinfo {author} {\bibfnamefont {A.~J.}\ \bibnamefont
  {Kruchkov}}, \bibinfo {author} {\bibfnamefont {G.}~\bibnamefont
  {Tarnopolsky}},\ and\ \bibinfo {author} {\bibfnamefont {A.}~\bibnamefont
  {Vishwanath}},\ }\bibfield  {title} {\bibinfo {title} {Magic angle hierarchy
  in twisted graphene multilayers},\ }\href@noop {} {\bibfield  {journal}
  {\bibinfo  {journal} {Physical Review B}\ }\textbf {\bibinfo {volume}
  {100}},\ \bibinfo {pages} {085109} (\bibinfo {year} {2019})}\BibitemShut
  {NoStop}%
\bibitem [{\citenamefont {Wong}\ \emph {et~al.}(2020)\citenamefont {Wong},
  \citenamefont {Nuckolls}, \citenamefont {Oh}, \citenamefont {Lian},
  \citenamefont {Xie}, \citenamefont {Jeon}, \citenamefont {Watanabe},
  \citenamefont {Taniguchi}, \citenamefont {Bernevig},\ and\ \citenamefont
  {Yazdani}}]{wong2020cascade}%
  \BibitemOpen
  \bibfield  {author} {\bibinfo {author} {\bibfnamefont {D.}~\bibnamefont
  {Wong}}, \bibinfo {author} {\bibfnamefont {K.~P.}\ \bibnamefont {Nuckolls}},
  \bibinfo {author} {\bibfnamefont {M.}~\bibnamefont {Oh}}, \bibinfo {author}
  {\bibfnamefont {B.}~\bibnamefont {Lian}}, \bibinfo {author} {\bibfnamefont
  {Y.}~\bibnamefont {Xie}}, \bibinfo {author} {\bibfnamefont {S.}~\bibnamefont
  {Jeon}}, \bibinfo {author} {\bibfnamefont {K.}~\bibnamefont {Watanabe}},
  \bibinfo {author} {\bibfnamefont {T.}~\bibnamefont {Taniguchi}}, \bibinfo
  {author} {\bibfnamefont {B.~A.}\ \bibnamefont {Bernevig}},\ and\ \bibinfo
  {author} {\bibfnamefont {A.}~\bibnamefont {Yazdani}},\ }\bibfield  {title}
  {\bibinfo {title} {Cascade of electronic transitions in magic-angle twisted
  bilayer graphene},\ }\href@noop {} {\bibfield  {journal} {\bibinfo  {journal}
  {Nature}\ }\textbf {\bibinfo {volume} {582}},\ \bibinfo {pages} {198}
  (\bibinfo {year} {2020})}\BibitemShut {NoStop}%
\bibitem [{\citenamefont {Cao}\ \emph {et~al.}(2021)\citenamefont {Cao},
  \citenamefont {Park}, \citenamefont {Watanabe}, \citenamefont {Taniguchi},\
  and\ \citenamefont {Jarillo-Herrero}}]{inplane_fields}%
  \BibitemOpen
  \bibfield  {author} {\bibinfo {author} {\bibfnamefont {Y.}~\bibnamefont
  {Cao}}, \bibinfo {author} {\bibfnamefont {J.~M.}\ \bibnamefont {Park}},
  \bibinfo {author} {\bibfnamefont {K.}~\bibnamefont {Watanabe}}, \bibinfo
  {author} {\bibfnamefont {T.}~\bibnamefont {Taniguchi}},\ and\ \bibinfo
  {author} {\bibfnamefont {P.}~\bibnamefont {Jarillo-Herrero}},\ }\bibfield
  {title} {\bibinfo {title} {Large pauli limit violation and reentrant
  superconductivity in magic-angle twisted trilayer graphene},\ }\href@noop {}
  {\bibfield  {journal} {\bibinfo  {journal} {arXiv preprint arXiv:2103.12083}\
  } (\bibinfo {year} {2021})}\BibitemShut {NoStop}%
\bibitem [{\citenamefont {Liu}\ \emph {et~al.}(2022)\citenamefont {Liu},
  \citenamefont {Zhang}, \citenamefont {Watanabe}, \citenamefont {Taniguchi},\
  and\ \citenamefont {Li}}]{liu2022isospin}%
  \BibitemOpen
  \bibfield  {author} {\bibinfo {author} {\bibfnamefont {X.}~\bibnamefont
  {Liu}}, \bibinfo {author} {\bibfnamefont {N.~J.}\ \bibnamefont {Zhang}},
  \bibinfo {author} {\bibfnamefont {K.}~\bibnamefont {Watanabe}}, \bibinfo
  {author} {\bibfnamefont {T.}~\bibnamefont {Taniguchi}},\ and\ \bibinfo
  {author} {\bibfnamefont {J.}~\bibnamefont {Li}},\ }\bibfield  {title}
  {\bibinfo {title} {Isospin order in superconducting magic-angle twisted
  trilayer graphene},\ }\href@noop {} {\bibfield  {journal} {\bibinfo
  {journal} {Nature Physics}\ ,\ \bibinfo {pages} {1}} (\bibinfo {year}
  {2022})}\BibitemShut {NoStop}%
\bibitem [{\citenamefont {Arora}\ \emph {et~al.}(2020)\citenamefont {Arora},
  \citenamefont {Polski}, \citenamefont {Zhang}, \citenamefont {Thomson},
  \citenamefont {Choi}, \citenamefont {Kim}, \citenamefont {Lin}, \citenamefont
  {Wilson}, \citenamefont {Xu}, \citenamefont {Chu} \emph
  {et~al.}}]{arora2020superconductivity}%
  \BibitemOpen
  \bibfield  {author} {\bibinfo {author} {\bibfnamefont {H.~S.}\ \bibnamefont
  {Arora}}, \bibinfo {author} {\bibfnamefont {R.}~\bibnamefont {Polski}},
  \bibinfo {author} {\bibfnamefont {Y.}~\bibnamefont {Zhang}}, \bibinfo
  {author} {\bibfnamefont {A.}~\bibnamefont {Thomson}}, \bibinfo {author}
  {\bibfnamefont {Y.}~\bibnamefont {Choi}}, \bibinfo {author} {\bibfnamefont
  {H.}~\bibnamefont {Kim}}, \bibinfo {author} {\bibfnamefont {Z.}~\bibnamefont
  {Lin}}, \bibinfo {author} {\bibfnamefont {I.~Z.}\ \bibnamefont {Wilson}},
  \bibinfo {author} {\bibfnamefont {X.}~\bibnamefont {Xu}}, \bibinfo {author}
  {\bibfnamefont {J.-H.}\ \bibnamefont {Chu}}, \emph {et~al.},\ }\bibfield
  {title} {\bibinfo {title} {Superconductivity in metallic twisted bilayer
  graphene stabilized by wse 2},\ }\href@noop {} {\bibfield  {journal}
  {\bibinfo  {journal} {Nature}\ }\textbf {\bibinfo {volume} {583}},\ \bibinfo
  {pages} {379} (\bibinfo {year} {2020})}\BibitemShut {NoStop}%
\bibitem [{\citenamefont {Morissette}\ \emph {et~al.}(2022)\citenamefont
  {Morissette}, \citenamefont {Lin}, \citenamefont {Liu}, \citenamefont
  {Rhodes}, \citenamefont {Watanabe}, \citenamefont {Taniguchi}, \citenamefont
  {Hone}, \citenamefont {Scheurer}, \citenamefont {Lilly}, \citenamefont
  {Mounce} \emph {et~al.}}]{morissette2022electron}%
  \BibitemOpen
  \bibfield  {author} {\bibinfo {author} {\bibfnamefont {E.}~\bibnamefont
  {Morissette}}, \bibinfo {author} {\bibfnamefont {J.-X.}\ \bibnamefont {Lin}},
  \bibinfo {author} {\bibfnamefont {S.}~\bibnamefont {Liu}}, \bibinfo {author}
  {\bibfnamefont {D.}~\bibnamefont {Rhodes}}, \bibinfo {author} {\bibfnamefont
  {K.}~\bibnamefont {Watanabe}}, \bibinfo {author} {\bibfnamefont
  {T.}~\bibnamefont {Taniguchi}}, \bibinfo {author} {\bibfnamefont
  {J.}~\bibnamefont {Hone}}, \bibinfo {author} {\bibfnamefont {M.~S.}\
  \bibnamefont {Scheurer}}, \bibinfo {author} {\bibfnamefont {M.}~\bibnamefont
  {Lilly}}, \bibinfo {author} {\bibfnamefont {A.}~\bibnamefont {Mounce}}, \emph
  {et~al.},\ }\bibfield  {title} {\bibinfo {title} {Electron spin resonance and
  collective excitations in magic-angle twisted bilayer graphene},\ }\href@noop
  {} {\bibfield  {journal} {\bibinfo  {journal} {arXiv preprint
  arXiv:2206.08354}\ } (\bibinfo {year} {2022})}\BibitemShut {NoStop}%
\bibitem [{\citenamefont {Lake}\ and\ \citenamefont
  {Senthil}(2021)}]{lake2021reentrant}%
  \BibitemOpen
  \bibfield  {author} {\bibinfo {author} {\bibfnamefont {E.}~\bibnamefont
  {Lake}}\ and\ \bibinfo {author} {\bibfnamefont {T.}~\bibnamefont {Senthil}},\
  }\bibfield  {title} {\bibinfo {title} {Reentrant superconductivity through a
  quantum lifshitz transition in twisted trilayer graphene},\ }\href@noop {}
  {\bibfield  {journal} {\bibinfo  {journal} {Physical Review B}\ }\textbf
  {\bibinfo {volume} {104}},\ \bibinfo {pages} {174505} (\bibinfo {year}
  {2021})}\BibitemShut {NoStop}%
\bibitem [{\citenamefont {Qin}\ and\ \citenamefont
  {MacDonald}(2021)}]{qin2021plane}%
  \BibitemOpen
  \bibfield  {author} {\bibinfo {author} {\bibfnamefont {W.}~\bibnamefont
  {Qin}}\ and\ \bibinfo {author} {\bibfnamefont {A.~H.}\ \bibnamefont
  {MacDonald}},\ }\bibfield  {title} {\bibinfo {title} {In-plane critical
  magnetic fields in magic-angle twisted trilayer graphene},\ }\href@noop {}
  {\bibfield  {journal} {\bibinfo  {journal} {Physical Review Letters}\
  }\textbf {\bibinfo {volume} {127}},\ \bibinfo {pages} {097001} (\bibinfo
  {year} {2021})}\BibitemShut {NoStop}%
\bibitem [{\citenamefont {Ledwith}\ \emph {et~al.}(2021)\citenamefont
  {Ledwith}, \citenamefont {Khalaf}, \citenamefont {Zhu}, \citenamefont {Carr},
  \citenamefont {Kaxiras},\ and\ \citenamefont {Vishwanath}}]{ledwith2021tb}%
  \BibitemOpen
  \bibfield  {author} {\bibinfo {author} {\bibfnamefont {P.~J.}\ \bibnamefont
  {Ledwith}}, \bibinfo {author} {\bibfnamefont {E.}~\bibnamefont {Khalaf}},
  \bibinfo {author} {\bibfnamefont {Z.}~\bibnamefont {Zhu}}, \bibinfo {author}
  {\bibfnamefont {S.}~\bibnamefont {Carr}}, \bibinfo {author} {\bibfnamefont
  {E.}~\bibnamefont {Kaxiras}},\ and\ \bibinfo {author} {\bibfnamefont
  {A.}~\bibnamefont {Vishwanath}},\ }\bibfield  {title} {\bibinfo {title} {Tb
  or not tb? contrasting properties of twisted bilayer graphene and the
  alternating twist n-layer structures (n= 3, 4, 5, ...)},\ }\href@noop {}
  {\bibfield  {journal} {\bibinfo  {journal} {arXiv preprint arXiv:2111.11060}\
  } (\bibinfo {year} {2021})}\BibitemShut {NoStop}%
\bibitem [{\citenamefont {Cornfeld}\ \emph {et~al.}(2020)\citenamefont
  {Cornfeld}, \citenamefont {Rudner},\ and\ \citenamefont
  {Berg}}]{cornfeld2020spin}%
  \BibitemOpen
  \bibfield  {author} {\bibinfo {author} {\bibfnamefont {E.}~\bibnamefont
  {Cornfeld}}, \bibinfo {author} {\bibfnamefont {M.~S.}\ \bibnamefont
  {Rudner}},\ and\ \bibinfo {author} {\bibfnamefont {E.}~\bibnamefont {Berg}},\
  }\bibfield  {title} {\bibinfo {title} {Spin-polarized superconductivity:
  order parameter topology, current dissipation, and double-period josephson
  effect},\ }\href@noop {} {\bibfield  {journal} {\bibinfo  {journal} {arXiv
  preprint arXiv:2006.10073}\ } (\bibinfo {year} {2020})}\BibitemShut {NoStop}%
\bibitem [{\citenamefont {Bultinck}\ \emph
  {et~al.}(2020{\natexlab{a}})\citenamefont {Bultinck}, \citenamefont {Khalaf},
  \citenamefont {Liu}, \citenamefont {Chatterjee}, \citenamefont {Vishwanath},\
  and\ \citenamefont {Zaletel}}]{bultinck2020ground}%
  \BibitemOpen
  \bibfield  {author} {\bibinfo {author} {\bibfnamefont {N.}~\bibnamefont
  {Bultinck}}, \bibinfo {author} {\bibfnamefont {E.}~\bibnamefont {Khalaf}},
  \bibinfo {author} {\bibfnamefont {S.}~\bibnamefont {Liu}}, \bibinfo {author}
  {\bibfnamefont {S.}~\bibnamefont {Chatterjee}}, \bibinfo {author}
  {\bibfnamefont {A.}~\bibnamefont {Vishwanath}},\ and\ \bibinfo {author}
  {\bibfnamefont {M.~P.}\ \bibnamefont {Zaletel}},\ }\bibfield  {title}
  {\bibinfo {title} {Ground state and hidden symmetry of magic-angle graphene
  at even integer filling},\ }\href@noop {} {\bibfield  {journal} {\bibinfo
  {journal} {Physical Review X}\ }\textbf {\bibinfo {volume} {10}},\ \bibinfo
  {pages} {031034} (\bibinfo {year} {2020}{\natexlab{a}})}\BibitemShut
  {NoStop}%
\bibitem [{\citenamefont {Kang}\ and\ \citenamefont
  {Vafek}(2019)}]{kang2019strong}%
  \BibitemOpen
  \bibfield  {author} {\bibinfo {author} {\bibfnamefont {J.}~\bibnamefont
  {Kang}}\ and\ \bibinfo {author} {\bibfnamefont {O.}~\bibnamefont {Vafek}},\
  }\bibfield  {title} {\bibinfo {title} {Strong coupling phases of partially
  filled twisted bilayer graphene narrow bands},\ }\href@noop {} {\bibfield
  {journal} {\bibinfo  {journal} {Physical review letters}\ }\textbf {\bibinfo
  {volume} {122}},\ \bibinfo {pages} {246401} (\bibinfo {year}
  {2019})}\BibitemShut {NoStop}%
\bibitem [{\citenamefont {Lian}\ \emph {et~al.}(2021)\citenamefont {Lian},
  \citenamefont {Song}, \citenamefont {Regnault}, \citenamefont {Efetov},
  \citenamefont {Yazdani},\ and\ \citenamefont {Bernevig}}]{lian2021twisted}%
  \BibitemOpen
  \bibfield  {author} {\bibinfo {author} {\bibfnamefont {B.}~\bibnamefont
  {Lian}}, \bibinfo {author} {\bibfnamefont {Z.-D.}\ \bibnamefont {Song}},
  \bibinfo {author} {\bibfnamefont {N.}~\bibnamefont {Regnault}}, \bibinfo
  {author} {\bibfnamefont {D.~K.}\ \bibnamefont {Efetov}}, \bibinfo {author}
  {\bibfnamefont {A.}~\bibnamefont {Yazdani}},\ and\ \bibinfo {author}
  {\bibfnamefont {B.~A.}\ \bibnamefont {Bernevig}},\ }\bibfield  {title}
  {\bibinfo {title} {Twisted bilayer graphene. iv. exact insulator ground
  states and phase diagram},\ }\href@noop {} {\bibfield  {journal} {\bibinfo
  {journal} {Physical Review B}\ }\textbf {\bibinfo {volume} {103}},\ \bibinfo
  {pages} {205414} (\bibinfo {year} {2021})}\BibitemShut {NoStop}%
\bibitem [{\citenamefont {Zhang}\ \emph
  {et~al.}(2019{\natexlab{a}})\citenamefont {Zhang}, \citenamefont {Mao},
  \citenamefont {Cao}, \citenamefont {Jarillo-Herrero},\ and\ \citenamefont
  {Senthil}}]{zhang2019nearly}%
  \BibitemOpen
  \bibfield  {author} {\bibinfo {author} {\bibfnamefont {Y.-H.}\ \bibnamefont
  {Zhang}}, \bibinfo {author} {\bibfnamefont {D.}~\bibnamefont {Mao}}, \bibinfo
  {author} {\bibfnamefont {Y.}~\bibnamefont {Cao}}, \bibinfo {author}
  {\bibfnamefont {P.}~\bibnamefont {Jarillo-Herrero}},\ and\ \bibinfo {author}
  {\bibfnamefont {T.}~\bibnamefont {Senthil}},\ }\bibfield  {title} {\bibinfo
  {title} {Nearly flat chern bands in moir{\'e} superlattices},\ }\href@noop {}
  {\bibfield  {journal} {\bibinfo  {journal} {Physical Review B}\ }\textbf
  {\bibinfo {volume} {99}},\ \bibinfo {pages} {075127} (\bibinfo {year}
  {2019}{\natexlab{a}})}\BibitemShut {NoStop}%
\bibitem [{\citenamefont {Bultinck}\ \emph
  {et~al.}(2020{\natexlab{b}})\citenamefont {Bultinck}, \citenamefont
  {Chatterjee},\ and\ \citenamefont {Zaletel}}]{bultinck2020mechanism}%
  \BibitemOpen
  \bibfield  {author} {\bibinfo {author} {\bibfnamefont {N.}~\bibnamefont
  {Bultinck}}, \bibinfo {author} {\bibfnamefont {S.}~\bibnamefont
  {Chatterjee}},\ and\ \bibinfo {author} {\bibfnamefont {M.~P.}\ \bibnamefont
  {Zaletel}},\ }\bibfield  {title} {\bibinfo {title} {Mechanism for anomalous
  hall ferromagnetism in twisted bilayer graphene},\ }\href@noop {} {\bibfield
  {journal} {\bibinfo  {journal} {Physical review letters}\ }\textbf {\bibinfo
  {volume} {124}},\ \bibinfo {pages} {166601} (\bibinfo {year}
  {2020}{\natexlab{b}})}\BibitemShut {NoStop}%
\bibitem [{\citenamefont {Zhang}\ \emph
  {et~al.}(2019{\natexlab{b}})\citenamefont {Zhang}, \citenamefont {Mao},\ and\
  \citenamefont {Senthil}}]{zhang2019twisted}%
  \BibitemOpen
  \bibfield  {author} {\bibinfo {author} {\bibfnamefont {Y.-H.}\ \bibnamefont
  {Zhang}}, \bibinfo {author} {\bibfnamefont {D.}~\bibnamefont {Mao}},\ and\
  \bibinfo {author} {\bibfnamefont {T.}~\bibnamefont {Senthil}},\ }\bibfield
  {title} {\bibinfo {title} {Twisted bilayer graphene aligned with hexagonal
  boron nitride: Anomalous hall effect and a lattice model},\ }\href@noop {}
  {\bibfield  {journal} {\bibinfo  {journal} {Physical Review Research}\
  }\textbf {\bibinfo {volume} {1}},\ \bibinfo {pages} {033126} (\bibinfo {year}
  {2019}{\natexlab{b}})}\BibitemShut {NoStop}%
\bibitem [{\citenamefont {Repellin}\ \emph {et~al.}(2020)\citenamefont
  {Repellin}, \citenamefont {Dong}, \citenamefont {Zhang},\ and\ \citenamefont
  {Senthil}}]{repellin2020ferromagnetism}%
  \BibitemOpen
  \bibfield  {author} {\bibinfo {author} {\bibfnamefont {C.}~\bibnamefont
  {Repellin}}, \bibinfo {author} {\bibfnamefont {Z.}~\bibnamefont {Dong}},
  \bibinfo {author} {\bibfnamefont {Y.-H.}\ \bibnamefont {Zhang}},\ and\
  \bibinfo {author} {\bibfnamefont {T.}~\bibnamefont {Senthil}},\ }\bibfield
  {title} {\bibinfo {title} {Ferromagnetism in narrow bands of moir{\'e}
  superlattices},\ }\href@noop {} {\bibfield  {journal} {\bibinfo  {journal}
  {Physical Review Letters}\ }\textbf {\bibinfo {volume} {124}},\ \bibinfo
  {pages} {187601} (\bibinfo {year} {2020})}\BibitemShut {NoStop}%
\bibitem [{\citenamefont {Liu}\ and\ \citenamefont
  {Dai}(2021)}]{liu2021theories}%
  \BibitemOpen
  \bibfield  {author} {\bibinfo {author} {\bibfnamefont {J.}~\bibnamefont
  {Liu}}\ and\ \bibinfo {author} {\bibfnamefont {X.}~\bibnamefont {Dai}},\
  }\bibfield  {title} {\bibinfo {title} {Theories for the correlated insulating
  states and quantum anomalous hall effect phenomena in twisted bilayer
  graphene},\ }\href@noop {} {\bibfield  {journal} {\bibinfo  {journal}
  {Physical Review B}\ }\textbf {\bibinfo {volume} {103}},\ \bibinfo {pages}
  {035427} (\bibinfo {year} {2021})}\BibitemShut {NoStop}%
\bibitem [{\citenamefont {Khalaf}\ \emph
  {et~al.}(2020{\natexlab{a}})\citenamefont {Khalaf}, \citenamefont {Ledwith},\
  and\ \citenamefont {Vishwanath}}]{khalaf2020symmetry}%
  \BibitemOpen
  \bibfield  {author} {\bibinfo {author} {\bibfnamefont {E.}~\bibnamefont
  {Khalaf}}, \bibinfo {author} {\bibfnamefont {P.}~\bibnamefont {Ledwith}},\
  and\ \bibinfo {author} {\bibfnamefont {A.}~\bibnamefont {Vishwanath}},\
  }\bibfield  {title} {\bibinfo {title} {Symmetry constraints on
  superconductivity in twisted bilayer graphene: Fractional vortices, $4 e $
  condensates or non-unitary pairing},\ }\href@noop {} {\bibfield  {journal}
  {\bibinfo  {journal} {arXiv preprint arXiv:2012.05915}\ } (\bibinfo {year}
  {2020}{\natexlab{a}})}\BibitemShut {NoStop}%
\bibitem [{\citenamefont {Christos}\ \emph {et~al.}(2021)\citenamefont
  {Christos}, \citenamefont {Sachdev},\ and\ \citenamefont
  {Scheurer}}]{christos2021correlated}%
  \BibitemOpen
  \bibfield  {author} {\bibinfo {author} {\bibfnamefont {M.}~\bibnamefont
  {Christos}}, \bibinfo {author} {\bibfnamefont {S.}~\bibnamefont {Sachdev}},\
  and\ \bibinfo {author} {\bibfnamefont {M.~S.}\ \bibnamefont {Scheurer}},\
  }\bibfield  {title} {\bibinfo {title} {Correlated insulators, semimetals, and
  superconductivity in twisted trilayer graphene},\ }\href@noop {} {\bibfield
  {journal} {\bibinfo  {journal} {arXiv preprint arXiv:2106.02063}\ } (\bibinfo
  {year} {2021})}\BibitemShut {NoStop}%
\bibitem [{\citenamefont {Scheurer}\ and\ \citenamefont
  {Samajdar}(2020)}]{scheurer2020pairing}%
  \BibitemOpen
  \bibfield  {author} {\bibinfo {author} {\bibfnamefont {M.~S.}\ \bibnamefont
  {Scheurer}}\ and\ \bibinfo {author} {\bibfnamefont {R.}~\bibnamefont
  {Samajdar}},\ }\bibfield  {title} {\bibinfo {title} {Pairing in
  graphene-based moir{\'e} superlattices},\ }\href@noop {} {\bibfield
  {journal} {\bibinfo  {journal} {Physical Review Research}\ }\textbf {\bibinfo
  {volume} {2}},\ \bibinfo {pages} {033062} (\bibinfo {year}
  {2020})}\BibitemShut {NoStop}%
\bibitem [{\citenamefont {de~Vries}\ \emph {et~al.}(2021)\citenamefont
  {de~Vries}, \citenamefont {Portoles}, \citenamefont {Zheng}, \citenamefont
  {Taniguchi}, \citenamefont {Watanabe}, \citenamefont {Ihn}, \citenamefont
  {Ensslin},\ and\ \citenamefont {Rickhaus}}]{de2021gate}%
  \BibitemOpen
  \bibfield  {author} {\bibinfo {author} {\bibfnamefont {F.~K.}\ \bibnamefont
  {de~Vries}}, \bibinfo {author} {\bibfnamefont {E.}~\bibnamefont {Portoles}},
  \bibinfo {author} {\bibfnamefont {G.}~\bibnamefont {Zheng}}, \bibinfo
  {author} {\bibfnamefont {T.}~\bibnamefont {Taniguchi}}, \bibinfo {author}
  {\bibfnamefont {K.}~\bibnamefont {Watanabe}}, \bibinfo {author}
  {\bibfnamefont {T.}~\bibnamefont {Ihn}}, \bibinfo {author} {\bibfnamefont
  {K.}~\bibnamefont {Ensslin}},\ and\ \bibinfo {author} {\bibfnamefont
  {P.}~\bibnamefont {Rickhaus}},\ }\bibfield  {title} {\bibinfo {title}
  {Gate-defined josephson junctions in magic-angle twisted bilayer graphene},\
  }\href@noop {} {\bibfield  {journal} {\bibinfo  {journal} {Nature
  Nanotechnology}\ }\textbf {\bibinfo {volume} {16}},\ \bibinfo {pages} {760}
  (\bibinfo {year} {2021})}\BibitemShut {NoStop}%
\bibitem [{\citenamefont {Rodan-Legrain}\ \emph {et~al.}(2021)\citenamefont
  {Rodan-Legrain}, \citenamefont {Cao}, \citenamefont {Park}, \citenamefont
  {de~la Barrera}, \citenamefont {Randeria}, \citenamefont {Watanabe},
  \citenamefont {Taniguchi},\ and\ \citenamefont
  {Jarillo-Herrero}}]{rodan2021highly}%
  \BibitemOpen
  \bibfield  {author} {\bibinfo {author} {\bibfnamefont {D.}~\bibnamefont
  {Rodan-Legrain}}, \bibinfo {author} {\bibfnamefont {Y.}~\bibnamefont {Cao}},
  \bibinfo {author} {\bibfnamefont {J.~M.}\ \bibnamefont {Park}}, \bibinfo
  {author} {\bibfnamefont {S.~C.}\ \bibnamefont {de~la Barrera}}, \bibinfo
  {author} {\bibfnamefont {M.~T.}\ \bibnamefont {Randeria}}, \bibinfo {author}
  {\bibfnamefont {K.}~\bibnamefont {Watanabe}}, \bibinfo {author}
  {\bibfnamefont {T.}~\bibnamefont {Taniguchi}},\ and\ \bibinfo {author}
  {\bibfnamefont {P.}~\bibnamefont {Jarillo-Herrero}},\ }\bibfield  {title}
  {\bibinfo {title} {Highly tunable junctions and non-local josephson effect in
  magic-angle graphene tunnelling devices},\ }\href@noop {} {\bibfield
  {journal} {\bibinfo  {journal} {Nature Nanotechnology}\ }\textbf {\bibinfo
  {volume} {16}},\ \bibinfo {pages} {769} (\bibinfo {year} {2021})}\BibitemShut
  {NoStop}%
\bibitem [{\citenamefont {Po}\ \emph {et~al.}(2018)\citenamefont {Po},
  \citenamefont {Zou}, \citenamefont {Vishwanath},\ and\ \citenamefont
  {Senthil}}]{po2018origin}%
  \BibitemOpen
  \bibfield  {author} {\bibinfo {author} {\bibfnamefont {H.~C.}\ \bibnamefont
  {Po}}, \bibinfo {author} {\bibfnamefont {L.}~\bibnamefont {Zou}}, \bibinfo
  {author} {\bibfnamefont {A.}~\bibnamefont {Vishwanath}},\ and\ \bibinfo
  {author} {\bibfnamefont {T.}~\bibnamefont {Senthil}},\ }\bibfield  {title}
  {\bibinfo {title} {Origin of mott insulating behavior and superconductivity
  in twisted bilayer graphene},\ }\href@noop {} {\bibfield  {journal} {\bibinfo
   {journal} {Physical Review X}\ }\textbf {\bibinfo {volume} {8}},\ \bibinfo
  {pages} {031089} (\bibinfo {year} {2018})}\BibitemShut {NoStop}%
\bibitem [{\citenamefont {Choi}\ \emph {et~al.}(2019)\citenamefont {Choi},
  \citenamefont {Kemmer}, \citenamefont {Peng}, \citenamefont {Thomson},
  \citenamefont {Arora}, \citenamefont {Polski}, \citenamefont {Zhang},
  \citenamefont {Ren}, \citenamefont {Alicea}, \citenamefont {Refael} \emph
  {et~al.}}]{choi2019electronic}%
  \BibitemOpen
  \bibfield  {author} {\bibinfo {author} {\bibfnamefont {Y.}~\bibnamefont
  {Choi}}, \bibinfo {author} {\bibfnamefont {J.}~\bibnamefont {Kemmer}},
  \bibinfo {author} {\bibfnamefont {Y.}~\bibnamefont {Peng}}, \bibinfo {author}
  {\bibfnamefont {A.}~\bibnamefont {Thomson}}, \bibinfo {author} {\bibfnamefont
  {H.}~\bibnamefont {Arora}}, \bibinfo {author} {\bibfnamefont
  {R.}~\bibnamefont {Polski}}, \bibinfo {author} {\bibfnamefont
  {Y.}~\bibnamefont {Zhang}}, \bibinfo {author} {\bibfnamefont
  {H.}~\bibnamefont {Ren}}, \bibinfo {author} {\bibfnamefont {J.}~\bibnamefont
  {Alicea}}, \bibinfo {author} {\bibfnamefont {G.}~\bibnamefont {Refael}},
  \emph {et~al.},\ }\bibfield  {title} {\bibinfo {title} {Electronic
  correlations in twisted bilayer graphene near the magic angle},\ }\href@noop
  {} {\bibfield  {journal} {\bibinfo  {journal} {Nature Physics}\ }\textbf
  {\bibinfo {volume} {15}},\ \bibinfo {pages} {1174} (\bibinfo {year}
  {2019})}\BibitemShut {NoStop}%
\bibitem [{\citenamefont {Jiang}\ \emph {et~al.}(2019)\citenamefont {Jiang},
  \citenamefont {Lai}, \citenamefont {Watanabe}, \citenamefont {Taniguchi},
  \citenamefont {Haule}, \citenamefont {Mao},\ and\ \citenamefont
  {Andrei}}]{jiang2019charge}%
  \BibitemOpen
  \bibfield  {author} {\bibinfo {author} {\bibfnamefont {Y.}~\bibnamefont
  {Jiang}}, \bibinfo {author} {\bibfnamefont {X.}~\bibnamefont {Lai}}, \bibinfo
  {author} {\bibfnamefont {K.}~\bibnamefont {Watanabe}}, \bibinfo {author}
  {\bibfnamefont {T.}~\bibnamefont {Taniguchi}}, \bibinfo {author}
  {\bibfnamefont {K.}~\bibnamefont {Haule}}, \bibinfo {author} {\bibfnamefont
  {J.}~\bibnamefont {Mao}},\ and\ \bibinfo {author} {\bibfnamefont {E.~Y.}\
  \bibnamefont {Andrei}},\ }\bibfield  {title} {\bibinfo {title} {Charge order
  and broken rotational symmetry in magic-angle twisted bilayer graphene},\
  }\href@noop {} {\bibfield  {journal} {\bibinfo  {journal} {Nature}\ }\textbf
  {\bibinfo {volume} {573}},\ \bibinfo {pages} {91} (\bibinfo {year}
  {2019})}\BibitemShut {NoStop}%
\bibitem [{\citenamefont {Kerelsky}\ \emph {et~al.}(2019)\citenamefont
  {Kerelsky}, \citenamefont {McGilly}, \citenamefont {Kennes}, \citenamefont
  {Xian}, \citenamefont {Yankowitz}, \citenamefont {Chen}, \citenamefont
  {Watanabe}, \citenamefont {Taniguchi}, \citenamefont {Hone}, \citenamefont
  {Dean} \emph {et~al.}}]{kerelsky2019maximized}%
  \BibitemOpen
  \bibfield  {author} {\bibinfo {author} {\bibfnamefont {A.}~\bibnamefont
  {Kerelsky}}, \bibinfo {author} {\bibfnamefont {L.~J.}\ \bibnamefont
  {McGilly}}, \bibinfo {author} {\bibfnamefont {D.~M.}\ \bibnamefont {Kennes}},
  \bibinfo {author} {\bibfnamefont {L.}~\bibnamefont {Xian}}, \bibinfo {author}
  {\bibfnamefont {M.}~\bibnamefont {Yankowitz}}, \bibinfo {author}
  {\bibfnamefont {S.}~\bibnamefont {Chen}}, \bibinfo {author} {\bibfnamefont
  {K.}~\bibnamefont {Watanabe}}, \bibinfo {author} {\bibfnamefont
  {T.}~\bibnamefont {Taniguchi}}, \bibinfo {author} {\bibfnamefont
  {J.}~\bibnamefont {Hone}}, \bibinfo {author} {\bibfnamefont {C.}~\bibnamefont
  {Dean}}, \emph {et~al.},\ }\bibfield  {title} {\bibinfo {title} {Maximized
  electron interactions at the magic angle in twisted bilayer graphene},\
  }\href@noop {} {\bibfield  {journal} {\bibinfo  {journal} {Nature}\ }\textbf
  {\bibinfo {volume} {572}},\ \bibinfo {pages} {95} (\bibinfo {year}
  {2019})}\BibitemShut {NoStop}%
\bibitem [{\citenamefont {Hu}(1994)}]{hu1994midgap}%
  \BibitemOpen
  \bibfield  {author} {\bibinfo {author} {\bibfnamefont {C.-R.}\ \bibnamefont
  {Hu}},\ }\bibfield  {title} {\bibinfo {title} {Midgap surface states as a
  novel signature for d x a 2-x b 2-wave superconductivity},\ }\href@noop {}
  {\bibfield  {journal} {\bibinfo  {journal} {Physical review letters}\
  }\textbf {\bibinfo {volume} {72}},\ \bibinfo {pages} {1526} (\bibinfo {year}
  {1994})}\BibitemShut {NoStop}%
\bibitem [{\citenamefont {L{\"o}fwander}\ \emph {et~al.}(2001)\citenamefont
  {L{\"o}fwander}, \citenamefont {Shumeiko},\ and\ \citenamefont
  {Wendin}}]{lofwander2001andreev}%
  \BibitemOpen
  \bibfield  {author} {\bibinfo {author} {\bibfnamefont {T.}~\bibnamefont
  {L{\"o}fwander}}, \bibinfo {author} {\bibfnamefont {V.}~\bibnamefont
  {Shumeiko}},\ and\ \bibinfo {author} {\bibfnamefont {G.}~\bibnamefont
  {Wendin}},\ }\bibfield  {title} {\bibinfo {title} {Andreev bound states in
  high-tc superconducting junctions},\ }\href@noop {} {\bibfield  {journal}
  {\bibinfo  {journal} {Superconductor Science and Technology}\ }\textbf
  {\bibinfo {volume} {14}},\ \bibinfo {pages} {R53} (\bibinfo {year}
  {2001})}\BibitemShut {NoStop}%
\bibitem [{\citenamefont {Kashiwaya}\ \emph {et~al.}(1996)\citenamefont
  {Kashiwaya}, \citenamefont {Tanaka}, \citenamefont {Koyanagi},\ and\
  \citenamefont {Kajimura}}]{stm_slab}%
  \BibitemOpen
  \bibfield  {author} {\bibinfo {author} {\bibfnamefont {S.}~\bibnamefont
  {Kashiwaya}}, \bibinfo {author} {\bibfnamefont {Y.}~\bibnamefont {Tanaka}},
  \bibinfo {author} {\bibfnamefont {M.}~\bibnamefont {Koyanagi}},\ and\
  \bibinfo {author} {\bibfnamefont {K.}~\bibnamefont {Kajimura}},\ }\bibfield
  {title} {\bibinfo {title} {Theory for tunneling spectroscopy of anisotropic
  superconductors},\ }\href {https://doi.org/10.1103/PhysRevB.53.2667}
  {\bibfield  {journal} {\bibinfo  {journal} {Phys. Rev. B}\ }\textbf {\bibinfo
  {volume} {53}},\ \bibinfo {pages} {2667} (\bibinfo {year}
  {1996})}\BibitemShut {NoStop}%
\bibitem [{\citenamefont {Yamashiro}\ \emph {et~al.}(1998)\citenamefont
  {Yamashiro}, \citenamefont {Tanaka}, \citenamefont {Tanuma},\ and\
  \citenamefont {Kashiwaya}}]{Kashiwaya_spin_triplet_tunneling}%
  \BibitemOpen
  \bibfield  {author} {\bibinfo {author} {\bibfnamefont {M.}~\bibnamefont
  {Yamashiro}}, \bibinfo {author} {\bibfnamefont {Y.}~\bibnamefont {Tanaka}},
  \bibinfo {author} {\bibfnamefont {Y.}~\bibnamefont {Tanuma}},\ and\ \bibinfo
  {author} {\bibfnamefont {S.}~\bibnamefont {Kashiwaya}},\ }\bibfield  {title}
  {\bibinfo {title} {Theory of tunneling conductance for normal
  metal/insulator/ triplet superconductor junction},\ }\href@noop {} {\bibfield
   {journal} {\bibinfo  {journal} {Journal of the Physical Society of Japan}\
  }\textbf {\bibinfo {volume} {67}},\ \bibinfo {pages} {3224} (\bibinfo {year}
  {1998})}\BibitemShut {NoStop}%
\bibitem [{\citenamefont {Tsuei}\ and\ \citenamefont
  {Kirtley}(2000)}]{rev_mod_phys_pairing_sym}%
  \BibitemOpen
  \bibfield  {author} {\bibinfo {author} {\bibfnamefont {C.~C.}\ \bibnamefont
  {Tsuei}}\ and\ \bibinfo {author} {\bibfnamefont {J.~R.}\ \bibnamefont
  {Kirtley}},\ }\bibfield  {title} {\bibinfo {title} {Pairing symmetry in
  cuprate superconductors},\ }\href {https://doi.org/10.1103/RevModPhys.72.969}
  {\bibfield  {journal} {\bibinfo  {journal} {Rev. Mod. Phys.}\ }\textbf
  {\bibinfo {volume} {72}},\ \bibinfo {pages} {969} (\bibinfo {year}
  {2000})}\BibitemShut {NoStop}%
\bibitem [{\citenamefont {Kimura}\ \emph {et~al.}(2009)\citenamefont {Kimura},
  \citenamefont {Barber~Jr}, \citenamefont {Ono}, \citenamefont {Ando},\ and\
  \citenamefont {Dynes}}]{kimura2009josephson}%
  \BibitemOpen
  \bibfield  {author} {\bibinfo {author} {\bibfnamefont {H.}~\bibnamefont
  {Kimura}}, \bibinfo {author} {\bibfnamefont {R.}~\bibnamefont {Barber~Jr}},
  \bibinfo {author} {\bibfnamefont {S.}~\bibnamefont {Ono}}, \bibinfo {author}
  {\bibfnamefont {Y.}~\bibnamefont {Ando}},\ and\ \bibinfo {author}
  {\bibfnamefont {R.~C.}\ \bibnamefont {Dynes}},\ }\bibfield  {title} {\bibinfo
  {title} {Josephson scanning tunneling microscopy: A local and direct probe of
  the superconducting order parameter},\ }\href@noop {} {\bibfield  {journal}
  {\bibinfo  {journal} {Physical Review B}\ }\textbf {\bibinfo {volume} {80}},\
  \bibinfo {pages} {144506} (\bibinfo {year} {2009})}\BibitemShut {NoStop}%
\bibitem [{\citenamefont {Hamidian}\ \emph {et~al.}(2016)\citenamefont
  {Hamidian}, \citenamefont {Edkins}, \citenamefont {Joo}, \citenamefont
  {Kostin}, \citenamefont {Eisaki}, \citenamefont {Uchida}, \citenamefont
  {Lawler}, \citenamefont {Kim}, \citenamefont {Mackenzie}, \citenamefont
  {Fujita} \emph {et~al.}}]{hamidian2016detection}%
  \BibitemOpen
  \bibfield  {author} {\bibinfo {author} {\bibfnamefont {M.}~\bibnamefont
  {Hamidian}}, \bibinfo {author} {\bibfnamefont {S.}~\bibnamefont {Edkins}},
  \bibinfo {author} {\bibfnamefont {S.~H.}\ \bibnamefont {Joo}}, \bibinfo
  {author} {\bibfnamefont {A.}~\bibnamefont {Kostin}}, \bibinfo {author}
  {\bibfnamefont {H.}~\bibnamefont {Eisaki}}, \bibinfo {author} {\bibfnamefont
  {S.}~\bibnamefont {Uchida}}, \bibinfo {author} {\bibfnamefont
  {M.}~\bibnamefont {Lawler}}, \bibinfo {author} {\bibfnamefont {E.-A.}\
  \bibnamefont {Kim}}, \bibinfo {author} {\bibfnamefont {A.}~\bibnamefont
  {Mackenzie}}, \bibinfo {author} {\bibfnamefont {K.}~\bibnamefont {Fujita}},
  \emph {et~al.},\ }\bibfield  {title} {\bibinfo {title} {Detection of a
  cooper-pair density wave in bi2sr2cacu2o8+ x},\ }\href@noop {} {\bibfield
  {journal} {\bibinfo  {journal} {Nature}\ }\textbf {\bibinfo {volume} {532}},\
  \bibinfo {pages} {343} (\bibinfo {year} {2016})}\BibitemShut {NoStop}%
\bibitem [{\citenamefont {Liu}\ \emph {et~al.}(2021)\citenamefont {Liu},
  \citenamefont {Wang}, \citenamefont {Watanabe}, \citenamefont {Taniguchi},
  \citenamefont {Vafek},\ and\ \citenamefont {Li}}]{liu2021tuning}%
  \BibitemOpen
  \bibfield  {author} {\bibinfo {author} {\bibfnamefont {X.}~\bibnamefont
  {Liu}}, \bibinfo {author} {\bibfnamefont {Z.}~\bibnamefont {Wang}}, \bibinfo
  {author} {\bibfnamefont {K.}~\bibnamefont {Watanabe}}, \bibinfo {author}
  {\bibfnamefont {T.}~\bibnamefont {Taniguchi}}, \bibinfo {author}
  {\bibfnamefont {O.}~\bibnamefont {Vafek}},\ and\ \bibinfo {author}
  {\bibfnamefont {J.}~\bibnamefont {Li}},\ }\bibfield  {title} {\bibinfo
  {title} {Tuning electron correlation in magic-angle twisted bilayer graphene
  using coulomb screening},\ }\href@noop {} {\bibfield  {journal} {\bibinfo
  {journal} {Science}\ }\textbf {\bibinfo {volume} {371}},\ \bibinfo {pages}
  {1261} (\bibinfo {year} {2021})}\BibitemShut {NoStop}%
\bibitem [{\citenamefont {Cao}\ \emph {et~al.}(2020{\natexlab{b}})\citenamefont
  {Cao}, \citenamefont {Chowdhury}, \citenamefont {Rodan-Legrain},
  \citenamefont {Rubies-Bigorda}, \citenamefont {Watanabe}, \citenamefont
  {Taniguchi}, \citenamefont {Senthil},\ and\ \citenamefont
  {Jarillo-Herrero}}]{cao2020strange}%
  \BibitemOpen
  \bibfield  {author} {\bibinfo {author} {\bibfnamefont {Y.}~\bibnamefont
  {Cao}}, \bibinfo {author} {\bibfnamefont {D.}~\bibnamefont {Chowdhury}},
  \bibinfo {author} {\bibfnamefont {D.}~\bibnamefont {Rodan-Legrain}}, \bibinfo
  {author} {\bibfnamefont {O.}~\bibnamefont {Rubies-Bigorda}}, \bibinfo
  {author} {\bibfnamefont {K.}~\bibnamefont {Watanabe}}, \bibinfo {author}
  {\bibfnamefont {T.}~\bibnamefont {Taniguchi}}, \bibinfo {author}
  {\bibfnamefont {T.}~\bibnamefont {Senthil}},\ and\ \bibinfo {author}
  {\bibfnamefont {P.}~\bibnamefont {Jarillo-Herrero}},\ }\bibfield  {title}
  {\bibinfo {title} {Strange metal in magic-angle graphene with near planckian
  dissipation},\ }\href@noop {} {\bibfield  {journal} {\bibinfo  {journal}
  {Physical review letters}\ }\textbf {\bibinfo {volume} {124}},\ \bibinfo
  {pages} {076801} (\bibinfo {year} {2020}{\natexlab{b}})}\BibitemShut
  {NoStop}%
\bibitem [{\citenamefont {Jaoui}\ \emph {et~al.}(2021)\citenamefont {Jaoui},
  \citenamefont {Das}, \citenamefont {Di~Battista}, \citenamefont
  {D{\'\i}ez-M{\'e}rida}, \citenamefont {Lu}, \citenamefont {Watanabe},
  \citenamefont {Taniguchi}, \citenamefont {Ishizuka}, \citenamefont
  {Levitov},\ and\ \citenamefont {Efetov}}]{jaoui2021quantum}%
  \BibitemOpen
  \bibfield  {author} {\bibinfo {author} {\bibfnamefont {A.}~\bibnamefont
  {Jaoui}}, \bibinfo {author} {\bibfnamefont {I.}~\bibnamefont {Das}}, \bibinfo
  {author} {\bibfnamefont {G.}~\bibnamefont {Di~Battista}}, \bibinfo {author}
  {\bibfnamefont {J.}~\bibnamefont {D{\'\i}ez-M{\'e}rida}}, \bibinfo {author}
  {\bibfnamefont {X.}~\bibnamefont {Lu}}, \bibinfo {author} {\bibfnamefont
  {K.}~\bibnamefont {Watanabe}}, \bibinfo {author} {\bibfnamefont
  {T.}~\bibnamefont {Taniguchi}}, \bibinfo {author} {\bibfnamefont
  {H.}~\bibnamefont {Ishizuka}}, \bibinfo {author} {\bibfnamefont
  {L.}~\bibnamefont {Levitov}},\ and\ \bibinfo {author} {\bibfnamefont {D.~K.}\
  \bibnamefont {Efetov}},\ }\bibfield  {title} {\bibinfo {title}
  {Quantum-critical continuum in magic-angle twisted bilayer graphene},\
  }\href@noop {} {\bibfield  {journal} {\bibinfo  {journal} {arXiv preprint
  arXiv:2108.07753}\ } (\bibinfo {year} {2021})}\BibitemShut {NoStop}%
\bibitem [{\citenamefont {Lyu}\ \emph {et~al.}(2021)\citenamefont {Lyu},
  \citenamefont {Tuchfeld}, \citenamefont {Verma}, \citenamefont {Tian},
  \citenamefont {Watanabe}, \citenamefont {Taniguchi}, \citenamefont {Lau},
  \citenamefont {Randeria},\ and\ \citenamefont {Bockrath}}]{lyu2021strange}%
  \BibitemOpen
  \bibfield  {author} {\bibinfo {author} {\bibfnamefont {R.}~\bibnamefont
  {Lyu}}, \bibinfo {author} {\bibfnamefont {Z.}~\bibnamefont {Tuchfeld}},
  \bibinfo {author} {\bibfnamefont {N.}~\bibnamefont {Verma}}, \bibinfo
  {author} {\bibfnamefont {H.}~\bibnamefont {Tian}}, \bibinfo {author}
  {\bibfnamefont {K.}~\bibnamefont {Watanabe}}, \bibinfo {author}
  {\bibfnamefont {T.}~\bibnamefont {Taniguchi}}, \bibinfo {author}
  {\bibfnamefont {C.~N.}\ \bibnamefont {Lau}}, \bibinfo {author} {\bibfnamefont
  {M.}~\bibnamefont {Randeria}},\ and\ \bibinfo {author} {\bibfnamefont
  {M.}~\bibnamefont {Bockrath}},\ }\bibfield  {title} {\bibinfo {title}
  {Strange metal behavior of the hall angle in twisted bilayer graphene},\
  }\href@noop {} {\bibfield  {journal} {\bibinfo  {journal} {Physical Review
  B}\ }\textbf {\bibinfo {volume} {103}},\ \bibinfo {pages} {245424} (\bibinfo
  {year} {2021})}\BibitemShut {NoStop}%
\bibitem [{\citenamefont {Khalaf}\ \emph
  {et~al.}(2020{\natexlab{b}})\citenamefont {Khalaf}, \citenamefont
  {Chatterjee}, \citenamefont {Bultinck}, \citenamefont {Zaletel},\ and\
  \citenamefont {Vishwanath}}]{khalaf2020charged}%
  \BibitemOpen
  \bibfield  {author} {\bibinfo {author} {\bibfnamefont {E.}~\bibnamefont
  {Khalaf}}, \bibinfo {author} {\bibfnamefont {S.}~\bibnamefont {Chatterjee}},
  \bibinfo {author} {\bibfnamefont {N.}~\bibnamefont {Bultinck}}, \bibinfo
  {author} {\bibfnamefont {M.~P.}\ \bibnamefont {Zaletel}},\ and\ \bibinfo
  {author} {\bibfnamefont {A.}~\bibnamefont {Vishwanath}},\ }\bibfield  {title}
  {\bibinfo {title} {Charged skyrmions and topological origin of
  superconductivity in magic angle graphene},\ }\href@noop {} {\bibfield
  {journal} {\bibinfo  {journal} {arXiv preprint arXiv:2004.00638}\ } (\bibinfo
  {year} {2020}{\natexlab{b}})}\BibitemShut {NoStop}%
\bibitem [{\citenamefont {Tseng}\ \emph {et~al.}(2022)\citenamefont {Tseng},
  \citenamefont {Ma}, \citenamefont {Liu}, \citenamefont {Watanabe},
  \citenamefont {Taniguchi}, \citenamefont {Chu},\ and\ \citenamefont
  {Yankowitz}}]{tseng2022anomalous}%
  \BibitemOpen
  \bibfield  {author} {\bibinfo {author} {\bibfnamefont {C.-C.}\ \bibnamefont
  {Tseng}}, \bibinfo {author} {\bibfnamefont {X.}~\bibnamefont {Ma}}, \bibinfo
  {author} {\bibfnamefont {Z.}~\bibnamefont {Liu}}, \bibinfo {author}
  {\bibfnamefont {K.}~\bibnamefont {Watanabe}}, \bibinfo {author}
  {\bibfnamefont {T.}~\bibnamefont {Taniguchi}}, \bibinfo {author}
  {\bibfnamefont {J.-H.}\ \bibnamefont {Chu}},\ and\ \bibinfo {author}
  {\bibfnamefont {M.}~\bibnamefont {Yankowitz}},\ }\bibfield  {title} {\bibinfo
  {title} {Anomalous hall effect at half filling in twisted bilayer graphene},\
  }\href@noop {} {\bibfield  {journal} {\bibinfo  {journal} {arXiv preprint
  arXiv:2202.01734}\ } (\bibinfo {year} {2022})}\BibitemShut {NoStop}%
\bibitem [{\citenamefont {Diez-Merida}\ \emph {et~al.}(2021)\citenamefont
  {Diez-Merida}, \citenamefont {D{\'\i}ez-Carl{\'o}n}, \citenamefont {Yang},
  \citenamefont {Xie}, \citenamefont {Gao}, \citenamefont {Watanabe},
  \citenamefont {Taniguchi}, \citenamefont {Lu}, \citenamefont {Law},\ and\
  \citenamefont {Efetov}}]{diez2021magnetic}%
  \BibitemOpen
  \bibfield  {author} {\bibinfo {author} {\bibfnamefont {J.}~\bibnamefont
  {Diez-Merida}}, \bibinfo {author} {\bibfnamefont {A.}~\bibnamefont
  {D{\'\i}ez-Carl{\'o}n}}, \bibinfo {author} {\bibfnamefont {S.}~\bibnamefont
  {Yang}}, \bibinfo {author} {\bibfnamefont {Y.-M.}\ \bibnamefont {Xie}},
  \bibinfo {author} {\bibfnamefont {X.-J.}\ \bibnamefont {Gao}}, \bibinfo
  {author} {\bibfnamefont {K.}~\bibnamefont {Watanabe}}, \bibinfo {author}
  {\bibfnamefont {T.}~\bibnamefont {Taniguchi}}, \bibinfo {author}
  {\bibfnamefont {X.}~\bibnamefont {Lu}}, \bibinfo {author} {\bibfnamefont
  {K.~T.}\ \bibnamefont {Law}},\ and\ \bibinfo {author} {\bibfnamefont {D.~K.}\
  \bibnamefont {Efetov}},\ }\bibfield  {title} {\bibinfo {title} {Magnetic
  josephson junctions and superconducting diodes in magic angle twisted bilayer
  graphene},\ }\href@noop {} {\bibfield  {journal} {\bibinfo  {journal} {arXiv
  preprint arXiv:2110.01067}\ } (\bibinfo {year} {2021})}\BibitemShut {NoStop}%
\bibitem [{\citenamefont {Po}\ \emph {et~al.}(2019)\citenamefont {Po},
  \citenamefont {Zou}, \citenamefont {Senthil},\ and\ \citenamefont
  {Vishwanath}}]{po2019faithful}%
  \BibitemOpen
  \bibfield  {author} {\bibinfo {author} {\bibfnamefont {H.~C.}\ \bibnamefont
  {Po}}, \bibinfo {author} {\bibfnamefont {L.}~\bibnamefont {Zou}}, \bibinfo
  {author} {\bibfnamefont {T.}~\bibnamefont {Senthil}},\ and\ \bibinfo {author}
  {\bibfnamefont {A.}~\bibnamefont {Vishwanath}},\ }\bibfield  {title}
  {\bibinfo {title} {Faithful tight-binding models and fragile topology of
  magic-angle bilayer graphene},\ }\href@noop {} {\bibfield  {journal}
  {\bibinfo  {journal} {Physical Review B}\ }\textbf {\bibinfo {volume} {99}},\
  \bibinfo {pages} {195455} (\bibinfo {year} {2019})}\BibitemShut {NoStop}%
\bibitem [{\citenamefont {Ioffe}\ and\ \citenamefont
  {Millis}(2002)}]{ioffe2002d}%
  \BibitemOpen
  \bibfield  {author} {\bibinfo {author} {\bibfnamefont {L.}~\bibnamefont
  {Ioffe}}\ and\ \bibinfo {author} {\bibfnamefont {A.}~\bibnamefont {Millis}},\
  }\bibfield  {title} {\bibinfo {title} {d-wave superconductivity in doped mott
  insulators},\ }\href@noop {} {\bibfield  {journal} {\bibinfo  {journal}
  {Journal of Physics and Chemistry of Solids}\ }\textbf {\bibinfo {volume}
  {63}},\ \bibinfo {pages} {2259} (\bibinfo {year} {2002})}\BibitemShut
  {NoStop}%
\bibitem [{\citenamefont {Lee}\ and\ \citenamefont
  {Wen}(1997)}]{lee1997unusual}%
  \BibitemOpen
  \bibfield  {author} {\bibinfo {author} {\bibfnamefont {P.~A.}\ \bibnamefont
  {Lee}}\ and\ \bibinfo {author} {\bibfnamefont {X.-G.}\ \bibnamefont {Wen}},\
  }\bibfield  {title} {\bibinfo {title} {Unusual superconducting state of
  underdoped cuprates},\ }\href@noop {} {\bibfield  {journal} {\bibinfo
  {journal} {Physical review letters}\ }\textbf {\bibinfo {volume} {78}},\
  \bibinfo {pages} {4111} (\bibinfo {year} {1997})}\BibitemShut {NoStop}%
\bibitem [{\citenamefont {Bruder}(1990)}]{bruder_andreev_scattering}%
  \BibitemOpen
  \bibfield  {author} {\bibinfo {author} {\bibfnamefont {C.}~\bibnamefont
  {Bruder}},\ }\bibfield  {title} {\bibinfo {title} {Andreev scattering in
  anisotropic superconductors},\ }\href
  {https://doi.org/10.1103/PhysRevB.41.4017} {\bibfield  {journal} {\bibinfo
  {journal} {Phys. Rev. B}\ }\textbf {\bibinfo {volume} {41}},\ \bibinfo
  {pages} {4017} (\bibinfo {year} {1990})}\BibitemShut {NoStop}%
\bibitem [{\citenamefont {Pals}\ \emph {et~al.}(1977)\citenamefont {Pals},
  \citenamefont {Van~Haeringen},\ and\ \citenamefont
  {Van~Maaren}}]{pals1977josephson}%
  \BibitemOpen
  \bibfield  {author} {\bibinfo {author} {\bibfnamefont {J.}~\bibnamefont
  {Pals}}, \bibinfo {author} {\bibfnamefont {W.}~\bibnamefont
  {Van~Haeringen}},\ and\ \bibinfo {author} {\bibfnamefont {M.}~\bibnamefont
  {Van~Maaren}},\ }\bibfield  {title} {\bibinfo {title} {Josephson effect
  between superconductors in possibly different spin-pairing states},\
  }\href@noop {} {\bibfield  {journal} {\bibinfo  {journal} {Physical Review
  B}\ }\textbf {\bibinfo {volume} {15}},\ \bibinfo {pages} {2592} (\bibinfo
  {year} {1977})}\BibitemShut {NoStop}%
\bibitem [{\citenamefont {Cuevas}\ \emph {et~al.}(1996)\citenamefont {Cuevas},
  \citenamefont {Mart{\'\i}n-Rodero},\ and\ \citenamefont
  {Yeyati}}]{cuevas1996hamiltonian}%
  \BibitemOpen
  \bibfield  {author} {\bibinfo {author} {\bibfnamefont {J.}~\bibnamefont
  {Cuevas}}, \bibinfo {author} {\bibfnamefont {A.}~\bibnamefont
  {Mart{\'\i}n-Rodero}},\ and\ \bibinfo {author} {\bibfnamefont {A.~L.}\
  \bibnamefont {Yeyati}},\ }\bibfield  {title} {\bibinfo {title} {Hamiltonian
  approach to the transport properties of superconducting quantum point
  contacts},\ }\href@noop {} {\bibfield  {journal} {\bibinfo  {journal}
  {Physical Review B}\ }\textbf {\bibinfo {volume} {54}},\ \bibinfo {pages}
  {7366} (\bibinfo {year} {1996})}\BibitemShut {NoStop}%
\bibitem [{\citenamefont {Mahan}(1990)}]{mahan_many_body}%
  \BibitemOpen
  \bibfield  {author} {\bibinfo {author} {\bibfnamefont {G.}~\bibnamefont
  {Mahan}},\ }\href@noop {} {\emph {\bibinfo {title} {Many-Particle
  Physics}}},\ \bibinfo {edition} {2nd}\ ed.,\ Physics of Solids and Liquids\
  (\bibinfo  {publisher} {Springer US},\ \bibinfo {year} {1990})\BibitemShut
  {NoStop}%
\end{thebibliography}%

		\begin{widetext}

			\appendix 
			
			\section{Modeling the superconductor}\label{app:details} 
			
			In this appendix we collect some formulae relevant for the discussions about pairing in the main text. We first frame our discussion within standard weak-coupling BCS theory, and then discuss how things can be treated within an approach better suited for a strongly coupled nodal superconductor. 
			
			\ss{Weak coupling} 
			
			The approach we will take in this appendix will be to focus on the weak-coupling regime near $\nu=-3$, where TBG presumably possesses a well-defined Fermi surface out of which the SC forms. Our treatment will only describe electrons within the partially-filled active flat band, whose electrons are labeled by spin and valley indices. In this treatment the physics of the transformation between the sublattice and band bases --- as well as the nontrivial fragile topology of the flat bands \cite{po2019faithful} --- will be completely swept under the rug, and we will simply work directly with operators that create electrons in the active band. Ignoring the nontrivial fragile topology of the Bloch wavefunctions is allowed for the present purposes, as all of the action will take place at the Fermi surface, and we may always fix a gauge in which any singularities in the Bloch functions are pushed out to regions far away the Fermi surface. With this discussion in mind then,
			the Bogoliubov Hamiltonian projected into the flavor-polarized subspace is 
			\bea  \label{ham}H_\bfk & = \bpm \Xi_{\tau^z\bfk}\mcp  & \wh \De_\bfk \\ \wh \De^\da_{\bfk} & - \Xi_{-\tau^z\bfk} \mcp^T \epm,\eea 
			where the function $\Xi_\bfk$ is 
			\be \Xi_{\bfk} = \xi_{\bfk} + \tau^z \d_\bfk\ee
			with $\xi_\bfk$ the $K$-valley dispersion and with $\d_\bfk$ the depairing energy induced by the in-plane field.
			While $\d_\bfk$ can be computed within e.g. the BM model as 
			\be \d_\bfk = \frac{c_0}2 \bfB_\prl \cdot \lan u_{K\bfk} | (-\s^x, \s^y) \mu^z |  u_{K\bfk}\ran\ee
			(here $u_{K\bfk}$ are the Bloch functions of the valence band in valley $K$ and $\s^a,\mu^a$ are Pauli matrices in sublattice and layer space, respectively), the bands of the state that the SC arises from will generically be modified significantly by the flavor polarization, and hence more work is required to obtain a good unbiased estimate for $\d_\bfk$. For now we will simply content ourselves with using the $C_3$ symmetry of the $K$ valley dispersion to write $\d_\bfk$ at a given angle $\t$ on the Fermi surface as 
			\be \label{deltadef_app} \d_\t = B_\prl \(\d_0 \cos(\t - \t_B) + \d_3 \cos(2\t + \t_B)\),\ee 
			where we have kept only the two lowest harmonics in the angle $\t$ (the subscripts on the constants $\d_0,\d_3$ denote the angular momentum channel). 
			
			The quasiparticle Green's function projected into $\mch_{SVL}$ is defined as the matrix $\mcg_{\o,\bfk}$ satisfying
			\be (\mcp\oplus \mcp^T\o -H_\bfk)\mcg_{\o,\bfk} = \mcp \oplus \mcp^T.\ee 
			Using $\Xi_{\tau^z\bfk} \wh \De_\bfk = \wh \De_\bfk \tau^x \Xi_{\tau^z\bfk} \tau^x$, we find 
			\be \label{qpgreens} \mcg_{\o,\bfk} = \bpm (\o+\tau^x\Xi_{-\tau^z\bfk}\tau^x)\mcp 
			& - \wh \De_\bfk \\ -\wh \De^\da_\bfk & (\o-\tau^x \Xi_{\tau^z\bfk}\tau^x)\mcp^T \epm 
			[(\o - \l^z \tau^z\d_{\l^z\tau^z\bfk})^2 - E_{\l^z\tau^z\bfk}^2 ]\inv, \ee 
			where $\l^\mu$ are Pauli matrices in Nambu space, and we have made the usual definition 
			\be E_\bfk \equiv \sqrt{\xi^2_\bfk + |\De_\bfk|^2}.\ee

			\ss{Strong coupling regime} \label{sec:strong} 

			In this section we describe how the traditional weak-coupling analysis of the rest of the paper can be modified to reflect a strong coupling situation not describable within the purview of standard BCS theory. By a `strong coupling situation', we mean one in which electrons in the superconductor are very tightly bound in Cooper pairs, with the Cooper pair condensate and fermionic quasiparticles treated independently, rather than linked together through some mean-field treatment. In this approach the magnitude of the gap is fixed, and the variation of the properties of the SC with $T,\bfB_\prl$ are controlled solely by how these parameters affect the nodal quasiparticles, which are modeled as massless Dirac fermions. 
			
			Our motivation for considering this description is based on several factors. One is that as discussed, TBG is likely in a regime of fairly strong coupling close to $\nu=-2$. Relatedly --- although we prefer flavor polarization as an explanation for the `pseudogap' observed above $T_c$ in STM --- it is possible that some of this gap is coming from Cooper pair binding energy. Finally, experimentally one observes a fairly large (factor of $\sim 2$) separation between $T_{MF}$ (defined as e.g. the temperature where the resistance is 50\% of its extrapolated normal-state value) and $T_{BKT}$. 
			These facts mean the suppression of the SC with $T$ and $\bfB_\prl$ may be due in  part to loss of phase coherence, rather than just Cooper pair unbinding (particularly near $\nu=-2$). 
			
			In the following we will briefly examine the response of the SC to $\bfB_\prl$, assuming that $\De$ is fixed and that we can get away with restricting our attention to the nodal quasiparticles right near the nodes. The appropriate Hamiltonian to employ in this scenario is \cite{ioffe2002d}
			\bea \label{strong_ham} H&=  \frac{\r_0}2 (\D \phi)^2 + \sum_n \(\Psi_n^\da H^{qp}_n\Psi_n + \frac 12 \D\phi \cdot \Psi_n^\da \l^z\tau^z\bfv_{F,\l^z\tau^zn} \Psi_n\),\eea 
			where $\phi$ is the phase of the condensate, $\Psi_n = (\psi_n, \psi^\da_{-n})$ destroys a quasiparticle with current along the $\uvth_n$ direction with $-n$ denoting the node at angle $\t_n + \pi$, $\r_0$ is the $T=0$ phase stiffness, and where $\Psi_n^\da \bfv_{F,n} \Psi_n$ is the current carried by the $\Psi_n$ quasiparticles. The quasiparticle Hamiltonian has the Dirac form 
			\bea H^{qp}_n & = \l^z (\mcp \oplus \mcp^T) (v_{F,\l^z\tau^zn} k_\prl  + \tau^z \d_{\t_{\l^z\tau^zn}}) + (\l^+-p\l^-)M_\eta v_{\De,n} k_\perp  \eea 
			where $v_{\De,n}$ is as in \eqref{delta_expansion}, $v_{F,n}$ is the $K$-valley Fermi velocity (always positive), $k_\prl$ ($k_\perp$) are the momentum components along (normal to) the node direction, and where for convenience we have defined the matrix 
			\be M_\eta = -p\tau^+|\eta\ran\lan \eta'| + \tau^- |\eta'\ran\lan \eta|,\ee
			which satisfies $M_\eta^\da = -p M_\eta$. This yields the Green's function
			\be \label{strong_qpgreens} \mcg_{\o,\bfk,n} = \( \o-\a^z\d_{\a^z n}+k_\prl \l^z  v_{F,\a^zn} - (\l^+M_\eta - p\l^- M_\eta) v_{\De,n} k_\perp \) \frac1{(\o-\a^z \d_{\a^zn})^2 - (k_\prl^2 v^2_{F,\a^zn} + k_\perp^2 v_{\De,n}^2)},\ee
			from which one can calculate a DOS of the same form as the $\o \ll \De$ limit of \eqref{sc_dos}.

			Both $T$ and $\bfB_\prl$ have the effect of making a finite density of nodal quasiparticles present in the ground state. Since the quasiparticles carry a definite current, they can then provide a `counterflow' mechanism for reducing the supercurrent and superfluid stiffness, thereby giving a way of suppressing the SC even with $|\De|$ held fixed (as in this section). 
			A calculation along the lines of that done in \cite{lee1997unusual} gives the superfluid stiffness (in the clean limit) 
			\bea \r^{ij} = \r_0\d^{ij} - 2T\sum_n \frac{v_{F,n}}{v_{\De,n}} \t^i_n \t^j_n \ln [2\cosh(\d_{\t_n}/2T)].\eea 
			Again following \cite{lee1997unusual}, the supercurrent density $\bfJ$ in the presence of a uniform applied vector potential $\bfA$ is obtained as  
			\bea \label{strong_current} \bfJ & = 4 \r_0 \bfA - \sum_{n,s=\pm1} \frac{s\uvth_n }{v_{\De,n}}\int_0^\infty d\nu\, f(\nu - s(\d_{\t_n} + \bfA\cdot\bfv_{F,n})) \nu 
			,\eea 
			where $f$ is the Fermi function
			(note that our conventions are such that $(\d J^i / \d A^j)|_{\bfA=0} = 4\r^{ij}$). 
			To find the critical current along a particular direction, we simply maximize the above expression with respect to $A$. 

			First consider the $p$-wave case. At zero field, increasing $T$ has the effect of suppressing $\r^{yy}$ (for $\De_\t = \De\cos(\t)$) by an amount linear in $T$; in this situation a BKT transition would occur when $\sqrt{\r_{xx}\r_{yy}} = 2T/\pi$. The critical current $J_c$ obtained from \eqref{strong_current} has a strong nematic dependence on the current direction and on $\t_B$, and in fact when $\bfJ$ is normal to the nodes and $T=0,B_\prl=0$, the critical current is formally infinite within this model (since the quasiparticles only provide a backflow current along the directions $\uvth_n$). Now consider the $d$-wave case. At zero field, both $\r_{xx}$ and $\r_{yy}$ are suppressed by an amount linear in $T$. The dependence of the critical current on $\t_B$ is much less anisotropic. 
			
			The extent to which the loss of phase coherence induced by quasiparticle backflow can explain the nematicity observed in \cite{cao2020nematicity} (as well as the possibility of using this effect to help distinguish $p$ from $d$-wave pairing) depends on whether or not the primary effect of the in-plane field is to decrease the SC gap, or to increase the number of excited quasiparticles present in the ground state. If the latter effect were dominant, the observed nematicity would then be an indication that the gap is $p$-wave, rather than $d$-wave. However, since the nematicity is observed even in the weak-coupling regime (where we expect gap suppression to be the dominant effect of the field), having the nematicity originate from this quasiparticle backflow mechanism seems fairly unlikely.

			\section{Andreev bound states at SN interfaces} 
			
			\label{app_andreev_bound_states}

			Andreev bound states, which occur at superconductor-normal metal (SN) interfaces and at vacuum-terminated edges of superconductors, are a manifestation of the quantum interference effects of electron and hole quasiparticles scattering off the superconducting order parameter.
			The existence of these bound states are evident from the appearance of a sharp conductance peak occurring at zero energy (i.e. the zero-bias conductance, ZBC) and can be probed either by lateral tunneling into the edge from a normal metal, or by using STM. 
			These bound states shine remarkable insight into the nodal nature of the superconducting order parameter, and as such can potentially be used to distinguish between different orbital characters \cite{bruder_andreev_scattering}.
			We focus on the geometrical setup of TBG gated such that a normal and superconducting region are adjacent (being separated by a narrow insulating interface) to each other in the $x$-$y$ plane, as depicted in Fig. \ref{fig:andreev}.
			Such a setup permits an application of the (extension of \cite{stm_slab}) Blonder–Tinkham–Klapwijk (BTK) tunneling theory, where unlike the $c$-axis STM tunneling discussed in Sec. \ref{sec:stm}, the quantum-mechanical boundary-value problem approach of BTK theory is valid for in-plane (i.e. two-dimensional) tunneling phenomena.
			
			\begin{figure}
				\centering
				\includegraphics[scale=0.5]{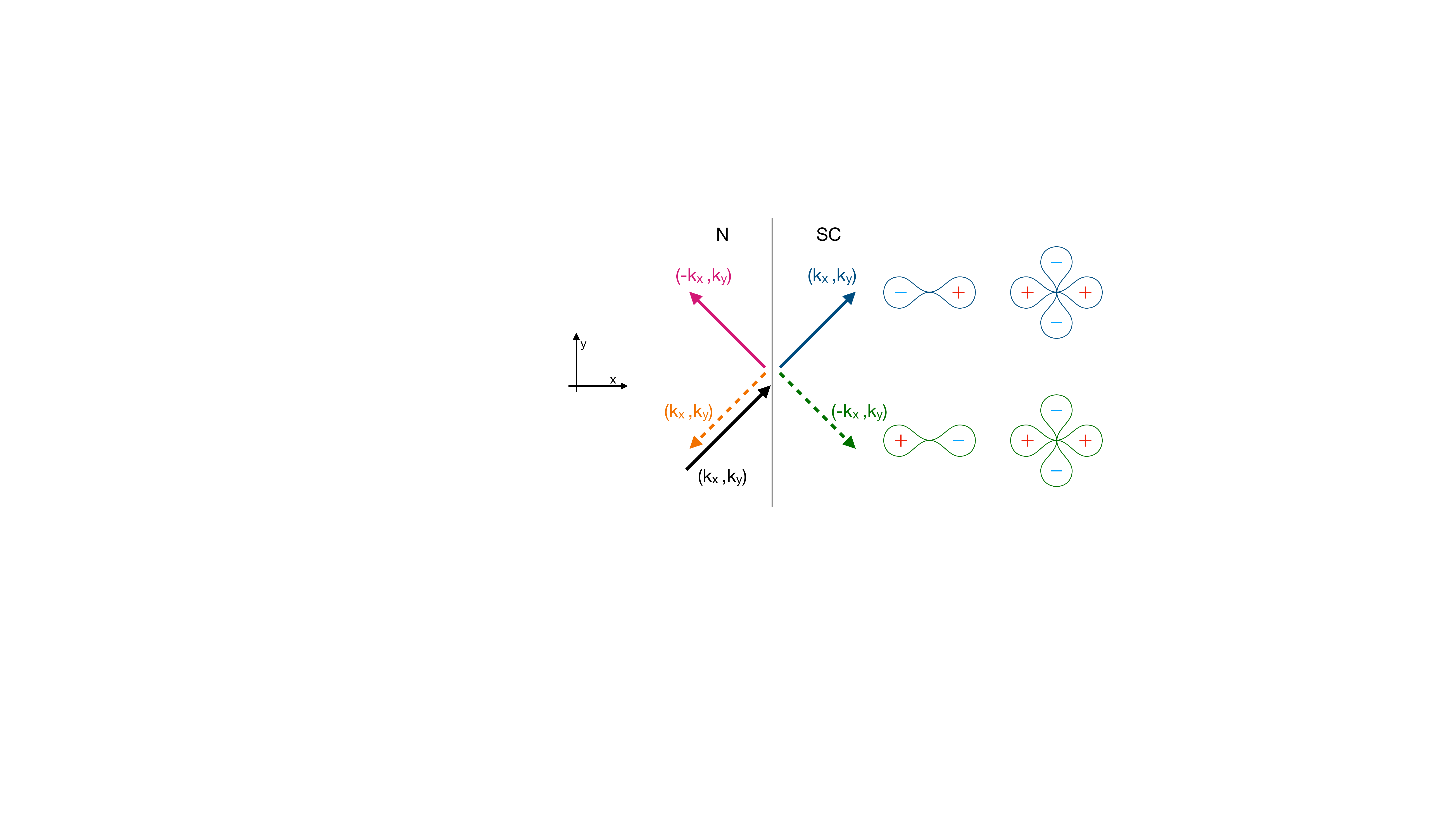}
				\caption{\label{fig_andreev} Schematic of tunneling and reflection of electron on a normal metal (N) and superconductor (SC) interface [100], with the electron and hole velocities depicted by solid and dashed lines, respectively (adapted from \cite{stm_slab}).
					The incident electron (solid black arrow) can be normal reflected as an electron (solid pink arrow), or Andreev reflected as a hole (dashed orange arrow), or transmitted as an electron quasi-particle (EQP) in the SC (solid blue arrow) or transmitted as a hole quasi-particle (HQP) in the SC (dashed green arrow). The corresponding momenta for the electrons and holes are presented beside each arrow, where for electrons (holes) the velocity and momenta are (anti-)parallel. The momenta parallel to the interface ($y$-direction) is conserved.
					Due to the differing momenta for the EQP and HQP, they experience differing superconducting pairing potentials (depicted for strain selected orientation $\vt_n=0$) of this interface.}
			\end{figure}
			Adapting the approach proposed in Ref. \cite{Kashiwaya_spin_triplet_tunneling}, electrons incident on a SN interface aligned along the [100] axis with momentum ($k_x, k_y$) are either: (i) normal reflected as electrons with momentum ($-k_x, k_y$), (ii) Andreev reflected as a hole with momentum ($k_x, k_y$), (iii) transmitted into the superconductor as an electron-quasiparticle with momentum ($k_x, k_y$), or (iv) transmitted into the superconductor as a hole-quasiparticle with momentum ($-k_x, k_y$).
			We note that the interface preserves translation symmetry along the $y$-direction.
			We present in Fig. \ref{fig_andreev} a schematic of the transmission and reflection of an incident electron.
			Due to the electron and hole quasiparticles in the superconductors being realized at different momenta, they experience pairing order parameters/potentials $\Delta_+ = \Delta(k_x, k_y)$ and $\Delta_- = \Delta(-k_x, k_y)$, respectively.
			For the simple s-wave order parameter, these potentials are identical. 
			
			The unnormalized tunneling conductance is given by $\sigma_S = \int_{\phi} \sum_{\alpha} \mathcal{C}_{\alpha} \sigma_{S,\alpha}$, where $\alpha$ denotes the spin-valley flavor of the electrons in TBG (e.g. $K \uparrow$ and $K' \downarrow$ in the case of anti-parallel SVL), $\mathcal{C}_{\alpha}$ denotes the fraction of the electrons with flavor $\alpha$ contributing to the tunneling conductance, and an integral over the azimuthal angle ($\phi$) is explicitly written. 
			Using the normal ($b_{\mu \alpha}$)and Andreev ($a_{\mu \alpha}$) coefficients between flavors $\mu, \alpha$, the conductance (in units of $\frac{e^2}{h}$) from a single flavour conductance can be recast into,
			\begin{align}
				\sigma_{S,\alpha} & = 1 + \sum_\mu \left( | a_{\mu, \alpha} |^2 - | b_{\mu, \alpha} |^2 \right), %
			\end{align}
			where $\mu$ runs over the aforementioned SVL subspace degrees of freedom (e.g. over $K \uparrow$ and $K' \downarrow$ in the case of anti-parallel SVL).
			Indeed, within the framework of BTK theory (in particular the quasi-classical approximated Bogoliubov-de-Gennes equations), the tunneling between different electron ($\alpha$) and hole ($\mu$) quasiparticles is permitted in the presence of a non-vanishing order parameter, $(\hat{\De}_\bfk)^{\alpha \mu}$ i.e. pairs of fermion flavors inter-linked via the associated pairing order parameter. 
			Each contribution to the conductance of a single flavor $\alpha$ is of the form \cite{stm_slab,Kashiwaya_spin_triplet_tunneling},
			\begin{align}
				\sigma_{S,\alpha} = \sigma_N \frac{1+ \sigma_N | \Gamma_+ ^{\mu \alpha}   | ^2 + (\sigma_N -1) | \Gamma_+ ^{\mu \alpha} \Gamma_- ^{\mu \alpha} | ^2 } {| 1 + (\sigma_N -1) \Gamma_+ ^{\mu \alpha} \Gamma_- ^{\mu \alpha} e^{i (\varphi_- - \varphi_+)} |^2},
			\end{align}
			where $\Delta_{\bfk, \pm}^{\mu \alpha} = | \Delta_{\bfk, \pm}^{\mu \alpha} | e^{i \varphi_{\pm}}$ denotes the paring potential experienced by electrons ($+$) and holes ($-$), $\Gamma_{\pm} ^{\mu \alpha} = \frac{(eV) - \Omega_{\pm} ^{\mu \alpha}}{ | \Delta_{\bfk, \pm}^{\mu \alpha} | }$, $\Omega_{\pm} ^{\mu \alpha} = \sqrt{(eV)^2 -  | \Delta_{\bfk, \pm}^{\mu \alpha} |^2}$, $V$ is the applied bias on the normal metal-superconductor interface, and $\sigma_N = \sigma(\lambda, Z)$ is the normal state conductance.
			Here $\lambda =\frac{k_{FS}}{k_{FN}} \frac{\cos{\theta_S}}{\cos{\theta_N}}$ is the ratio of the Fermi wavevector $k_{FN}$ in the normal metal ($k_{FS}$ in the superconductor) and $\theta_N$ ($\theta_S$) is the angle of incidence (transmission) of the incident electron (electron quasiparticle).
			The BTK parameter $Z$ is proportional to the barrier height i.e. $Z \ll 1$ for a highly transparent interface (low barrier), and $Z \gg 1$ for a weakly transparent interface (high barrier).
			We note that for a given incoming $\alpha$ SVL state, we have the corresponding pairing function $\Delta_{\bfk, \pm}^{\mu \alpha}$ that relates to the transmitted/reflected SVL state $\mu$.
			In the limit of low transmission (i.e. a high barrier interface, $\sigma_N \rightarrow 0$), the conductance contribution acquires an appreciable contribution in the zero energy limit ($E\rightarrow 0$) only when $ \varphi_- - \varphi_+ = -1$, i.e. when the pairing potential is of opposite sign for the electron and hole-quasiparticles.
			We note that due to tunneling conductance dependence on the amount of singlet and triplet pairings as well as the contribution of each fermion flavor, to generically obtain a ZBC it is sufficient for such a pole in any of the $\alpha, \mu$ combinations.
			As alluded to above, for isotropic $s$- wave order parameters, $\phi_- = \phi_+$ and as such $s$-wave superconductors do not exhibit an appreciable ZBC. 
			
			For the $d$- and $p$-wave order parameters considered here, the requirement to obtain an appreciable ZBC requires the associated strain angle $\gamma$ to be $\pm \frac{\pi}{4}$ and $(0, \pm \pi)$ for $d$- and $p$-waves respectively; and a vanishing ZBC occurs for $(0, \pm \frac{\pi}{2})$ and $ \pm \frac{\pi}{2}$ for $d$- and $p$-waves respectively.
			Indeed varying the angular strain offset is equivalent to varying the interface orientation, and as such 
			performing a zero bias-conductance measurement with varying interface orientations (for instance from [100] to [$\overline{1}$00]) would indicate a $d$-wave ($p$-wave) superconductor with the observance of two (one) sharp peaks over the range of interface orientations (i.e. rotated over a range of 180 degrees).

			\section{Josephson experiments} \label{app:josephson} 
			
			In this appendix we will discuss various types of Josephson experiments, which are a standard way of revealing phase-sensitive information about the pairing symmetry. 
			
			\ss{In-plane junctions}
			
			Focusing on a single lead whose junction has unit normal vector $\bfn = (\cos(\t_n),\sin(\t_n))$, the tunneling Hamiltonian is
			\be \label{htunnel_inplane} H_{tun} = \sum_{\bfG_n} \int_{\bfk,\bfk'} c^\da_{S;\bfk,\tau,\s}T_{n,\bfk,\bfk';\tau,\tau';s, s'}(\bfn) c_{T;\bfK_{\tau'}+\bfG_n +\bfk', s'},\ee 
			where $c_S$ ($c_T$) destroys electrons in the TBG sample ($s$-wave lead), and all repeated valley and spin indices are summed. Here $\bfk,\bfk'$ run over the TBG BZ, while the sum over $\bfG_n$ is over (monolayer graphene and moire) reciprocal lattice vectors, with $\bfK_{\pm}$ the momenta of the monolayer graphene $K$ ($K'$) points. To account for SVL, the tunneling matrix $T$ is taken to conserve spin, and is proportional to the projector $\mcp$. 
			
			The tunneling matrix must be invariant under rotations, invariant under time reversal, and be proportional to the projector $\mcp$. We will ignore the dependence of the tunneling matrix on the momentum of the $s$-wave lead (due to the rotation invariance of the $s$-wave gap), and will assume that any induced SOC at the junction is negligible. We then parametrize the tunneling matrix as 
			\be T_{\bfk} = (t_0 + t_2 (\bfk\cdot\bfn)^2)\mcp + \cdots,\ee 
			where $\cdots$ are higher order in $\bfk\cdot\bfn$. In the following we will assume that the momentum non-conserving $\bfn$-independent part of $T_\bfk$ is largest, and will work to linear order in $t_2/t_0$. 
			The free energy of the Josephson junction coming from $2e$ tunneling events is then
			\be F_J = \Re \sum_\o \int_{\bfk}\Tr[ \mcf^S_{\o,\bfk}  (\mcf^L_{\o})^* ](t_0 + t_2 k^2\cos(\t_\bfk - \t_n))^2,\ee
			where $\mcf^{S/L}$ are the anomalous Green's functions of the TBG sample and $s$-wave lead, respectively.  $\mcf^S_{\o,\bfk}$ is (see appendix \ref{app:details})
			\bea 	\mcf^S_{\o,\bfk} & = e^{i\phi^S}  \De^S_\bfk \(\tau^- \frac{|\eta'\ran\lan \eta |}{D_{-\bfk}} - p \tau^+ \frac{|\eta\ran\lan \eta'|}{D_{\bfk}}\),\eea 
			where $\De^{S}_\bfk$ is real and $D_\bfk \equiv \o^2 - (\xi_\bfk^2 + |\De^S_\bfk|^2)$, with $\xi_\bfk$ the dispersion in the $K$ valley as before. $\mcf^L$ on the other hand is simply proportional to $is^y$ and the gap $\De^L e^{i\phi^L}$ of the $s$-wave lead, and carries no important momentum dependence. We then have 
			\bea F_J &= \Re \sum_\o \int_{\bfk,\bfk'}  e^{i(\phi^S-\phi^L)} \De^L\De^S_\bfk  \mcd^p_\bfk(t_0 + t_2 k^2\cos(\t_\bfk - \t_n))^2,\eea  
			where we have defined the functions 
			\be \mcd^p_\bfk \equiv D_{-\bfk}\inv + p D_\bfk\inv.\ee  

			Consider first the $d$-wave case, where the orbital parity is $p=1$. Since $\mcd_\bfk^+$ is invariant under $C_6$ rotations, and as the gap is $\De^S_\bfk = \De^S\cos(2\t_\bfk - \g)$ which contains only angular momentum $l=\pm2$ harmonics, the integration over $\t_\bfk$ produces (to leading order in $t_2/t_0$)
			\be F^{l=2}_{J} =  t_0t_2 B_2 \cos(\phi^S-\phi^L)\cos(2\t_n-2\g),\ee  %
			where $B_2$ is an unimportant constant proportional to $\De^S\De^L$ and the zeroth angular harmonic of $\mcd^+_\bfk$ (the next leading harmonic is the sixth, which produces a subleading term proportional to $t_2^2\cos(4\t_n-\g)$). 
			
			Now consider the $p$-wave case, so that $p=-1$. The lowest nonzero harmonic of $\mcd^-_\bfk$ has angular momentum 3, 
			and so with $\De^S_\bfk = \De^S \cos(\t_\bfk - \g)$ we obtain 
			\be F_{J}^{l=1} = t_0t_2 B_1 \cos(\phi^S-\phi^L)   \cos(2\t_n+\g),\ee 
			where $B_1$ is yet another unimportant constant proportional to $\De^S\De^L$ and the third angular momentum of $\mcd^-_\bfk$.
			Note that the Josepshon current $I_{2e}$ is generically nonzero for both $p$ and $d$-wave pairing.  
			
			Now consider a setup where a loop of a conventional $s$-wave SC is attached to TBG at two leads, with normal vectors $\bfn_a$, $a=1,2$. Let the lead order parameter at the two junctions be $\De^L e^{i\phi^L_a}$. 
			The combined free energy for both Josephson junctions is then
			\be F^l_J = t_0t_2 B_l\sum_{a=1,2} \cos(\phi^S-\phi^L_a) \cos(2\t_{n_a}-pl\g).\ee
			The important thing to note here is that the dependence on $\t_{n_a}$ is the {\it same} for both $l=1$ and $l=2$ --- therefore within the context of the approximations made above, we {\it cannot} use these types of phase-sensitive experiments to distinguish $p$ from $d$-wave gaps. Note however that in the $p$-wave case, the nonvanishing of $I_{2e}$ owes its existence entirely to the fact that $\mcd^-_\bfk\neq0$, i.e. to the difference in the dispersions in the $K$, $K'$ valleys (in a conventional SC without valley indices this effect would not arise, and the $p$-wave case would yield $I_{2e}=0$ in the absence of SOC). Intuitively, the $C_3$ anisotropy of the $K$-valley dispersion and the fact that $C_3$ only possesses a single 2d irrep combine to prevent this and similar experiments from distinguishing $p$ and $d$ pairing. We can however use SQUID experiments to distinguish $p$ and $d$ from $s$: in the usual corner SQUID geometry, where $\t_{n_1} = \t_{n_2} + \pi/2$, an $s$-wave gap has a minimum in the free energy for $\phi^L_1 -\phi^L_2 \in \twp \zz$, while $p$ and $d$-wave gaps have minima for $\phi^L_1 - \phi^L_2 \in \pi + 2\pi\zz$. 
			
			\ss{$c$-axis junctions}

			We now discuss situations where a Josephson junction is formed along the $c$-axis, with a probe SC either directly stacked on top of the TBG sample, or embedded on a superconducting STM tip. In what follows we will assume that the probe is a singlet SC
			with either $s$ or $d$-wave pairing. 
			
			If the tunneling matrix $T$ is completely independent of momentum, the contribution of $2e$ tunneling events to the Josephson current will vanish as long as the TBG order parameter $\De^S_\t$ is nodal and has a single harmonic ($\int_\t \De^S_\t = 0$). While the total current will be rendered nonzero by $4e$ Josephson tunneling events, this case can be distinguished by a doubling of the Josephson frequency, $d\phi/dt = 4V$ \cite{pals1977josephson} (and regardless, the current in this case is likely to be quite small). Thus in this limit we can distinguish between $s$-wave and order parameters with nonzero angular momentum $l$, but cannot distinguish between different values of $l$. 
			
			Consider then the limit in which the tunneling is momentum conserving to a good approximation, with the dominant tunneling events being those containing a relatively small number of Umklapp scatterings. This means that we can effectively model a TBG electron in valley $\tau$ as only coupling to electrons in the probe SC with momenta $\bfK_{\tau,a} + \bfk'$ with $k' < \L$, where $\L \ll |K|$ is some cutoff much less than the size of the monolayer graphene BZ, and where $a=1,2,3$ labels the locations of the monolayer graphene $\bfK_\tau$ points when projected into the BZ of the probe SC (with the probe Fermi surface assumed for simplicity to lie fairly close to all of the $\bfK_{\tau,a}$).

			Calculations along the lines of those performed for in-plane junctions then yield 
			\bea F_J &= \Re \sum_\o \int_{\bfk,\bfk'}  e^{i(\phi^S-\phi^P)}\sfT_{\bfk,\bfk'} \De^P\De^S_\bfk  \( \frac{\lan \eta | s^y|\eta'\ran}{D_{-\bfk}} - p \frac{\lan \eta'|s^y |\eta\ran}{D_\bfk}\),\eea  
			where $\sfT$ is a function proportional the square of the bare tunneling strength and we have taken the effective probe order parameter felt by the TBG electrons as $s^y\De^Pe^{i\phi^P}$, where the momentum-dependent quantity $\De^P$ is an average of the probe order parameter over the projection of the monolayer $\bfK_\tau$ points into the probe BZ. For the $S^z \neq 0$ triplet state where $\k\eta \propto \k{\eta'}$ (which we have already argued is disfavored by transport experiments) we have $\lan \eta | s^y | \eta'\ran = 0$, and the charge-$2e$ Josephson current vanishes, $I_{2e}=0$.\footnote{$I_{2e}$ also vanishes for an $S^z=0$ triplet SC whose polarization occurs in valley space, as in that case $F_J\propto \Tr[s^ys^x] = 0$).} On the other hand, for the anti-parallel SVL state we have advocated for, we have 
			\bea  \label{caxisfj} F_J &= \Re \sum_\o \int_{\bfk,\bfk'}  e^{i(\phi^S-\phi^P)}\sfT_{\bfk,\bfk'} \De^P\De^S_\bfk  \mcd^p_\bfk,\eea  
			where $\mcd^p_\bfk \equiv D_{-\bfk}\inv + p D_\bfk\inv$. 
			Due to the $C_3$ symmetry of the $K$-valley dispersion, $\mcd^+_\bfk$ is invariant under $C_6$ rotations, while $\mcd^-_\bfk$ is only invariant under $C_3$ rotations. In any case, since the tunneling $\sfT_{\bfk,\bfk'}$ presumably does not break $C_6$ rotations, the integral over $\bfk$ will lead to $I_{2e}=0$, {\it regardless} of the orbital angular momentum of the probe and TBG superconductors (as long as the latter is nonzero). This is essentially due to the fact that the TBG electrons tunnel only into a small region of the tip BZ near the $\bfK,\bfK'$ points; this prevents e.g. the angular anisotropy of a $d$-wave gap in the probe SC from compensating that of a $d$-wave gap in the TBG sample. Therefore in all of the $c$-axis scenarios we have considered, $I_{2e}=0$. An observation of a sizable $I_{2e}\neq0$ in such a $c$-axis tunneling experiment would then point to an $s$-wave order parameter, and force us to re-examine our priors about the gap being nodal.

			\ss{Stacked bilayers} 
			
			Finally, consider a heterostructure consisting of {\it two} vertically-stacked TBG superconductors, with the top SC being able to be rotated relative to the bottom SC by an arbitrary angle $\vp$. In this case, the Josephson free energy is (assuming a tunneling matrix $T_{\bfk,\bfk'}= T \d_{\bfk,\bfk'}$ for simplicity) 
			\be F_J^l(\vp) \propto T^2 \De^t\De^b \cos(\phi^t-\phi^b) \cos(l[\g^t - \g^b +\vp])\ee
			where the order parameters on the top / bottom TBG layers are 
			\be \De^{t/b}_\t = \De^{t/b} e^{i\phi^{t/b}} \cos(l[\t + \g^{t/b} \pm \vp/2]).\ee  
			For $d$-wave pairing, $I_{2e}(\vp)$ would therefore vanish at four locations as $\vp$ is varied from $0$ to $\twp$, while for $p$-wave pairing $I_{2e}(\vp)$ would only vanish twice. Note that as in the context of the Andreev bound state experiment discussed in the main text, the ability to sample multiple different values of $\vp$ using the {\it same} two TBG superconductors is necessary, since $\g^t,\g^b$ are fixed by non-universal details in a given device, and as such the relative orientation between the nodes on the two layers is not known a priori. 
			
			\section{Tunneling conductance} \label{app:stm} 
			
			In this appendix we provide a discussion of the tunneling conductance $dI/dV$ from a normal tip into a TBG superconductor, both the simpler weak-tunneling limit relevant for standard STM studies, as well as the more involved strong-tunneling limit relevant for point-contact spectroscopy.

			\subsection{Weak tunneling limit: STM current}
			
			First consider the regime in which the tunneling strength is weak. In this regime the tunneling conductance $dI/dV$ is simply proportional to the local DOS in the SC, which is computed from $-\frac1\twp  \Im\int_\bfk \Tr[\mcg(\o+i0^+,\bfk) (1+\l^z)]$ as
			\bea\label{sc_dos} \r_S(\o) & =  \frac1\pi\sum_{s=\pm1}\int \frac{d\t}\twp \, m_\t \, \ct((\o+s\d_\t)^2 - |\De_\t|^2) \frac{|\o + s\d_\t|}{\sqrt{(\o + s\d_\t)^2 - |\De_\t|^2}},\eea
			where $m_\t$ is the (angle-resolved) effective mass in the $K$ valley, and $\d_\t$ is the (angle-resolved) depairing energy induced by the in-plane field. Due to the $C_3$ symmetry of the $K$ valley dispersion, we will parametrize $m_\t$ as 
			\be \label{mpara} m_\t = m(1+\z\cos(3\t)),\ee 
			and $\d_\t$ as 
			\be \label{deltadef} \d_\t = B_\prl \(\d_0 \cos(\t - \t_B) + \d_3 \cos(2\t + \t_B)\),\ee 
			where we have kept only the two lowest harmonics in the angle $\t$ (the subscripts on the constants $\d_0,\d_3$ denote the angular momentum channel). 
			x	The magnitudes of $\d_0,\d_3$ are set by the scale 
			\be e B_\prl v_D c_0/2 \approx 0.15 (B_\prl/1{\rm T}) \, {\rm meV},\ee 
			where $v_D$ is the monolayer Dirac velocity and $c_0$ is the interlayer separation. This has a numeric value of $\approx 0.15  B_\prl {\rm meV / T}$, similar to the depairing energies calculated in \cite{cao2020nematicity}.

			From \eqref{sc_dos}, we see that the magnetic field acts to create effective angle-dependent chemical potentials of strength $\pm\d_\t$ for the quasiparticles. For a nodal order parameter, this affects the DOS in two main ways: it splits the single coherence peak at $\o=\De$ into multiple sub-peaks, and fills in the DOS near zero bias. 
			
			Let us examine what happens near zero bias. 
			We denote the locations of the nodal points as $\{\bfk_n\} = \{k_{F,n}\t_n\}$, near which the order parameter may be expanded as 
			\be \label{delta_expansion} |\De_{\t_n + \t}|^2 \approx (k_{F,n} v_{\De,n} \t)^2 \ee 
			for some nodal velocities $v_{\De,n}$. At frequencies $|\o\pm \d_{\t_n}| \ll  k_{F,n}v_{\De,n}$, the density of states at zero bias is then 
			\be \label{zbc} \r_S(0) \approx \frac1\pi \sum_n \frac{ |\d_{\t_n} |}{v_{F,n}v_{\De,n}}.\ee 
			
			\begin{figure}
				\includegraphics[width=.4\textwidth]{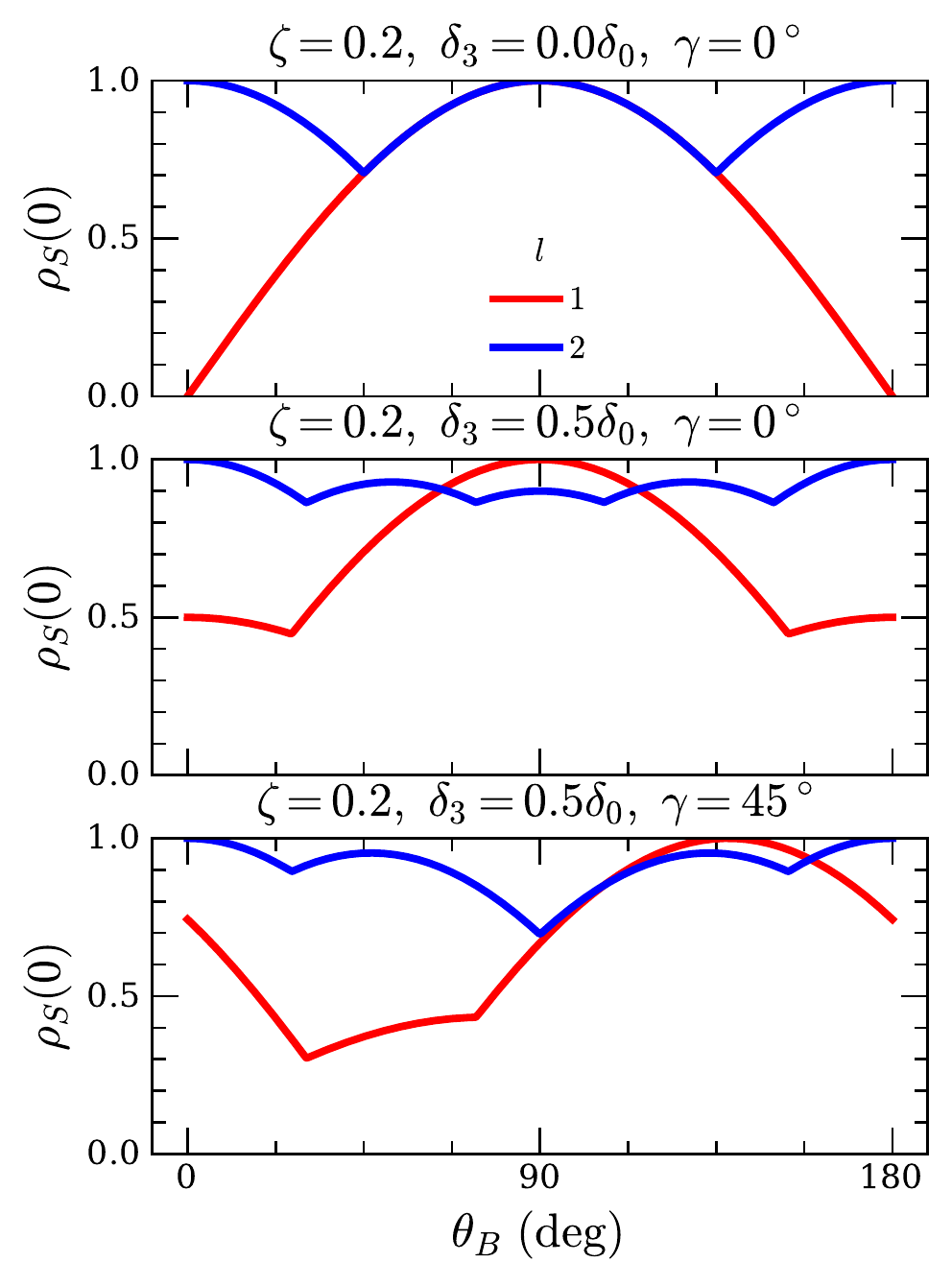}
				\caption{\label{fig:rhoszero} Density of states at zero frequency as a function of $\t_B$ for several different values of $\d_3,\g$, and with $\z=0.2$. In each plot we have normalized $\r_S(0)$ by its maximum value. The red curves $(l=1)$ are $p$-wave, and the blue curves $(l=2)$ are $d$-wave. }
			\end{figure}
			
			Since $\r_S(0)$ does not vanish at nonzero $\bfB_\prl = 0$ unless $\d_{\t_n}$ happens to vanish for all $n$, a small magnetic field will fill in the nodes by an amount which is linear in the field strength.
			While the dependence of $\r_S(0)$ on $\t_B$ depends on the set of nodal angles $\{\t_n\}$, it does not generically provide a robust way of distinguishing $p$ and $d$-wave pairing. If either $\d_3 = 0$, the two cases can be distinguished: in the $p$-wave case $\r_S(0)$ always has one maxima and vanishes one as $\t_B$ is varied from $0$ to $\pi$, while in the $d$-wave case $\r_S(0)$ always has two maxima and never vanishes. 
			However when $\d_3$ is nonzero, the $p$ and $d$-wave cases become harder to distinguish for a similar reason as in the above discussion of SQUID experiments; this is illustrated in figure \ref{fig:rhoszero}. While in a given experiment we may get lucky, with $\d_3$ happening to be small or $\g$ happening to be close to $0,\pi$ --- in which case we could (barring smearing effects from temperature / quasiparticle broadening) distinguish $p$ from $d$ --- in general the situation will likely not be very clear-cut.

			\ss{Strong tunneling limit: point-contact spectroscopy}
			
			We now discuss the limit where the tunneling strength is strong, which is relevant for describing point-contact spectroscopy experiments such as those of Ref. \cite{oh2021evidence}. 
			The usual approach \cite{stm_slab} for determining $dI/dV$ in the strong tunneling limit, which was used for the fitting analysis of \cite{oh2021evidence}, is to assume that the STM tip can be approximated by a large slab of normal metal separated from the SC by a delta function barrier. One then determines the current by solving a single-particle quantum mechanics problem, which is done by matching the wavefunctions and their normal derivatives at the interface. To us the conceptual correctness of such an approach is not so obvious in the present setting, where the STM tip may not a priori be well-approximated as a large slab of normal metal and where due to the two dimensional nature of the SC, there is no way to match normal derivatives of the wavefunctions. For these reasons we employ a more direct Keldysh approach to calculate the tunneling current. An expression for $dI/dV$ in this approach was in fact already stated in \cite{cuevas1996hamiltonian}; in the following we simply elaborate on the derivation of this formula and generalize it in a way which allows us to easily account for both spin and valley degrees of freedom. 
			
			We model the system through the Hamiltonian 
			\be H = \sum_{\bfq\tau s} C_{\bfq\tau s}^\da  \l^z \ep_{\bfK_\tau+\bfq} C_{\bfq\tau s} + \int_\bfk \psi^\da_\bfk H^{qp}_\bfk  \psi_\bfk + \sum_{\bfq\bfk\s s' s''\tau\tau'\tau''} (C^\da_{\bfq\tau s} \l^z  t_{\bfq\bfk\tau\tau's s'} \mcp_{\tau'\tau'' s' s''}\psi_{\bfk\tau'' s''} + h.c.),\ee 
			with $H^{qb}_\bfk$ as in \eqref{ham}, $C_{\bfq\tau s} = (c_{\bfq\tau s}, \, c^\da_{\bfq\tau s})^T$ with $c_{\bfq\tau s}$ destroying an electron on the tip at energy level $\ep_{\bfK_\tau+\bfq}$ (with $\bfK_\tau$ the projection of the monolayer graphene $\bfK_\tau$ point in the BZ of the tip) and with spin $s$, and where contraction in Nambu space is implied. As in the main text, $\mcp$ denotes the projector on to the SVL subspace. 
			
			The matrix $t_{\bfq\bfk\tau\tau'ss'}$ parametrizes the hopping between the tip and the superconductor, which for concreteness we will mostly take to conserve spin, $t_{\a\tau\bfk ss'}\propto \d_{ss'}$. We will consider two limiting regimes for the momentum dependence. The first occurs when the tip is to be modeled as a quantum dot: this is appropriate when the presence of the STM strongly breaks translation invariance in the plane, and when the only electrons which enter the sample are those near the (effectively zero-dimensional) STM tip. In this case the hopping matrix is independent of momentum and valley indices, and $t_{\bfq\bfk\tau\tau's s'} = t s^0$ (where we will take $t \in \rr$ for simplicity). The second regime occurs when the tip is to be modeled as a (three-dimensional) Fermi liquid, with the presence of the tip not essentially breaking translation symmetry in the plane, and with the hopping preserving in-plane momentum (for notational simplicity we have omitted sums over the out-of-plane component of the tip momenta; we will assume the tip Fermi surface is spherical and has a projection into the plane which encompasses the $\bfK_\tau$ points, meaning that for all $\tau$ and $\bfq$, $c_{\bfq\tau s}$ always destroys an electron somewhere on the tip Fermi surface). In this case $t_{\bfq\bfk\tau\tau's s'} = t s^0 \tau^0 \d_{\bfk,\bfq+\bfK_{\tau}}$.

			In the following we will use the Keldysh Green's function formalism to compute the current, which requires introducing the usual suspects 
			\be G^<_{IJ}(\o) = i\lan \Psi_I^\da(\o) \Psi_J(\o)\ran,\qq G_{IJ}(\o) = -i \int_t e^{i(\o+i0^+) t} \ct(t) \lan \{ \Psi_I(t),\Psi^\da_J(0) \} \ran,\ee
			where $I,J$ are composite labels labeling the device (either tip $T$ and sample $S$) and the indices conserved by the tunneling matrix (viz. spin and Nambu indices, and possibly momentum), and where the fields $\Psi_I$ are defined by
			\be \Psi_{TI'} = C_{I'}, \qq \Psi_{SI'} = \sum_{i\in NC} \psi_{I',i},\ee 
			where $NC$ is the set of indices which are {\it not} conserved by the tunneling matrix (viz. valley and momentum in the momentum non-conserving limit), and $I'$ is the composite index without the device (tip / sample) part. The device polarization implemented by $\mcp$ can be taken into account by conjugating all Green's functions by $\mcp\oplus \mcp^T$, where the direct sum is in Nambu space. In the following we will let this projection be implicit in the notation, with $G^<, G$ denoting the projected Green's functions. 
			Our notation will always be such that Green's functions without superscripts denote {\it retarded} correlation functions. Advanced correlators are defined by 
			\be G^A_{IJ}(\o) = +i\int_t e^{i(\o-i0^+)}\ct(-t)\lan \{ \Psi_I(t),\Psi_J^\da(0)\}\ran = [G^\da]_{IJ}.\ee 

			Letting $\mu^a$ denote Pauli matrices in device space, and taking the hopping amplitude $t$ to be real for simplicity, the current (following directly from the equation of motion of the rate of change of particle number in the tip) reads 
			\bea 
			\label{current}		I & = i \sum_{I'} \left[ t \langle \Psi_{T I'}^\dag (\tau) \Psi_{S I'} (\tau) \rangle - t \langle \Psi_{S I'}^\dag (\tau) \Psi_{T I'} (\tau) \rangle \right] \nonumber  \\
			&  = \frac12 \Tr[\lambda^z \mu^z \S G^<],
			\eea 
			where $t$ denotes a real hopping matrix element between the tip and sample, the trace includes both a sum over all degrees of freedom, as well as an integral over frequency, and where the self-energy matrix $\S$ is simply 
			\be \S = t\mu^x \l^z.\ee 
			We note that the corresponding $\Sigma^{<} =0$ due to a real tunneling matrix element chosen between the normal tip and superconducting sample.
			Note that to compute the current the lesser Green's function $G^<$ is employed, since the Green's functions with both times on the same half of the contour are equivalent to equilibrium ones. 
			
			To evaluate this trace, we will need to make use of the Dyson equations \cite{mahan_many_body} (suppressing all device indices and frequency dependence)
			\bea \label{gfunc_ids} (1-g\S)G^< & = g^<(1+\S G^\da)   \\ 
			G^< (1-\S g^\da) & = (1+G\S)g^< \\ 
			G^< & = (1+G\S)g^<(1+\S G^\da) \\  G & = \frac1{1-g \S} g = g \frac1{1-\S g},\eea 
			where lowercase letters denote free propagators, which are 
			\bea \label{free_props} g(\o) & =  g_T(\o) \oplus g_S(\o) \\ 
			g^<(\o) & = 2\pi i (\hat\r_Tf_T(\o) \oplus \hat\r_S(\o)f_S(\o))\\ 
			g^>(\o) & = g^<(\o) -\twp i (\hat\r_T(\o) \oplus \hat\r_S(\o))\eea 
			with $g_{\a}(\o)$ the free propagators for the tip and superconductor (which are matrices in Nambu and spin space), the $\oplus$ is in device space, and were the matrices $\hat\r_\a, f_\a$ are defined as 
			\be\hat \r_\a(\o) \equiv -\frac1{2\pi i}  (g_\a(\o) - g_\a(\o)^\da)\ee 
			and 
			\be f_T(\o) \equiv f(\o-\l^z V),\qq f_S(\o) \equiv f(\o)\l^0,\ee 
			with $f(\o)$ the Fermi function.  
			By taking the device off-diagonal components of the top two lines in \eqref{gfunc_ids}, we find  
			\bea G_{ST}^< &=t( G_{SS} \l^z g^<_T + G^<_{SS} \l^z g^\da_T) \\ 
			G_{TS}^< & =t ( g_T\l^z G^<_{SS} + g^<_T \l^z G^\da_{SS}),\eea  
			which gives the Green's functions appearing in the expression for the current in terms of the free tip Green's functions and the full dressed Green's functions of the superconductor (here and in the following, we will slightly abuse notation by writing $G^\da_{\a\b}$ for $[G^\da]_{\a\b}$). Plugging in to \eqref{current}, this yields 
			\bea I & = \frac{t^2}2 \Tr[\l^zg^<_T (G_{SS}-G^\da_{SS}) +\l^z (g^\da_T -g_T)G^<_{SS} ] \\ 
			& = \frac{t^2}2 \Tr[\l^z(
			g^<_T G^>_{SS} - g^>_T G^<_{SS})],\eea
			where we have used the identity $G^<-G^> = G-G^\da$ and the fact that the tip Green's functions, having no anomalous parts, commute with $\l^z$. The $G^{<,>}_{SS}$ can be evaluated with the third line of \eqref{gfunc_ids}, which gives
			\bea G^<_{SS} & = g^<_S + t^2 G_{SS} g^<_T G^\da_{SS} +t( G_{ST} \l^z g^<_S + g_S^< \l^z G^\da_{TS}) + t^2 G_{ST} \l^z g^<_S \l^z G^\da_{TS} \\ 
			&  = g_S^< + t^2(G_{SS}g_T g^<_S + g^<_S g^\da_T G^\da_{SS} + G_{SS}g^<_T G^\da_{SS}) + t^4 G_{SS} g_T g^<_S g^\da_T G^\da_{SS},\eea  
			where we used $G_{ST} = tG_{SS}\l^z g_T, G_{TS}^\da = t g_T^\da \l^z G_{SS}^\da$ in the second line. Making use of \eqref{free_props}, some algebra and the fact that $[\hat\r_T,f_T]=0$ (as the anomalous parts of the tip Green's function vanish) gives 
			\bea \label{curr} I & = \frac{(\twp t)^2}2 \Tr \Big[ \l^z(f_T-f_S)\( \hat\r_T \hat\r_S + t^2\hat\r_T(G_{SS}g_T\hat\r_S + \hat\r_S g_T^\da G_{SS}^\da) + t^4 \hat\r_T G_{SS} g_T \hat\r_S g_T^\da G_{SS}^\da \) +  t^2  \l^z\hat \r_T [f_T,G_{SS}] \hat\r_TG^\da_{SS}  \Big],\eea
			where again the trace includes an integral over frequency.  It now remains only to calculate $G_{SS}$, which can be done easily with the help of the last Dyson equation in \eqref{gfunc_ids}: 
			\be \label{dressed_greens} G_{SS} = (g\inv_{SS} - t^2 g_T)\inv.\ee 
			
			The first group of terms in \eqref{curr} vanishes when $\hat\r_S$ does; these terms are therefore responsible for normal tunneling processes that are only active outside of the superconducting gap. The last term on the other hand does not depend on the superconducting DOS, and can lead to a current even for biases inside the gap; this is therefore the term responsible for the in-gap Andreev current.
			
			The conductance is obtained by differentiating \eqref{curr} with respect to $V$, yielding 
			\bea \label{final_didv} \frac{dI}{dV}& = \frac{(\twp t)^2}2\int_\rr d\o \, f'(\o-V)  \Tr\[ \hat\r_T \hat\r_S + t^2 \hat\r_T \(G_{SS}g_T\hat\r_S + \hat\r_S g_T^\da G_{SS}^\da \) + t^4 \hat \r_T G_{SS}g_T\hat\r_S g_T^\da G_{SS}^\da +2t^2 \hat \r_T G_{SS,a} \hat\r_T G^\da_{SS,a}\](\o),\eea 
			where $G_{SS,a}$ is the anomalous part of $G_{SS}$ (viz. the part of $G_{SS}$ off-diagonal in Nambu indices), and where we have now written out the frequency integral explicitly.

			\sss{Momentum non-conserving tunneling} 
			
			We first consider the limit of `incoherent' tunneling, where the tunneling process does not conserve in-plane momentum. Therefore the tip Green's function is, treating the tip as being in the wide band limit with a constant density of states $\r_T$ and remembering the projection onto the SVL subspace, 
			\be \label{incoh_tip_greens} g_T(\o) = \int_\rr d\xi \, \frac{\r_T}{\o+ i\eta-\l^z \xi} \scp = -\pi i \hat\r_T \l^0\scp\ee
			where 
			\be \scp \equiv  \mcp \oplus \mcp^T,\ee 
			with the $\oplus$ in Nambu space.

			On the other hand, the free Green's function of the SC in the incoherent limit (a matrix in spin and Nambu space) is, reading off from \eqref{qpgreens} (and continuing to let $\k\eta, \k{\eta'}$ be real),
			\bea \label{incoh_gsc} g_S(\o) & = \sum_{\tau,\tau'}\int_\bfk   \[ \frac{\scp(\o  + \l^z (\xi_{\a^z\bfk} - \tau^z \d_{\a^z\bfk})) - \wh \De_\t \l^+ - \wh \De_\t^\da \l^- }{(\o -\a^z\d_{\a^z\bfk} )^2 - E^2_{\a^z\bfk} }\]_{\tau\tau'} \\ 
			& = -\frac1\fpi \int_\t \Big(\sum_{s=\pm1}  \frac{1+s\l^z}2 \( \U_{\t,s} (\o - s\d_{\t_s}) \k\eta\lan\eta| + \U_{\t,-s} (\o+s\d_{\t_{-s}}) \k{\eta'}\lan \eta'| \) \\  
			& \qq - \De_\t \l^+ (\U_{\t,-} \k{\eta'}\lan \eta| - p \U_{\t,+} \k\eta\lan \eta'|) - \De_\t^* \l^- (-p \U_{\t,+} \k{\eta'}\lan \eta| + \U_{\t,-} \k\eta\lan \eta'| ) \Big), \eea 
			where $\a^z \equiv \l^z \tau^z$ and we have defined 
			\be \U_{\t,\pm} \equiv \frac{m_{\t_\pm}}{\sqrt{|\De_\t|^2 - (\o \mp \d_{\t})^2}},\ee 
			with the notation $\t_{+} = \t ,\t_- = \t+\pi$.

			We will now use \eqref{incoh_gsc} to argue that either a) TBG is actually an $s$-wave SC (which for the reasons explained in the main text we are inclined to disfavor) or b) the tunneling experiments of \cite{oh2021evidence} are in fact rather well in the momentum-conserving tunneling regime (which given the experimental setup is perhaps slightly surprising). 
			
			To argue this, all we need to do is note the rather strong zero-bias peak in $dI/dV$ observed at $\bfB=0$ in \cite{oh2021evidence}; we claim that \eqref{incoh_gsc} cannot generically reproduce such a peak. Indeed, setting $\d_\t=0$ (and specializing to the case of AF SVL), we have 
			\bea g_S(\o)  & = -\int \frac{d\t}\twp\, \U_{\t,+}(\o + (\l^+ \De_{\t} - \l^- \De^*_{\t} )is^y).\eea 
			Since $m_\t$ is the mass in the $K$ valley, $C_3$ symmetry imposes $m_{\t} = m_{\t+2\pi /3}$, and so as in \eqref{mpara} we parametrize 
			\be m_\t = m (1 + \z \cos(3\t)),\ee 
			with $|\z|<1$ controlling the extent of the angular anisotropy. Now at $T=0$ the zero bias conductance only depends on $g_S(0)$, which is (taking $\De_\t \in \rr$ wolog) 
			\bea g_S(0) & = \int \frac{d\t}\twp \, m_\t \, {\rm sgn}(\De_\t) \l^y s^y \\ 
			& = m\int \frac{d\t}\twp \, \( \frac{1+p}2 + \z \frac{1-p}2 \cos(3\t)\) {\rm sgn}(\De_\t).\eea 
			This means that the strength of the Andreev reflection signal is proportional to $\z$ if $p=-1$, and is controlled by $m\int d\t \, {\rm sgn}(\De_\t)$ if $p=+1$. On the other hand, the conductance in the $V\ra \infty$ limit is determined only by $\int d\t \, m_\t /\twp  = m$. This leads to the normalized ZBC being depressed from the maximal value of 2 it takes in the s-wave case to something smaller, with the suppression being either by an integral like $\z \int d\t \cos(3\t) {\rm sgn}(\De_\t)$ in the $p=-1$ case, or $\int d\t \, {\rm sgn}(\De_\t)$ in the $p=1$ case.

			\begin{figure*}
				\subfloat{%
					\includegraphics[width=.45\linewidth]{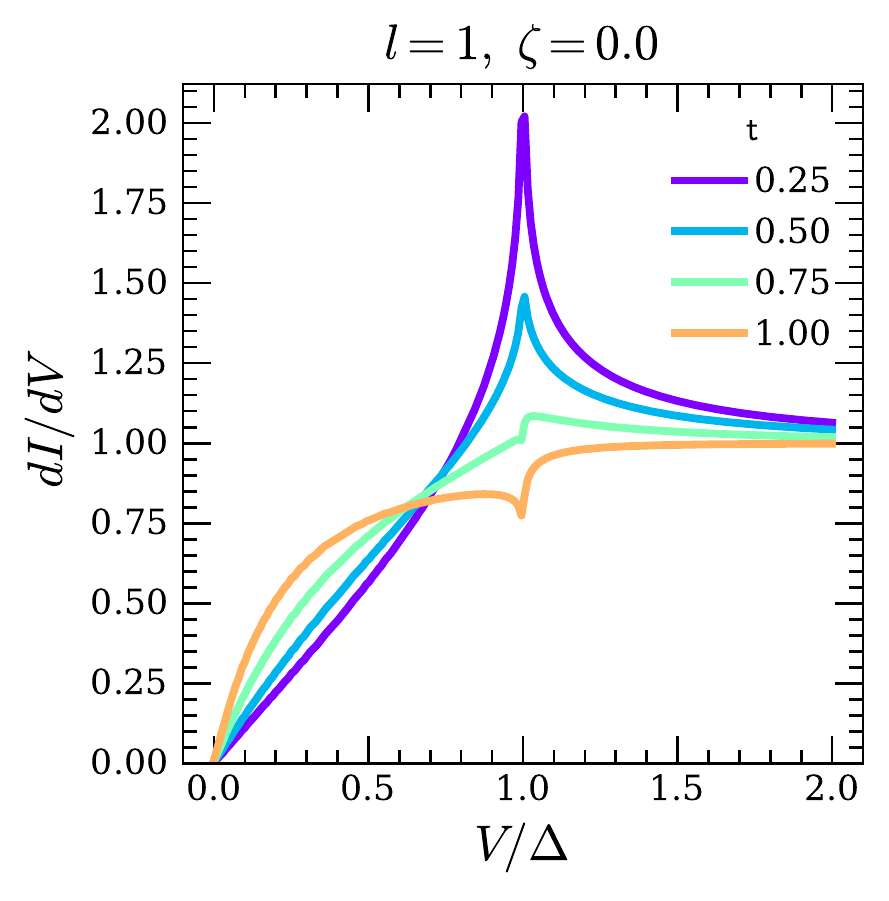}%
				}\hfill
				\subfloat{%
					\includegraphics[width=.45\linewidth]{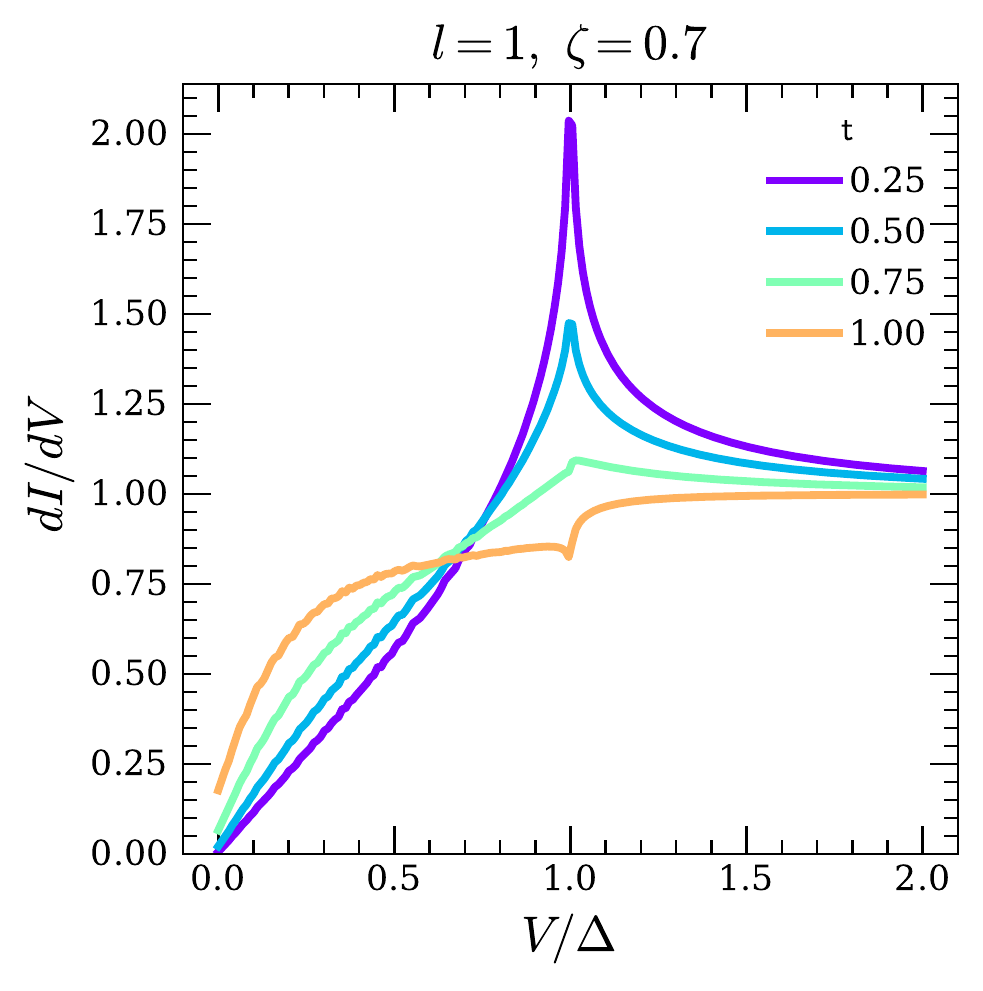}%
				}
				\caption{\label{fig:pwave_incoh} $dI/dV$ curves for $\De_\t = \De \cos(\t)$, normalized to their respective $V\ra\infty$ normal-state values. Here $\sft$ is the dimensionless tunneling strength defined by $\sft \equiv t \sqrt{m\pi\r_T}$. In terms of BTK theory, $\sft = 1$ corresponds to the case of a transparent barrier, $Z=0$ (and $\sft>1$ corresponds to an attractive barrier, with $Z<0$).   }
			\end{figure*}
			
			The net result of these effects is to produce tunneling curves that are in general significantly suppressed near zero bias. For a p-wave order parameter $\De_\t = \De \cos(\t)$, the normalized conductance is shown in figure \ref{fig:pwave_incoh}, for an isotropic effective mass ($\z=0$, left) and a large $C_3$ anisotropic mass ($\z=0.7$, right). The conductance in the two cases looks similar, but the anisotropy $\z\neq0$ is responsible for a (very small) nonzero ZBC in the latter case. 
			
			\begin{figure*}
				\subfloat{%
					\includegraphics[width=.45\linewidth]{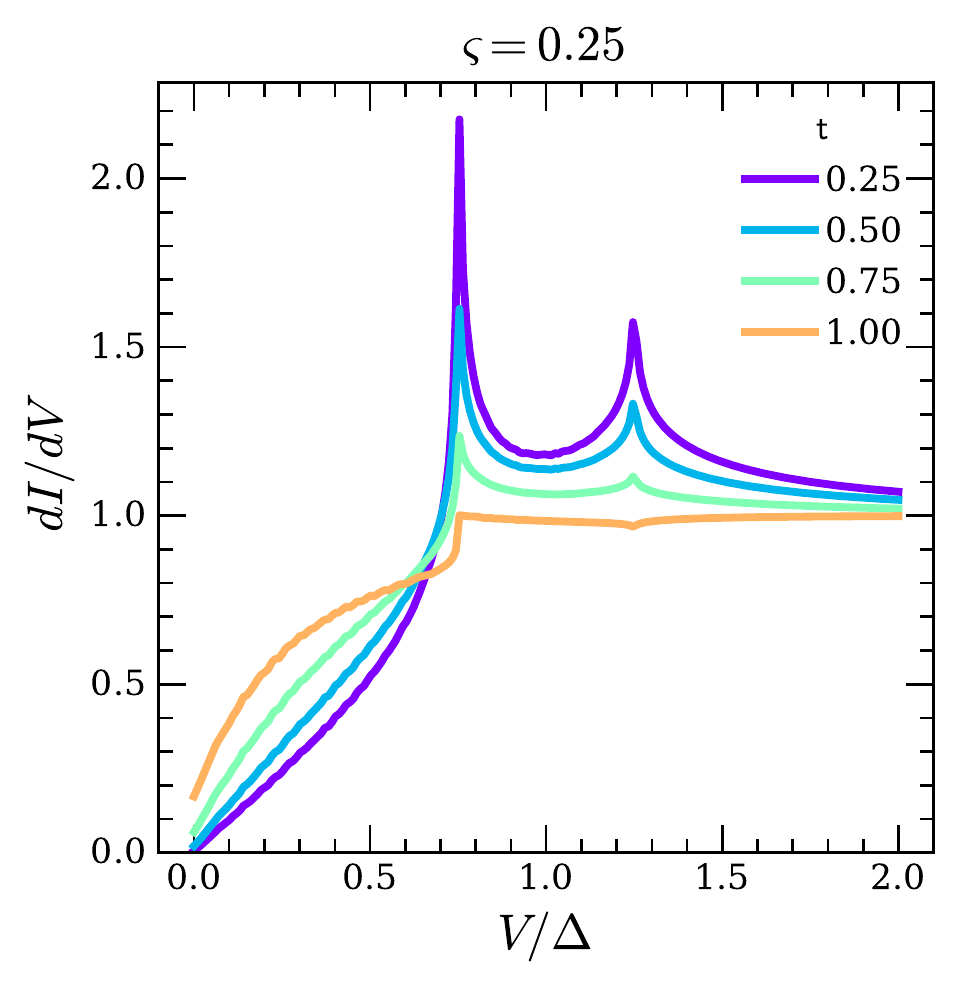}%
				}\hfill
				\subfloat{%
					\includegraphics[width=.45\linewidth]{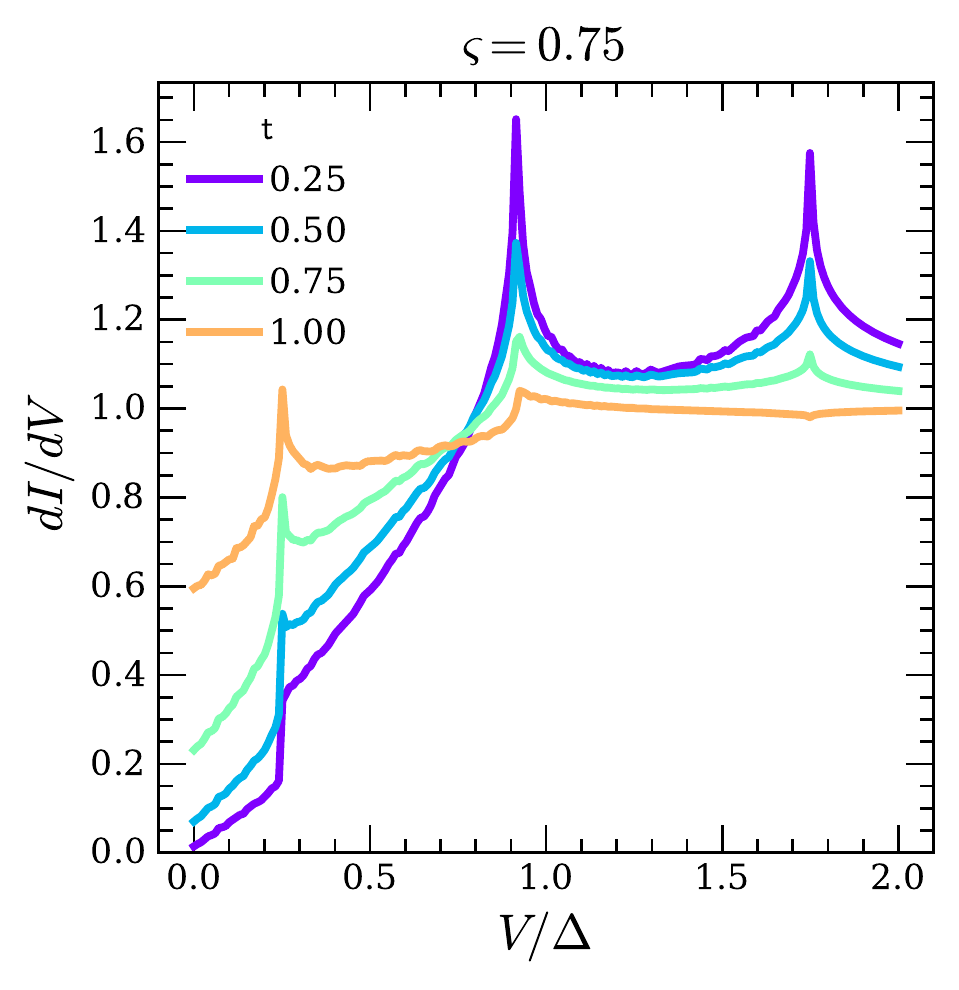}%
				}
				\caption{\label{fig:dwave_incoh} Same as in figure \ref{fig:pwave_incoh}, but with  t$\De_\t = \De(\cos(2\t) + \vs \cos(4\t))$. The multiple singularities are due to the fact that the gap possesses more than one harmonic. }
			\end{figure*}
			
			A nonzero ZBC is obtained in the $p=1$ case only by working with an order parameter for which $\int d\t \, {\rm sgn}(\De_\t) \neq 0$. A simple example of such an order parameter is $\De_\t = \De(\cos(2\t) + \vs \cos(4\t) )$, which has 4 nodes if $|\vs|<1$. 
			Examples of the normalized conductance for $\vs = 0.25, 0.75$ are shown in figure \ref{fig:dwave_incoh}. The nonzero value of $\vs$ allows the ZBC to be finite (though still suppressed relative to its normal state value), but the fact that $\De_\t$ includes multiple harmonics leads to multiple coherence peaks at larger biases, which are not observed in experiment. Even with $\vs = 0.75$, we are clearly a long way away from producing a peak in the ZBC.

			\sss{Momentum conserving tunneling} 
			
			Given the issues with incoherent tunneling we now consider the case of coherent tunneling, where the tunneling between the tip and the sample conserves in-plane momentum in the TBG BZ.\footnote{One could also consider an intermediate case, where the tunneling is incoherent in the TBG BZ but diagonal in valley; this leads to the same problems as in the valley-incoherent case as they originate from the integral over $\t$, rather than the sum over valleys.} To treat this case carefully, we would work with momentum-resolved Green's functions, and include a single integral over momentum in the trace appearing in \eqref{final_didv}. We will instead do something slightly simpler: for a given point on the Fermi surface at angle $\t$, we will take the tunneling to conserve momenta normal to $\wh\bfth$ (along the Fermi surface), but not parallel to $\wh\bfth$ (normal to the Fermi surface). This treatment nevertheless captures the essential way in which coherence affects the conductance, since in this treatment there is no interference between the signs of $\De_\t$ at different angles $\t$.

			\begin{figure}
				\subfloat{%
					\includegraphics[width=.5\linewidth]{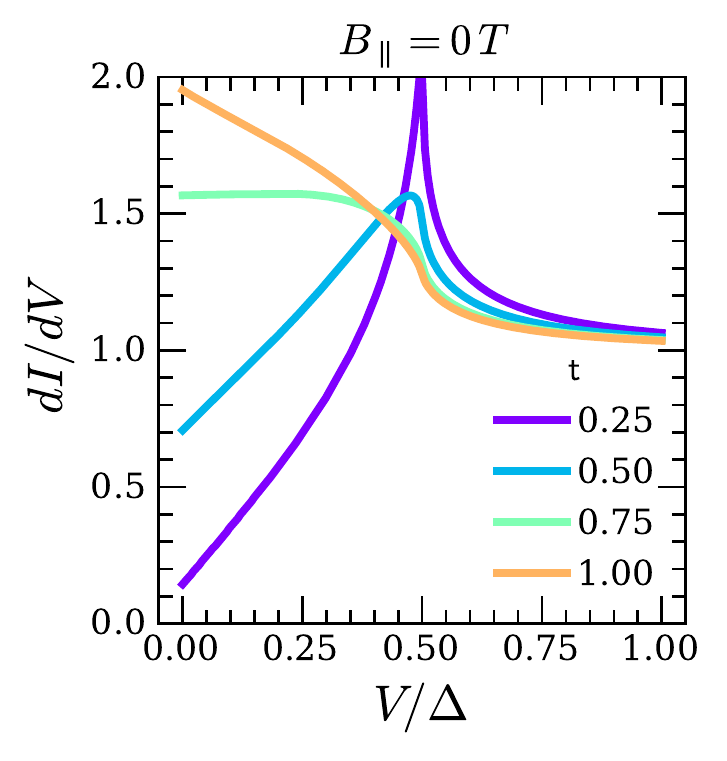}%
				}\hfill
				\subfloat{%
					\includegraphics[width=.5\linewidth]{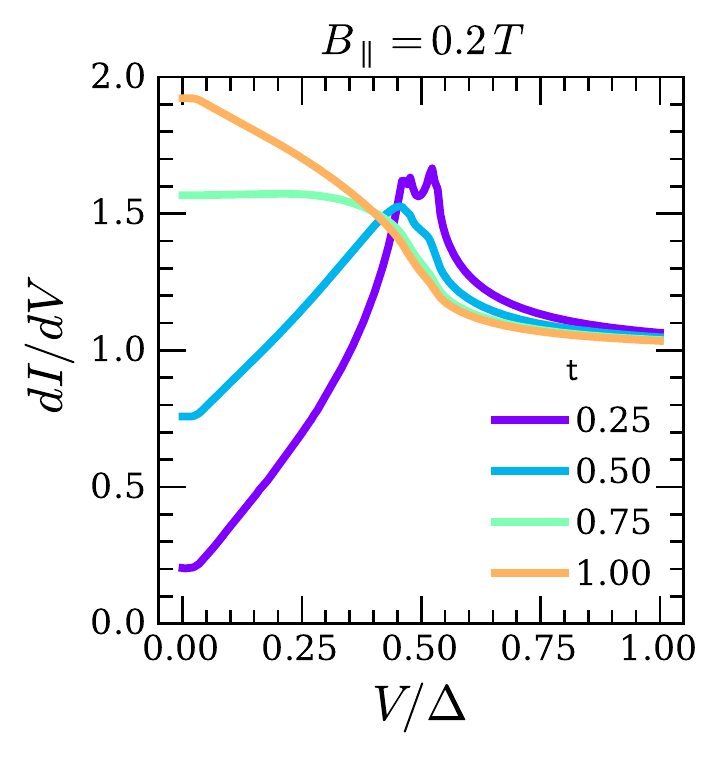}%
				}
				\caption{\label{fig:didv_vs_v} Tunneling conductance in the coherent tunneling model for a $p$-wave gap ($d$-wave is essentially the same), normalized to its normal-state value.  The mass anisotropy parameter is fixed as $\z=1/4$, and $\sft = t \sqrt{m\pi \r_T}$ is the dimensionless tunneling strength as in figure \ref{fig:pwave_incoh}. {\it Left:} zero magnetic field. {\it Right:} Field of magnitude $B_\prl = 0.2 T$ and $\t_B = \pi/4$, $\g=0$, and with the parameters in \eqref{deltadef} fixed at $\d_0 = 4 \d_3 = 0.15$meV/T.} 
			\end{figure}

			In this framework, the appropriate Green's function (a matrix in valley, spin, and Nambu space) to use when computing $dI/dV$ is 
			\bea \label{angular_gs} g_S(\o,\t) & = -\frac1\fpi \Big(\U_{\t,\a^z}(\o-\a^z\d_{\t_{\a^z}})\scp - \l^+ \De_\t \(\tau^-\U_{\t,-}  \k{\eta'}\lan\eta| - p\tau^+\U_{\t,{+}} \k{\eta}\lan{\eta'}|\) - \l^- \De_\t^* \(-p \tau^-\U_{\t,+} |\eta'\ran\lan \eta| + \tau^+\U_{\t,-}|\eta\ran\lan \eta'|\) \Big),  \eea  
			where $\t$ is integrated over as part of the trace in \eqref{final_didv}. 
			The conductance computed in this framework gives results much more in accordance with experiment, an example of which is shown in figure \ref{fig:didv_vs_v}.
			As an in-plane magnetic field modifies the conductance essentially by way of it acting as a chemical potential for quasiparticles, it has the effect of flattening out the conductance curves near zero bias, as well as splitting the coherence peak at low tunneling strength. 
			In the right panel of figure \ref{fig:didv_vs_v} we have taken $B_\prl = 0.2 $T, which in the experiment of \cite{cao2020nematicity} is rather comfortably below the minimum $B_{c,\prl}$ across the majority of the superconducting dome. The $\t_B$-dependence of $dI/dV$ at low bias is essentially the same as that of the $\t_B$-dependence of $\r_S(0)$ studied above.

		\end{widetext}

	\end{document}